\documentclass[twocolumn,bibyear]{aa}
\usepackage[varg]{txfonts}
\usepackage{natbib}
\usepackage{graphicx}
\usepackage{graphicx,epsfig,graphics}
\usepackage{graphics,url}
\usepackage{color}
\bibpunct{(}{)}{;}{a}{}{,} 
\def\nb{n_{b}}        
\def\rnum{n}          
\def\rtheo{n}         
\def\rcore{n_{b}}     
\def\rmcore{\epsilon} 
\def\rmatt{\epsilon}  

\begin{document}

\title{Unified equation of state for neutron stars on a microscopic basis}

\author {B. K. Sharma\inst{1,2}, M. Centelles\inst{1}, 
  X. Vi\~nas\inst{1}, M. Baldo\inst{3}, and G. F. Burgio\inst{3}}

\institute{
              $^1$ Departament d'Estructura i Constituents de la Mat\`eria
              and Institut de Ci\`encies del Cosmos (ICC),
              Facultat de F\'{\i}sica, Universitat de Barcelona,
              Diagonal 645, E-08028 Barcelona, Spain\\
              $^2$ Department of Sciences, Amrita Vishwa Vidyapeetham
              Ettimadai, Coimbatore- 641112, India\\
              $^3$ INFN Sezione di Catania,  Via Santa Sofia 64, 95123 Catania, Italy\\
}

\date{\today}

\abstract{
We derive a new equation of state (EoS) for neutron stars (NS) from the outer
crust to the core based on modern microscopic
calculations using the Argonne $v_{18}$ potential plus three-body forces
computed with the Urbana model. To deal with the inhomogeneous structures of 
matter in the NS crust, we use a recent 
nuclear energy density functional that is directly based on the same 
microscopic calculations, and which is able to reproduce the ground-state 
properties of nuclei along the periodic table. The EoS of the outer crust 
requires the masses of neutron-rich nuclei, which are obtained through 
Hartree-Fock-Bogoliubov calculations with the new functional when they are 
unknown experimentally. To compute the inner crust, Thomas-Fermi calculations 
in Wigner-Seitz cells are performed with the same functional. Existence of 
nuclear pasta is predicted in a range of average baryon densities between 
$\simeq$0.067 fm$^{-3}$ and $\simeq$0.0825 fm$^{-3}$, where the transition to 
the core takes place. The NS core is computed from the new nuclear EoS  
assuming non-exotic constituents (core of $npe\mu$ matter). In each 
region of the star, we discuss the comparison of the new EoS with previous EoSs 
for the complete NS structure,  widely used in astrophysical calculations. The new 
microscopically derived EoS fulfills at the same time a NS maximum mass of 
2~$M_\odot$ with a radius of 10 km, and a 1.5~$M_\odot$ NS with a radius of 11.6 
km.
}


\titlerunning{Unified equation of state for neutron stars  .....}
\authorrunning{B. K. Sharma et al.} 

\maketitle


\section{Introduction}

Neutron stars (NS) harbor unique conditions and phenomena that challenge the 
physical theories of matter. Beneath a thin stellar atmosphere, a NS interior 
consists of three main regions, namely, an outer crust, an inner crust, and a 
core, each one featuring a different physics 
\citep{shapiro,haenselbook,livingrev}. The core is the internal region at 
densities larger than 1.5$\times 10^{14}$ g/cm$^3$, where matter forms a 
homogeneous liquid composed of neutrons plus a certain fraction of protons, 
electrons, and muons that maintain the system in $\beta$ equilibrium. Deep in 
the core, at still higher densities, strange baryons and even deconfined quarks 
may appear \citep{shapiro,haenselbook}. Moving from the core to the exterior, 
density and pressure decrease. When the density becomes lower than approximately 
1.5$\times 10^{14}$ g/cm$^3$, matter inhomogeneities set in. The positive 
charges concentrate in individual clusters of charge $Z$ and form a solid 
lattice to minimize the Coulomb repulsion among them. The lattice is embedded in 
a gas of free neutrons and a background of electrons such that the whole system 
is charge neutral. This region of the star is called the inner crust, where the 
nuclear structures may adopt non-spherical shapes (generically referred to as 
nuclear pasta) in order to minimize their energy 
\citep{bbp71,ravenhall83,lorenz93,oyamatsu93}. 
At lower densities, neutrons are finally confined within the nuclear clusters 
and matter is made of a lattice of neutron-rich nuclei permeated by a degenerate 
electron gas. This region is known as the outer crust \citep{bps71} and extends,
from inside to outside, from a neutron drip density of about 4$\times
10^{11}$~g/cm$^3$ to a density of about $10^{4}$~g/cm$^3$. Most of the mass and 
size of a NS are accounted for by its core.  Although it is only a small fraction 
of the star mass and radius, the crust plays an important role in 
various observed astrophysical phenomena such as pulsar glitches, quasiperiodic 
oscillations in soft gamma-ray repeaters (SGR), and thermal relaxation in soft X-ray 
transients (SXT) \citep{haenselbook,livingrev,piro2005,strowatts06,steiner09,sotani12,
newton13,piekarewicz2014}, which depend on the departure of the star from the 
picture of a homogeneous fluid. Recent studies suggest that the existence of 
nuclear pasta layers in the NS crust may limit the rotational speed of pulsars 
and may be a possible origin of the lack of X-ray pulsars with long spin 
periods~\citep{pons13,horowitzpasta15}.

The equation of state (EoS) of neutron-rich matter is a basic input
needed to compute most properties of NSs. A large body of experimental data on 
nuclei, heavy ion collisions, and astrophysical observations has been gathered 
and used along the years to constrain the nuclear EoS and to understand the 
structure and properties of NS. Unfortunately, a direct link of measurements 
and observations with the underlying EoS is very difficult and a proper 
interpretation of the data necessarily needs some theoretical inputs. To reduce 
the uncertainty on these types of analyses, it is helpful to develop a 
microscopic theory of nuclear matter based on a sound many-body scheme and 
well-controlled basic interactions among nucleons. To this end, it is of 
particular relevance to have a unified theory able to describe on a microscopic 
level the complete structure of NS from the outer crust to the core.

There are just a few EoSs devised and used to describe the whole NS 
within a unified framework. It is usually assumed that the NS crust has 
the structure of a regular lattice that is treated in the Wigner-Seitz (WS) 
approximation. A partially phenomenological approach was developed by Lattimer 
and Swesty (LS) \citep{ls91}. The inner crust was computed using the 
compressible liquid drop model (CLDM) introduced by Baym, Bethe, and Pethick 
\citep{bbp71} to take into account the effect of the dripped neutrons. In the LS 
model the EoS is derived from a Skyrme nuclear effective force. There are 
different versions of this EOS \citep{ls91,LSweb}, each one having a different 
incompressibility. Another EOS was developed by Shen et al.\
\citep{shen98a,shen98b,Shenweb} based on a nuclear relativistic mean field (RMF) 
model. The crust was described in the Thomas-Fermi (TF) scheme using the 
variational method with trial profiles for the nucleon densities. The LS
and Shen EoSs are widely used in astrophysical calculations for both neutron 
stars and supernova simulations due to their numerical simplicity and the large 
range of tabulated densities and temperatures.

Douchin and Haensel (DH) \citep{douchin01} formulated a unified EoS for NS on 
the basis of the SLy4 Skyrme nuclear effective force \citep{chabanat98}, where 
some parameters of the Skyrme interaction were adjusted to reproduce the 
Wiringa et al.\ calculation of neutron matter \citep{wiringa88} above saturation 
density. Hence, the DH EoS contains certain microscopic input. In the DH model 
the inner crust was treated in the CLDM approach. More recently, unified EoSs 
for NS have been derived by the Brussels-Montreal group 
\citep{chamel11,pearson12,fantina13,potekhin13}. They are based on the BSk family 
of Skyrme nuclear effective forces \citep{goriely10}. 
Each force is fitted to the known masses of nuclei and adjusted among other 
constraints to reproduce a different microscopic EoS of neutron matter with 
different stiffness at high density. The inner crust is treated in the extended 
Thomas-Fermi approach with trial nucleon density profiles including 
perturbatively shell corrections for protons via the Strutinsky integral method. 
Analytical fits of these neutron-star EoSs have been constructed in order to 
facilitate their inclusion in astrophysical simulations \citep{potekhin13}. 
Quantal Hartree calculations for the NS crust have been systematically 
performed by \citep{Hor3,Hor4}. This approach uses a virial 
expansion at low density and a RMF effective interaction at intermediate and 
high densities, and the EoS of the whole NS has been tabulated for different 
RMF parameter sets. Also recently, a complete EoS for supernova matter has been 
developed within the statistical model \citep{hempel10}. We shall adopt here the 
EoS of the BSk21 model \citep{chamel11,pearson12,fantina13,potekhin13,goriely10} 
as a representative example of contemporary EoS for the complete NS structure, 
and a comparison with the other EoSs of the BSk family 
\citep{chamel11,pearson12,fantina13,potekhin13} and the RMF family 
\citep{Hor3,Hor4} shall be left for future study.

Our aim in this paper is to obtain a unified EoS for neutron stars based on a
microscopic many-body theory. The nuclear EoS of the model is derived in the 
Brueckner-Hartree-Fock (BHF) approach from the bare nucleon-nucleon (NN) 
interaction in free space (Argonne $v_{18}$ potential \citep{v18}) with 
inclusion of three-body forces (TBF) among nucleons. In the current state of 
the art of the Brueckner approach, the TBFs are reduced to a density-dependent 
two-body force by averaging over the third nucleon in the medium \citep{bbb}. We 
employ TBFs based on the Urbana model, consisting of an attractive term from 
two-pion exchange and a repulsive phenomenological central term, to reproduce 
the saturation point \citep{schia85,bbb,polls,Taranto:2013gya}. The corresponding 
nuclear EoS for symmetric and asymmetric nuclear matter fulfills several 
requirements imposed both by heavy ion collisions and astrophysical 
observations~\citep{Taranto:2013gya}.

Recently the connection between two-body and three-body forces within the
meson-nucleon theory of the nuclear interaction has been extensively discussed and
developed in \citep{tbfmic1,tbfmic2,zhli1,zhli2}. At present the theoretical
status of microscopically derived TBFs is still incipient, however a tentative
approach has been proposed using the same meson-exchange parameters as the
underlying NN potential. Results have been obtained with the Argonne $v_{18}$, the
Bonn B, and the Nijmegen 93 potentials \citep{zhli1,zhli2}. More recently the
chiral expansion theory to the nucleon interaction has been extensively developed
\citep{wein68,wein90, wein91, wein92,
Ent,Valder,Leut,Epel,Ca1,Ca2,Hebel,Diri,Hebel2,Rios,Mac,MacCor}.
This approach is based on a deeper level of the strong interaction theory, where
the QCD chiral symmetry is explicitly exploited in a low-momentum expansion of the
multi-nucleon interaction processes. In this approach multi-nucleon interactions
arise naturally and a hierarchy of the different orders can be established.
Despite some ambiguity in the parametrization of the force \citep{Mac} and some
difficulty in the treatment of many-body systems \citep{Rupak}, the method has
marked a great progress in the microscopic theory of nuclear systems. Indeed
it turns out \citep{MacCor,Ekst} that within this class of interactions a 
compatible treatment of few-nucleon systems and nuclear matter is possible.
However, this class of NN and TBF (or multi-body) interactions is devised on the
basis of a low-momentum expansion, where the momentum cut-off is fixed 
essentially by the mass of the $\rho$-meson.  As such they cannot be used at 
density well above saturation, where we are also going to test the proposed 
EoS. In any case, the strength of TBFs is model dependent. In particular, the 
role of TBFs appears to be marginal in the quark-meson model of the NN 
interaction \citep{fuku2014}. 

A many-body calculation of the inhomogeneous structures of a NS crust is 
currently out of reach at the level of the BHF approach that we can apply for 
the homogeneous matter of the core. In an attempt to maximize the use of the
same microscopic theory for the description of the complete \mbox{stellar}
structure, we employ in the crust calculations the recently developed BCPM 
(Barcelona-Catania-Paris-Madrid) nuclear energy density functional 
\citep{baldo08,baldo10,baldo13}. The BCPM functional has been obtained from the 
ab initio BHF calculations in nuclear matter within an approximation inspired 
by the Kohn-Sham formulation of density functional theory \citep{ks65}. Instead 
of starting from a certain effective interaction, the BCPM functional is built 
up with a bulk part obtained directly from the BHF results in symmetric and
neutron matter via the local density approximation. It is supplemented by 
a phenomenological surface part, which is absent in nuclear matter, together 
with the Coulomb, spin-orbit, and pairing contributions. This energy density 
functional constructed upon the BHF calculations has a reduced set of four 
adjustable parameters in total and describes the ground-state properties of 
finite nuclei similarly successfully as the Skyrme and Gogny 
forces~\citep{baldo08,baldo10,baldo13}.

We model the NS crust in the WS approximation. To compute the 
outer crust, we take the masses of neutron-rich nuclei from experiment if they 
are measured, and perform deformed Hartree-Fock-Bogoliubov (HFB) calculations 
with the BCPM energy density functional when the masses are unknown. To 
describe the inner crust, we perform self-consistent Thomas-Fermi calculations
with the BCPM functional in different periodic configurations (spheres, 
cylinders, slabs, cylindrical holes, and spherical bubbles). In this 
calculation of the inner crust, by construction of the BCPM functional, the 
low-density neutron gas and the bulk matter of the high-density nuclear 
structures are not only consistent with each other, but they both are given by 
the microscopic BHF calculation. We find that nuclear pasta shapes are the 
energetically most favourable configurations between a density of $\simeq 0.067$ 
fm$^{-3}$ (or 1.13$\times$10$^{14}$ g/cm$^3$) and the transition to the core, 
though the energy differences with the spherical solution are small. The NS 
core is assumed to be composed of $npe\mu$ matter. We compute the EoS of the 
core using the nuclear EoS from the BHF calculations. In the past years, the BHF 
approach has been extended in order to include the hyperon degrees of freedom 
\citep{hypns1,hypns2}, which play an important role in the study of neutron-star 
matter. However, in this work we are mainly interested on the properties of the 
nucleonic EoS, and therefore we do not consider cores with hyperons or other 
exotic components. In each region of the star we critically compare the new NS 
EoS with the results from various semi-microscopic approaches mentioned above,
where the crust and the core were calculated within the same theoretical scheme.
Lastly, we compute the mass-radius relation of neutron stars using the unified 
EoS for the crust and core that has been derived from the microscopic BHF 
calculations in nuclear matter. The predicted maximum mass and radii are 
compatible with the recent astrophysical observations and analyses. We reported 
some preliminary results about the new EoS recently in \citep{baldo13a}.

In Sec.~{\ref{sec:micro}} we summarize the microscopic nuclear input to our 
calculations and the derivation of the BCPM energy density functional for 
nuclei. We devote Sec.~{\ref{sec:outer}} to the calculations of the outer 
crust. In Sec.~{\ref{sec:inncrust}} we introduce the Thomas-Fermi formalism for 
the inner crust with the BCPM functional, and in Sec.~{\ref{sec:innres}} we 
discuss the results in the inner crust, including the pasta shapes. In 
Sec.~{\ref{sec:core}} we describe the calculation of the EoS in the core and 
obtain the mass-radius relation of neutron stars. We conclude with a summary 
and outlook in Sec.~{\ref{sec:sum}}.

\section{Microscopic input and energy density functional for nuclei}
\label{sec:micro}

As the microscopic BHF EoS of nuclear matter underlies our formulation,
we start by summarizing how the EoS of nuclear matter is
obtained in the BHF method and how it has been used to construct a 
suitable energy density functional for the description of finite nuclei.

The nuclear EoS of the model is derived in the framework of the Brueckner-Bethe-Goldstone
theory, which is based on a linked cluster expansion of the energy per nucleon of
nuclear matter (see \citep{book}, chapter 1 and references therein).
The basic ingredient in this many-body approach is the Brueckner reaction 
matrix $G$, which is the solution of the Bethe-Goldstone equation 
\begin{equation}
 G[\rtheo;\omega] = v + \sum_{k_a, k_b} v \frac
 { \left|k_a k_b\right\rangle Q \left\langle k_a k_b\right|}
  { \omega - e(k_a) - e(k_b) } G[\rtheo;\omega] \:, 
\label{e:bg}
\end{equation}
where $v$ is the bare NN interaction, $\rtheo$ is the nucleon 
number density, and $\omega$ is the starting energy.
The propagation of intermediate baryon pairs is determined by the Pauli operator 
$Q$ and the single-particle energy $e(k)$, given by
\begin{equation}
 e(k) = e(k;\rtheo) = \frac{k^2}{2m} + U(k;\rtheo) \:.
\label{e:en}
\end{equation}
We note that we assume natural units $\hbar= c= 1$ throughout the paper. The BHF 
approximation for the single-particle potential $U(k;n)$ using the continuous 
choice is
\begin{equation}
 U(k;\rtheo) = \sum _{k' \le k_F} 
 \big\langle k k'\big| G[\rtheo; e(k)+e(k')] \big|k k'\big\rangle_a \:,
\label{e:ukr}
\end{equation}
where the matrix element is antisymmetrized, as indicated by the ``$a$''
subscript.
Due to the occurrence of $U(k)$ in Eq.~(\ref{e:en}), the coupled system of
equations (\ref{e:bg}) to (\ref{e:ukr}) must be solved in a self-consistent manner
for several Fermi momenta of the particles involved. The corresponding BHF 
energy per nucleon is
\begin{equation}
 \frac{E}{A} =  \frac{3}{5} \frac{k_F^2}{2m} + \frac{1}{2\rtheo}  
 \sum_{k,k' \le k_F}  
 \big\langle k k'\big|G[\rtheo; e(k)+e(k')]\big|k k'\big\rangle_a. 
\end{equation}
In this scheme, the only input quantity we need is the bare NN interaction $v$ in 
the Bethe-Goldstone equation (\ref{e:bg}). 
The nuclear EoS can be calculated with good accuracy in the Brueckner 
two-hole-line approximation with the continuous choice for the 
single-particle potential, since the results in this scheme are 
quite close to the calculations which include also the three-hole-line
contribution \citep{song1,song2,baldobur01}. 
In the present work, we use the Argonne $v_{18}$ potential
\citep{v18} as the two-nucleon interaction. The dependence on the NN interaction,
as well as a comparison with other many-body approaches, has been systematically investigated
in \citep{polls}.

One of the well-known results of several studies, which lasted for about half a
century, is that non-relativistic calculations, based on purely two-body
interactions, fail to reproduce the correct saturation point of symmetric nuclear
matter, and three-body forces among nucleons are needed to correct this
deficiency. In our approach the TBF is reduced to a density-dependent two-body
force by averaging over the position of the third particle, assuming that the
probability of having two particles at a given distance is reduced 
according to the two-body correlation function \citep{bbb}.
In this work we use a phenomenological approach based on 
the so-called Urbana model, which consists of an attractive term due to two-pion exchange
with excitation of an intermediate $\Delta$ resonance, and a repulsive 
phenomenological central term \citep{schia85}. Those TBFs produce a shift of the
saturation point (the minimum) of about $+1$ MeV in energy. This adjustment was
obtained by tuning the two parameters contained in the TBFs, and was performed to
get an optimal saturation point (for details see \citep{bbb}).
The calculated nuclear EoS conforms to several constraints required by 
the phenomenology of heavy ion collisions and astrophysical 
observational data~\citep{Taranto:2013gya}.

For computational purposes, an educated polynomial fit is performed on top of the
microscopic calculation of the nuclear EoS including a fine tuning of the two
parameters of the three-body forces such that the saturation point be $E/A=-16$
MeV at a density $\rtheo_0=0.16$ fm$^{-3}$.
The interpolating polynomials for symmetric nuclear matter and pure
neutron matter are written as
\begin{eqnarray}
P_s(\rtheo) = \left( \frac{E}{A} \right)_{SNM} &=&  \sum_{k=1}^5 a_k \Big(\frac{\rtheo}{\rtheo_0}\Big)^k,
\nonumber \\
P_n(\rtheo) = \left( \frac{E}{A} \right)_{PNM}  &=& \sum_{k=1}^5 b_k \Big(\frac{\rtheo}{\rnum_{0n}}\Big)^k,
\label{poly}
\end{eqnarray}
where $\rnum_{0n}=0.155$ fm$^{-3}$.
The values of the coefficients of the interpolating polynomials
(BCP09 version \citep{baldo10} of the parameters) are given in Table~\ref{Table1}.
This fit is valid up to density around $\rtheo=0.4$ fm$^{-3}$, and it is used
only for convenience. For higher density, as the ones occurring in NS cores,
we use a direct numerical interpolation of the computed EoS, 
using the same procedure and functional form as in reference \citep{bs2010}
for a set of accurate BHF calculations. We found that the polynomial form of the EOS and the high density fit join smoothly in the interval 
$\rm 0.3 - 0.4~ fm^{-3 }$, for both the energy density and the pressure of the $\beta$ stable matter. The overall EOS so obtained will be reported in the corresponding Tables below and used for NS calculations. 
The properties of infinite nuclear matter at saturation are 
collected in Table~\ref{Table2}, and we see that their values agree very well  with the known empirical 
values. The symmetry energy and the corresponding slope parameter $L$  are two important quantities closely related to various
properties of neutron stars and to the thickness of the neutron skin of nuclei
\citep{horowitz01}. It can be noticed that the predicted values for $E_{\rm
sym}(\rtheo_0)$ and $L$ lie within the recent constraints derived from the
analysis of different astrophysical observations
\citep{newton13,steiner10,lattimer13,hebeler13} and nuclear
experiments \citep{chen10,tsang12,vinas14}.

\begin{table}[tb]
\caption{\label{Table1} Coefficients of the polynomial fits $E/A$  of the EoS of
symmetric matter and neutron matter, see Eq.~(\ref{poly}).}
\centering
\begin{tabular}{rrrrr}
\hline\hline
$k$ &  & $a_k$ (MeV) &  & $b_k$ (MeV) \\
\hline
 1  &  & $-$73.292028 & & $-$34.972615 \\
 2  &  & 49.964912     &  &  22.182307 \\
 3  &  & $-$18.037608 & & $-$7.151756 \\
 4  &  &  3.486179     &   &  1.790874 \\
 5  &  & $-$0.243552 &  & $-$0.169591 \\
\hline
\end{tabular}
\end{table}

\begin{table*}[bt]
\caption{\label{Table2} Properties of nuclear matter at saturation predicted 
by the EoS of this work in comparison with the empirical ranges 
\citep{newton13,steiner10,lattimer13,hebeler13,chen10,tsang12,vinas14}.
From left to right, the quantities are the energy per particle, density,
incompressibility, symmetry energy, and slope parameter of the symmetry energy: 
$L = 3\rtheo_0 \frac{\partial E_{\rm sym}(\rtheo)}{\partial \rtheo} \big|_{\rtheo_0}$.
The effective nucleon mass in the present model is $m^*=m$~\citep{baldo08}.
The nuclear matter properties of other EoSs that will be considered in the 
sections of results are also shown.
}
\centering
\begin{tabular}{lc c c c c}
\hline\hline
& $E/A$ & $\rtheo_0$ & $K$ & $E_{\rm sym}(\rtheo_0)$ & $L$ \\
& (MeV) & (fm$^{-3}$) & (MeV) & (MeV) & (MeV) \\
\hline
This work & $-$16.00 & 0.160 & 213.75 & 31.92 & 52.96 \\
Ska \citep{ls91}
          & $-$16.00 & 0.155 & 263.18 & 32.91 & 74.62 \\
SLy4 \citep{chabanat98}
          & $-$15.97 & 0.160 & 229.93 & 32.00 & 45.96 \\
BSk21 \citep{goriely10}
          & $-$16.05 & 0.158 & 245.8 \, & 30.0 \, & 46.6 \, \\
TM1 \citep{shen98b}
          & $-$16.26 & 0.145 & 281.14 & 36.89 & 110.79 \, \\
Empirical & $-16.0\pm0.1$ & $0.16\pm0.01$ & $220\pm30$ & $31\pm2$ &
$\sim$60$\pm25$ \\
\hline
\end{tabular}
\end{table*}

The BHF result can be directly employed for the calculations of the 
NS liquid core, where the nuclei have dissolved into their constituents, 
protons and neutrons. However, the description of finite nuclei and nuclear 
structures of a NS crust is not manageable on a fully microscopic level. 
Indeed, the only known tractable framework to solve the nuclear many-body 
problem in finite nuclei across the nuclear chart is provided by density 
functional theory. In order to describe finite nuclei, the BCPM
energy density functional was built \citep{baldo08,baldo10,baldo13} based on
the same microscopic BHF calculations presented before.
The BCPM functional is obtained within an approximation inspired by the 
Kohn-Sham method \citep{ks65}. In this approach the energy is split 
into two parts, the first one being the uncorrelated kinetic energy, while
the second one contains both the potential energy and the correlated part of the
kinetic energy. An auxiliary set of $A$ orthonormal orbitals $\varphi_i( {\bf
r},\sigma,q )$ is introduced (where $A$ is the nucleon number and $\sigma$ and 
$q$ are spin and isospin indices), allowing one to write the one-body density 
as if it were obtained from a Slater determinant as
$\rtheo( {\bf r} ) = \sum_{i,\sigma,q} | \varphi_i( {\bf r},\sigma,q ) |^2$, 
and the uncorrelated kinetic energy as
\begin{equation}
T_0 = \frac{1}{2m} \sum_{i,\sigma,q} 
\int |\nabla \varphi_i( {\bf r},\sigma,q ) |^2 d{\bf r} .
\end{equation}
To deal with the unknown form of the correlated part of the energy density
functional, a strategy often followed in atomic and molecular physics is to use 
accurate theoretical calculations performed in a uniform system which finally are 
parametrized in terms of the one-body density. We use a similar approach,
and we apply it to the nuclear many-body problem to obtain the BCPM functional.
For this purpose we split the interacting nuclear part of the energy ($E_{N}$) into 
bulk and surface contributions (i.e. $E_{N}= E_{N}^{\rm bulk} + E_{N}^{\rm 
surf}$). We obtain the bulk contribution $E_{N}^{\rm bulk}$ directly from the 
ab initio BHF calculations of the uniform nuclear matter system by a local 
density approximation. Namely, in our approach $E_{N}^{\rm bulk}$ depends 
locally on the nucleon density $\rtheo=\rnum_n + \rnum_p$ and the asymmetry 
parameter $\beta=(\rnum_n-\rnum_p)/(\rnum_n+\rnum_p)$ and reads as
\begin{equation}
E_{N}^{\rm bulk}[\rnum_p,\rnum_n] =
\int \big[P_s(\rtheo) (1 - \beta^2) + P_n (\rtheo)\beta^2 \big] \rtheo d{\bf r} , 
\label{bulk}
\end{equation}
where the polynomials $P_s(\rtheo)$ and $P_n(\rtheo)$ in powers of the one-body 
density have been introduced in Eq.~(\ref{poly}).

In addition to the bulk part, also surface, Coulomb, and spin-orbit contributions 
which are absent in nuclear matter are necessary in the interacting functional to 
properly describe finite nuclei. We make the simplest possible choice for the 
surface part of the functional by adopting the form
\begin{eqnarray}
E_{N}^{\rm surf}[\rnum_p,\rnum_n] &=&
\frac{1}{2}\sum_{q,q'}\int
\int \rnum_{q}({\bf r})
v_{q,q'}({\bf r}-{\bf r'}) \rnum_{q'}({\bf r'} )d{\bf r} d{\bf r'}
\nonumber \\
&-& \frac{1}{2}\sum_{q,q'}
\int {\rnum_{q}({\bf r})} \rnum_{q'}({\bf r}) d{\bf r}
\int v_{q,q'}({\bf r'}) d{\bf r'},
\label{surf}
\end{eqnarray}
where $q=n,p$ for neutrons and protons. The second term in (\ref{surf}) is
subtracted to not contaminate the bulk part determined from the microscopic
nuclear matter calculation. For the finite range form factors we take a
Gaussian shape $v_{q,q'}(r)=V_{q,q'}e^{-r^2/{\alpha}^2}$, with three adjustable
parameters: the range $\alpha$, the strength $V_{p,p}=V_{n,n}\equiv V_L$ for
like nucleons, and the strength $V_{n,p}=V_{p,n}\equiv V_U$ for unlike nucleons.
The Coulomb contribution to the functional is the sum of a direct term and
a Slater exchange term computed from the proton density:
\begin{equation}
E_{\rm coul}= \frac{e^2}{2}\!
\int\!\!\!\int\! \frac{\rnum_p({\bf r})\rnum_p({\bf r'})} {|{\bf r}-{\bf r'}|}
d{\bf r}d{\bf r'}
- \frac{3 e^2}{4}\Big(\frac{3}{\pi}\Big)^{1/3}\!\!\!
\int\! {\rnum_p^{4/3}({\bf r})} d{\bf r} .
\label{bcp3}
\end{equation}
As in Skyrme and Gogny forces, the spin-orbit term is a zero-range interaction
$v_{\rm s.o.}= i W_0({\mbox{\boldmath$\sigma$}}_i + {\mbox{\boldmath$\sigma$}}_j) 
\times [ {\bf k'} \times
\delta({\bf r}_i - {\bf r}_j) {\bf k}]$, whose contribution to the
energy reads
\begin{equation}
E_{\rm s.o.}= - \frac{W_0}{2}\!\int\! [ \rtheo({\bf r}) \nabla
{\bf J}({\bf r})+\rnum_n({\bf r}) \nabla {\bf J}_n({\bf r}) + \rnum_p({\bf r})
\nabla {\bf J}_p({\bf r})] d{\bf r} ,
\label{bcp4} 
\end{equation}
where the uncorrelated spin density ${\bf J}_q$ for each kind of nucleon is 
obtained using the auxiliary set of orbitals as
\begin{equation}
{\bf J}_q({\bf r})= -i \sum_{i,\sigma,\sigma'} \varphi_i^*({\bf
r},\sigma,q)
[\nabla \varphi_i({\bf r}, \sigma', q)\times
\langle \sigma \,\vert\, {\mbox{\boldmath$\sigma$}} \,\vert\, \sigma' \rangle ].
\label{bcp5} \end{equation}
Thus, the total energy of a finite nucleus is expressed as
\begin{equation}
E= T_0 + E_{N}^{\rm bulk} + E_{N}^{\rm surf} + E_{\rm coul} + E_{\rm s.o.} \,.
\label{eqfunct}
\end{equation}
In the calculations of open-shell nuclei we also take into account pairing 
correlations. The minimization procedure applied to the full functional gives a 
set of Hartree-like equations where the potential part includes in an effective 
way the overall exchange and correlation contributions. Further details of 
the functional can be found in \citep{baldo08,baldo10,baldo13}.

\begin{table}[tb]
\caption{\label{Table3} Parameters of the Gaussian form factors and
spin-orbit strength.}
\begin{center}
\begin{tabular}{r c c c c c c}
\hline\hline
$\alpha$ (fm) &  & $V_L$ (MeV) & & $V_U$ (MeV) & & $W_0$ (MeV) \\
\hline
0.90 &  &$-$137.024 & & $-$117.854 & & 95.43 \\
\hline
\end{tabular}
\end{center}
\end{table}

We note that the posited functional has {\sl only four} open parameters (the
strengths $V_L$ and $V_U$, the range $\alpha$ of the surface term, and the
spin-orbit strength $W_0$), while the rest of the functional is derived from the
microscopic BHF calculation. The four open parameters have been determined
by minimizing the root-mean-square (rms) deviation $\sigma_E$  between the
theoretical and experimental binding energies of a set of {\sl well-deformed}
nuclei in the rare-earth, actinide, and super-heavy mass regions of the nuclear 
chart \citep{baldo08}. The optimized values of the parameters are given in Table 
\ref{Table3}. The predictive power of the functional is then tested by computing 
the binding energies of 467 known spherical nuclei. A fairly reasonable rms 
deviation $\sigma_E = 1.3$ MeV between theory and experiment is found. Indeed, 
the rms deviations for binding energies and for charge radii of nuclei across 
the mass table with the BCPM functional are comparable to those obtained with 
highly successful nuclear mean field models such as the D1S Gogny force, the 
SLy4 Skyrme force, and the NL3 RMF parameter set, which for the same set of 
nuclei yield rms deviations $\sigma_E$ of 1.2--1.8 MeV, see 
\citep{baldo08,baldo10,baldo13}. The study of ground-state deformation
properties, fission barriers, and excited octupole states with the BCPM
functional \citep{robledo08,robledo10} shows that the deformation properties
of BCPM are similar to those predicted by the D1S Gogny force, which can
be considered as a benchmark to study deformed nuclei.

\section{The outer crust}
\label{sec:outer}

The outer crust of a NS is the region of the star that consists of neutron-rich 
nuclei and free electrons at densities approximately between 10$^4$ g/cm$^3$, 
where atoms become completely ionized, and 4$\times 10^{11}$ g/cm$^3$, where 
neutrons start to drip from the nuclei. 
The nuclei arrange themselves in a solid body-centreed cubic (bcc) lattice in 
order to minimize the Coulomb repulsion and are stabilized against 
$\beta$ decay by the surrounding electron gas. At the low densities of 
the beginning of the outer crust, the Coulomb lattice is populated by $^{56}$Fe 
nuclei. As the density of matter increases with increasing depth in the
crust, it becomes energetically favourable for the system to lower the proton 
fraction through electron captures with the energy excess carried away by 
neutrinos. The system progressively evolves towards a lattice of more and more
neutron-rich nuclei as it approaches the bottom of the outer crust, until the 
neutron drip density is reached and the inner crust of the NS begins.

\subsection{Formalism for the outer crust}

To describe the structure of the outer crust we follow the formalism developed 
by Baym, Pethick, and Sutherland (BPS) \citep{bps71}, as applied more recently 
in \citep{ruster06,roca08,roca12} (also see \citep{haenselbook,pearson11} and references 
therein). It is considered that the matter inside the star is cold and charge 
neutral, and that it is in its absolute ground state in complete thermodynamic 
equilibrium. This is a meaningful assumption for non-accreting neutrons stars 
that have lived long enough to cool down and to reach equilibrium after their 
creation. In the outer crust, the energy at average baryon number density 
$\nb$ consists of the nuclear plus electronic and lattice contributions:
\begin{equation}
E(A,Z,\nb) = E_{N} + E_{el} + E_l ,
\label{eq01}
\end{equation}
where $A$ is the number of nucleons in the nucleus, $Z$ is the atomic number, 
and $\nb=A/V$ (where $V$ stands for volume). The nuclear contribution to 
Eq.~(\ref{eq01}) corresponds to the mass of the nucleus, i.e. the rest mass 
energy of its neutrons and protons minus the nuclear binding energy:
\begin{equation}
E_{N}(A,Z) = M(A,Z) = (A-Z) m_n + Z m_p -B(A,Z) .
\label{eq02}
\end{equation}
The electronic contribution reads $E_{el}={\cal E}_{el}\,V$, where the 
energy density ${\cal E}_{el}$ of the electrons can be considered as a that of a 
degenerate relativistic free Fermi gas:
\begin{eqnarray}
{\cal E}_{el} &=& \frac{k_{Fe}}{8\pi^{2}}
(2k_{Fe}^{2}+m_{e}^{2}) \sqrt{k_{Fe}^{2}+m_{e}^{2}}
\nonumber\\
&& \mbox{} - \frac{m_{e}^{4}}{8\pi^{2}}
\ln\Big[ \Big(k_{Fe}+\sqrt{k_{Fe}^{2}+m_{e}^{2}}\Big) \,\Big/ m_{e} \Big] ,
\label{eq3}
\end{eqnarray}
with $k_{Fe}=\left(3\pi^{2}\rnum_{e}\right)^{1/3}$ being the Fermi momentum of
the electrons, $\rnum_{e}= (Z/A) \nb$ the electron number density, and $m_{e}$
the electron rest mass. The lattice energy can be written as
\begin{equation}
E_l = -C_l \frac{Z^2}{A^{1/3}} k_{Fb} ,
\label{eq04}
\end{equation}
where $k_{Fb}=\left(3\pi^{2}\nb\right)^{1/3} = (A/Z)^{1/3} k_{Fe}$ is the
average Fermi momentum and $C_l=3.40665 \times 10^{-3}$ for bcc lattices
\citep{roca08}. The lattice contribution has the form of the Coulomb
energy in the nuclear mass formula with a particular prefactor and corresponds to
the repulsion energy  among the nuclei distributed in the bcc lattice, their 
attraction with the electrons, and the repulsion energy among the electrons.

The basic assumption in the calculation is that thermal, hydrostatic, and 
chemical equilibrium is reached in each layer of the crust. As no pressure is 
exerted by the nuclei, only the electronic and lattice terms contribute to the 
pressure in the outer crust \citep{bps71}. Therefore, we have
\begin{equation}
P = -\left(\frac{\partial E}{\partial V}\right)_{T,A,Z}= 
\mu_e \rnum_e - {\cal E}_{el}
- \frac{\nb}{3}C_l \frac{Z^2}{A^{4/3}} k_{Fb} \,,
\label{eq05}
\end{equation}
where $\mu_e = \sqrt{k_{Fe}^2 + m_e^2}$ is the Fermi energy of the electrons 
including their rest mass. One has to find the nucleus that at a certain 
pressure minimizes the Gibbs free energy per particle, or chemical potential, 
$\mu= G/A= (E-T S)/A+P/\nb$ \citep{bps71}. As far as the system can be
assumed at zero temperature in good approximation, since the electronic 
Fermi energy is much larger that the temperature of the star, the quantity 
to be minimized is given by
\begin{eqnarray}
\mu(A,Z,P) &=& \frac{E(A,Z,\nb)}{A} + \frac{P}{\nb} \nonumber \\
&= & \frac{M(A,Z)}{A}
+ \frac{Z}{A} \mu_e - \frac{4}{3}C_l \frac{Z^2}{A^{4/3}} k_{Fb} . \;
\label{eq06}
\end{eqnarray}
For a fixed pressure, Eq.~(\ref{eq06}) is minimized with respect to the mass 
number $A$ and the atomic charge $Z$ of the nucleus in order to find the optimal 
configuration. The nuclear masses $M(A,Z)$ needed in Eq.~(\ref{eq06}) can be 
taken from experiment if they are available or calculated using nuclear models.

\subsection{Equation of state of the outer crust}
\label{eosouter}

Although thousands of nuclear masses are experimentally determined not far from 
stability, the masses of very neutron-rich nuclei are not known. We used in 
Eq.~(\ref{eq06}) the measured masses whenever available. For the unknown 
masses, we performed deformed Hartree-Fock-Bogoliubov calculations with the BCPM 
energy density functional, which has been constructed from the microscopic BHF 
calculations as described in Sec.~\ref{sec:micro}. We took the experimental data 
of masses from the most recent atomic mass evaluation, i.e. the AME2012 
evaluation \citep{ame2012}.
As the field of high-precision mass measurements of unstable neutron-rich 
nuclei continues to advance in the radioactive beam facilities worldwide 
\citep{thoennessen}, better constraints can be placed on the composition of the 
outer crust. Thus, we also included in our calculation the mass of $^{82}$Zn 
recently determined by a Penning-trap measurement at ISOLDE-CERN \citep{wolf12}. 
Being the most neutron-rich $N=50$ isotone known to date, $^{82}$Zn is an 
important nucleus in the study of the NS outer crust \citep{wolf12}.

\begin{figure}[t]
\includegraphics[width=1.00\columnwidth,clip=true]{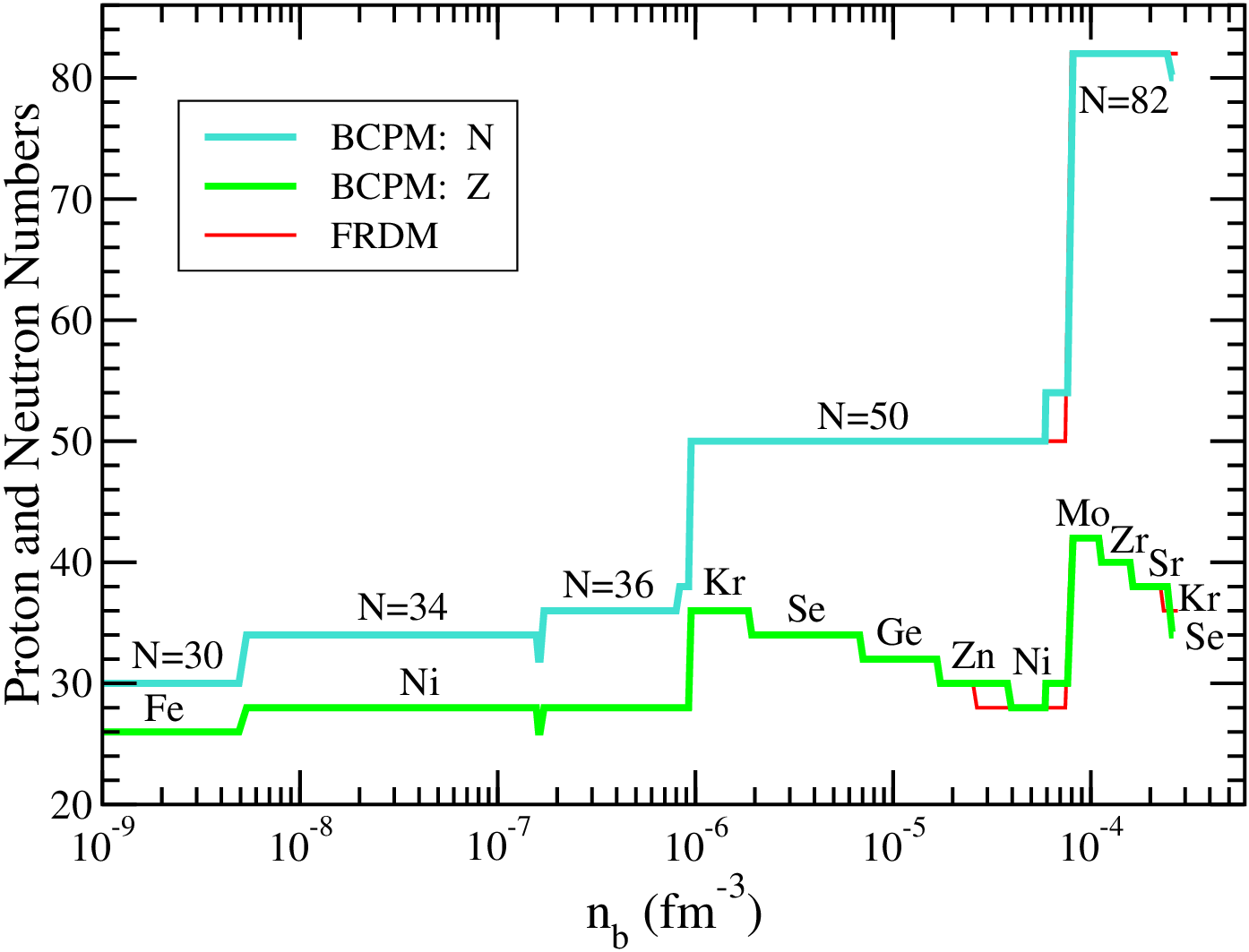}
\caption{\label{fig1} Neutron ($N$) and proton ($Z$) numbers of the predicted 
nuclei in the outer crust of a neutron star using the experimental nuclear
masses \citep{ame2012,wolf12} when available and the BCPM energy density
functional or the FRDM mass formula \citep{frdm95} for the unmeasured masses.}
\end{figure}

The calculated composition of the outer crust (neutron and proton numbers of the 
equilibrium nuclei) vs.. the average baryon density $\nb$ is displayed in 
Fig.~\ref{fig1}. After the $^{56}$Fe nucleus that populates the outer crust up 
to $\nb = 5\times 10^{-9}$~fm$^{-3}$ ($8\times 10^6$~g/cm$^3$), the composition 
profile exhibits a sequence of plateaus. The effect is related with the 
enhanced nuclear binding for magic nucleon numbers, i.e. $Z=28$ in the Ni 
plateau that occurs at intermediate densities in Fig.~\ref{fig1}, and 
$N=50$ and $N=82$ in the neutron plateaus that occur at higher 
densities. Along a neutron plateau, such as $N=50$, the nuclei experience 
electron captures that reduce the proton number, resulting in a staircase 
structure of shells of increasingly neutron-rich isotones, until the jump to the 
next neutron plateau ($N=82$) takes place. The composition of the crust is 
determined by the experimental masses up to densities around $2.5\times 10^{-5}$ 
fm$^{-3}$ ($4\times10^{10}$ g/cm$^3$). At higher densities, model masses need to 
be used because the relevant nuclei are more neutron rich and their masses are 
not measured.

The results for the composition in the high-density part of the outer crust from 
the microscopic BCPM model can be seen in Fig.~\ref{fig1}. They are shown along
with the results from the macroscopic-microscopic finite-range droplet model 
(FRDM) of M\"oller, Nix et al.\ \citep{frdm95} that was fitted with high 
accuracy to the thousands of experimental masses available at the time it was 
published. Despite BCPM has no more than four fitted parameters (for reference, 
the sophisticated macroscopic-microscopic models such as the FRDM 
\citep{frdm95} and the Duflo-Zuker model \citep{duflo95} have tens of 
adjustable parameters), the predictions of BCPM are overall quite in consonance 
with those of the FRDM. In particular, the jump to the $N=82$ plateau is 
predicted at practically the same density. Some differences are found,
though, in the width of the shell of $^{78}$Ni before the $N=82$ plateau, and in
the fact that BCPM reaches the neutron drip with a (very thin) shell of
$^{114}$Se, while the FRDM reaches the neutron drip with a shell of $^{118}$Kr. We
note that at the base of the outer crust (neutron drip) the baryon density, mass
density, pressure, and electron chemical potential of the equilibrium
configuration in BCPM are $\nb=2.62 \times 10^{-4}$ fm$^{-3}$, $\rmatt=4.37 \times
10^{11}$ g/cm$^3$, $P= 4.84 \times 10^{-4}$ MeV/fm$^3$, and $\mu_e=26.09$ MeV,
respectively. These values are close (cf.\ Table~I of \citep{roca08}) to the
results computed using the highly successful FRDM \citep{frdm95} and 
Duflo-Zuker \citep{duflo95} models, what gives us confidence that the BCPM 
energy density functional is well suited for extrapolating to the neutron-rich 
regions.

\begin{figure}
\includegraphics[width=1.00\columnwidth,clip=true]{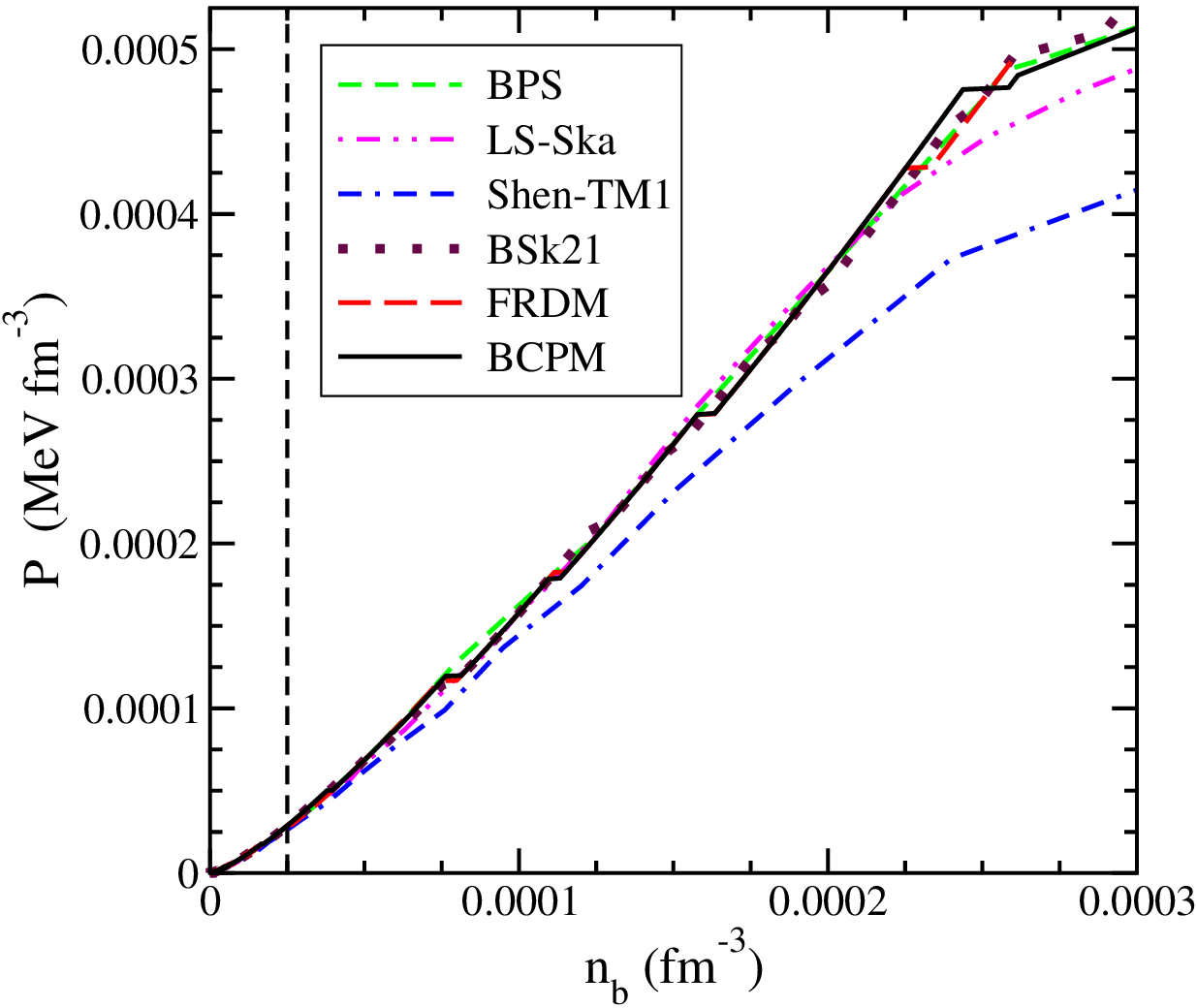}
\caption{\label{fig2} The pressure in the outer crust against the baryon density
using the experimental nuclear masses \citep{ame2012,wolf12} when available and
the BCPM energy density functional or the FRDM mass formula \citep{frdm95} for
the unmeasured masses. Also shown is the pressure from models BSk21, BPS,
Lattimer-Swesty (LS-Ska), and Shen et al. (Shen-TM1) (see text for details). The
dashed vertical line indicates the approximate end of the experimentally
constrained region.}
\end{figure}

\begin{table} 
\caption{\label{t:eos.outer} Composition and equation of state of the outer 
crust. The last nucleus in the table with known mass is $^{80}$Zn. The 
experimental masses determine the results up to densities around $2.5\times 
10^{-5}$ fm$^{-3}$, while at higher densities the calculation of the composition 
involves unmeasured masses.}
\small
\centering
\begin{tabular}{|c c r c c c|}
\hline\hline
$\nb$ &  $Z$ & $A$ \,\ & $\rmatt$ & $P$  & $\Gamma$ \\
(fm$^{-3}$) & & & (g cm$^{-3}$) & (erg cm$^{-3}$) & \\
\hline
 6.2203E$-$12  & 26 &  56 & 1.0317E+04 & 9.5393E+18 & 1.797 \\
 6.3129E$-$11  & 26 &  56 & 1.0471E+05 & 5.3379E+20 & 1.688 \\
 6.3046E$-$10  & 26 &  56 & 1.0457E+06 & 2.3241E+22 & 1.586 \\
 4.9516E$-$09  & 26 &  56 & 8.2138E+06 & 5.4155E+23 & 1.470 \\
 6.3067E$-$09  & 28 &  62 & 1.0462E+07 & 7.3908E+23 & 1.459 \\
 2.5110E$-$08  & 28 &  62 & 4.1659E+07 & 5.3113E+24 & 1.400 \\
 7.9402E$-$08  & 28 &  62 & 1.3176E+08 & 2.6112E+25 & 1.369 \\
 1.5828E$-$07  & 28 &  62 & 2.6269E+08 & 6.6859E+25 & 1.358 \\
 1.6400E$-$07  & 26 &  58 & 2.7220E+08 & 6.9610E+25 & 1.357 \\
 1.7778E$-$07  & 28 &  64 & 2.9508E+08 & 7.4978E+25 & 1.356 \\
 3.1622E$-$07  & 28 &  64 & 5.2496E+08 & 1.6340E+26 & 1.350 \\
 5.0116E$-$07  & 28 &  64 & 8.3212E+08 & 3.0390E+26 & 1.345 \\
 7.9431E$-$07  & 28 &  64 & 1.3191E+09 & 5.6433E+26 & 1.343 \\
 8.5093E$-$07  & 28 &  66 & 1.4132E+09 & 5.9393E+26 & 1.342 \\
 9.2239E$-$07  & 28 &  66 & 1.5319E+09 & 6.6181E+26 & 1.342 \\
 9.9998E$-$07  & 36 &  86 & 1.6609E+09 & 7.1858E+26 & 1.341 \\
 1.2587E$-$06  & 36 &  86 & 2.0908E+09 & 9.7829E+26 & 1.340 \\
 1.5845E$-$06  & 36 &  86 & 2.6324E+09 & 1.3318E+27 & 1.340 \\
 1.8587E$-$06  & 36 &  86 & 3.0881E+09 & 1.6492E+27 & 1.339 \\
 1.9952E$-$06  & 34 &  84 & 3.3151E+09 & 1.7369E+27 & 1.339 \\
 3.1620E$-$06  & 34 &  84 & 5.2552E+09 & 3.2161E+27 & 1.338 \\
 5.0116E$-$06  & 34 &  84 & 8.3320E+09 & 5.9526E+27 & 1.337 \\
 6.7858E$-$06  & 34 &  84 & 1.1285E+10 & 8.9241E+27 & 1.336 \\
 7.5849E$-$06  & 32 &  82 & 1.2615E+10 & 9.8815E+27 & 1.336 \\
 9.9996E$-$06  & 32 &  82 & 1.6635E+10 & 1.4293E+28 & 1.335 \\
 1.2589E$-$05  & 32 &  82 & 2.0947E+10 & 1.9437E+28 & 1.335 \\
 1.6595E$-$05  & 32 &  82 & 2.7622E+10 & 2.8107E+28 & 1.335 \\
 1.9053E$-$05  & 30 &  80 & 3.1718E+10 & 3.2111E+28 & 1.335 \\
 2.5118E$-$05  & 30 &  80 & 4.1828E+10 & 4.6433E+28 & 1.334 \\
 3.1621E$-$05  & 30 &  80 & 5.2673E+10 & 6.3132E+28 & 1.334 \\
 3.7973E$-$05  & 30 &  80 & 6.3269E+10 & 8.0596E+28 & 1.334 \\
 4.1685E$-$05  & 28 &  78 & 6.9462E+10 & 8.6285E+28 & 1.334 \\
 5.8754E$-$05  & 28 &  78 & 9.7955E+10 & 1.3639E+29 & 1.334 \\
 6.3093E$-$05  & 30 &  84 & 1.0520E+11 & 1.4867E+29 & 1.334 \\
 7.6207E$-$05  & 30 &  84 & 1.2711E+11 & 1.9125E+29 & 1.334 \\
 8.4137E$-$04  & 42 & 124 & 1.4035E+11 & 2.0101E+29 & 1.334 \\
 1.0964E$-$04  & 42 & 124 & 1.8299E+11 & 2.8616E+29 & 1.334 \\
 1.2022E$-$04  & 40 & 122 & 2.0067E+11 & 3.1040E+29 & 1.334 \\
 1.4125E$-$04  & 40 & 122 & 2.3584E+11 & 3.8485E+29 & 1.334 \\
 1.5667E$-$04  & 40 & 122 & 2.6163E+11 & 4.4187E+29 & 1.334 \\
 1.6982E$-$04  & 38 & 120 & 2.8364E+11 & 4.7062E+29 & 1.334 \\
 2.0416E$-$04  & 38 & 120 & 3.4112E+11 & 6.0164E+29 & 1.334 \\
 2.4265E$-$04  & 38 & 120 & 4.0556E+11 & 7.5746E+29 & 1.334 \\
 2.6155E$-$04  & 34 & 114 & 4.3721E+11 & 7.7582E+29 & 1.334 \\
\hline
\end{tabular}
\end{table}

In Fig.~\ref{fig2} we display the EoS, or pressure-density relationship, of the 
outer crust obtained from the experimental masses plus BCPM, and from the 
experimental masses plus the FRDM. One observes small jumps in the density for 
particular values of the pressure. They are associated with the change from an 
equilibrium nucleus to another in the composition. During this change the 
pressure and the chemical potential remain constant, implying a finite 
variation of the baryon density \citep{bps71,ruster06,haenselbook,pearson11}. 
After the region constrained by the experimental masses (marked by the dashed 
vertical line in Fig.~\ref{fig2}), the pressures of BCPM (black solid line) and 
the FRDM (red dashed line) extrapolate very similarly, with only some 
differences appreciated by the end of the outer crust. Our results for the 
composition and EoS of the outer crust are given in Table~\ref{t:eos.outer}. 
The quantities $\rmatt$ and $\Gamma$ in this table are, respectively, the mass 
density of matter and the adiabatic index $\Gamma= (\nb/P)\,(d P/d \nb$).

Also plotted in Fig.~\ref{fig2} is the pressure in the outer crust from some
popular EoSs that model the complete structure of the NS. The figure is drawn 
up to $\nb=3 \times 10^{-4}$ fm$^{-3}$, thus comprising the change from the 
outer crust to the inner crust in order to 
allow comparison of the EoSs also in this region (notice, however, that 
inner-crust results are not available for the FRDM). We show in Fig.~\ref{fig2} 
the EoS from the recent BSk21 Skyrme nuclear effective force 
\citep{pearson12,fantina13,potekhin13,goriely10} tabulated in \citep{potekhin13}. 
The parameters of this force were fitted \citep{goriely10} to reproduce with high 
accuracy almost all known nuclear masses, and to various physical conditions 
including the neutron matter EoS from microscopic calculations. We see in 
Fig.~\ref{fig2} that after the experimentally constrained region, the BSk21 
pressure is similar to the BCPM and FRDM results, with just some displacement 
around the densities where the composition changes from a nucleus to the next 
one. In the seminal work of BPS \citep{bps71} the nuclear masses for the outer 
crust were provided by an early semi-empirical mass table. The corresponding 
EoS is seen to display a similar pattern with the BCPM, FRDM, and BSk21 results 
in Fig.~\ref{fig2}. The EoS by Lattimer and Swesty \citep{ls91}, taken here in 
its Ska version \citep{LSweb} (LS-Ska), and the EoS by Shen et al.\
\citep{shen98a,shen98b,Shenweb} (Shen-TM1) were computed with, respectively,
the Skyrme force Ska and the relativistic mean-field model TM1. In the two cases
the calculations of masses are of semiclassical type and $A$ and $Z$ vary in a
continuous way. Therefore, these EoSs do not present jumps at the densities
associated with a change of nucleus in the crust. Beyond this feature, the 
influence of shell effects in the EoS is rather moderate because to the extent 
that the pressure at the densities of interest is largely determined by the 
electrons, small changes of the atomic number $Z$ compared with its 
semiclassical estimate modify only marginally the electron density and, 
consequently, the pressure. The LS-Ska EoS shows good agreement with the 
previously discussed models, with some departure from them in the transition to 
the inner crust. The largest discrepancies in Fig.~\ref{fig2} are observed with 
the Shen-TM1 EoS that in this region predicts a softer crustal pressure with 
density than the other models.

\section{The inner crust: Formalism}
\label{sec:inncrust}

Below the outer crust, the inner crust starts at a density about 4$\times 
10^{11}$ g/cm$^3$, where nuclei have become so neutron rich that neutrons drip 
in the environment, and extends until the NS core. The structure of the inner 
crust consists of a Coulomb lattice of nuclear clusters permeated by the gases 
of free neutrons and free electrons. This is a unique system that is not
accessible to experiment because of the presence of the free neutron gas. The 
fraction of free neutrons increases with growing density of matter. At the 
bottom layers of the inner crust the equilibrium nuclear shape may change from 
sphere, to cylinder, slab, tube (cylindrical hole), and bubble (spherical hole) 
before going into uniform matter. These shapes are generically known as 
``nuclear pasta''. 

Full quantal Hartree-Fock (HF) calculations of the nuclear structures in the 
inner crust are complicated by the treatment of the neutron gas and the 
eventual need to deal with different geometries. As a result, large scale 
calculations of the inner crust and nuclear pasta have been performed very often 
with semiclassical methods such as the CLDM \citep{ls91,douchin01} or the 
Thomas-Fermi (TF) approximation and their variants, employing effective forces 
for the nuclear interaction, see~\citep{haenselbook,livingrev} for 
reviews. Calculations of TF type, including pasta phases, have been carried out 
both in the non-relativistic \citep{oyamatsu93,gogelein07,onsi08,pearson12} and
in the relativistic \citep{cheng97,shen98a,shen98b,avancini08,grill12} nuclear
mean-field theories.

In this work we apply the self-consistent TF approximation to describe the inner 
crust for two reasons. First, our main aim is to obtain the EoS of the neutron 
star, which is largely driven by the neutron gas and where the contribution of 
shell corrections is to some extent marginal. Second, the
semiclassical methods, as far as they do not contain shell 
effects, are well suited to describe non-spherical shapes, i.e. pasta phases, as 
we shall discuss below. Nevertheless, it is to be mentioned that in the case of 
spherical symmetry, shell corrections for protons in the inner crust may be 
introduced perturbatively on top of the semiclassical results via the Strutinsky
shell-correction method. These corrections have been included in the calculations
of the inner crust by the Brussels-Montreal group 
\citep{chamel11,pearson12,fantina13,potekhin13} with the BSk19--BSk21 Skyrme 
forces \citep{goriely10}. Shell effects can be taken into account 
self-consistently using the HF method. Indeed, HF calculations of inner crust 
matter were performed in spherical WS cells in the pioneering work of Negele 
and Vautherin \citep{negele73}. Pairing effects in the inner crust can have 
important consequences to describe, e.g. pulsar glitches phenomena or the 
cooling of NS. Pairing correlations have been included in BCS and HFB 
calculations of the inner crust in the spherical WS approach, see for instance 
\citep{pizzo02,sandu04,baldo07,grill11,pastore11} and references therein, and 
also in the BCS theory of anisotropic multiband superconductivity beyond the WS 
approach \citep{chamel10}. These works are mainly devoted to investigate the 
superfluid properties of NS inner crusts and only, to our knowledge, in 
\citep{baldo07} the EoS of the inner crust including pairing correlations 
was reported. 

It must be mentioned as well that the formation and the properties of nuclear 
pasta have been investigated in three-dimensional HF 
\citep{gogelein07,pais12,schuetrumpf14} and TF
\citep{okamoto13} calculations in cubic boxes that avoid any assumptions on 
the geometry of the system. Generally speaking, the results of these 
calculations observe the usual pasta shapes but additional complex morphologies 
can appear as stable or metastable states at the transitions between shapes 
\citep{gogelein07,pais12,schuetrumpf14,okamoto13}. 
Techniques based on Monte Carlo and molecular dynamics simulations, which do not 
impose a periodicity or symmetry of the system unlike the WS approximation, 
have also been applied to study nuclear pasta, 
see~\citep{horowitz04,watanabe09,horowitz13,horowitzpasta15} and references 
therein. (We note that some of the quoted three-dimensional and simulation 
calculations are for pasta in supernova matter rather than in cold neutron-star 
matter.) These calculations are highly time consuming and a detailed 
pressure-vs.-density relation for the whole inner crust is not yet available.

\subsection{Self-consistent Thomas-Fermi description of the inner
crust of a neutron star}
\label{sec:tfeqs}

In this section we describe the method used for computing the structure and the
EoS of the inner crust with the BCPM energy density functional. Although a summary
of the approach has been presented elsewhere \citep{baldo13a}, here we provide a
more complete report of the formalism.

The total energy of an ensemble of $A-Z$ neutrons, $Z$~protons, and $Z$ 
electrons in a spherical Wigner-Seitz (WS) cell of volume $V_{c}= 4\pi R_c^3 /3$ 
can be expressed as
\begin{eqnarray}
E &=& {E(A,Z,R_{c})} \,=\,
\int_{V_{c}}\!\! \Big[{\cal H}\left(\rnum_{n},\rnum_{p}\right)
+ m_{n}\rnum_{n} + m_{p}\rnum_{p}
\nonumber \\[2mm]
&& \mbox{}
+ {\cal E}_{el}\left(\rnum_{e}\right)
+ {\cal E}_{\rm coul}\left(\rnum_{p},\rnum_{e}\right)
+ {\cal E}_{ex}\left(\rnum_{p},\rnum_{e}\right)
\Big] d{\bf r} .
\label{eq1}
\end{eqnarray}
The term ${\cal H}\left(\rnum_{n},\rnum_{p}\right)$ is the contribution of the 
nuclear energy density, without the nucleon rest masses. In our approach it reads
\begin{eqnarray}
{\cal H}\left(\rnum_{n} ,\rnum_{p}\right) &=&
\frac{3}{5}\frac{\left(3\pi^{2}\right)^{2/3}}{2m_{n}}
\rnum_{n}^{5/3}({\bf r}) 
+ \frac{3}{5}\frac{\left(3\pi^{2}\right)^{2/3}}{2m_{p}}
\rnum_{p}^{5/3}({\bf r})  \nonumber \\
&& \mbox{} +{\cal V}\left(\rnum_{n}({\bf r}),\rnum_{p}({\bf r})\right) ,
\label{eq2}
\end{eqnarray}
where the neutron and proton kinetic energy densities are written in the 
TF approximation and ${\cal V}\left(\rnum_{n},\rnum_{p}\right)$ is 
the interacting part provided by the BCPM nuclear energy density functional 
(see \ Sec.~\ref{sec:micro}), which contains bulk and surface contributions. 
The term ${\cal E}_{el}\left(\rnum_{e}\right)$ in Eq.~(\ref{eq1}) is the 
relativistic energy density due to the motion of the electrons, including their 
rest mass. Since at the densities prevailing in neutron-star inner crusts the 
Fermi energy of the electrons is much higher than the Coulomb energy, we 
computed ${\cal E}_{el}$ using the energy density of a uniform relativistic 
Fermi gas given by Eq.~(\ref{eq3}).
The term ${\cal E}_{\rm coul}\left(\rnum_{p},\rnum_{e}\right)$ in Eq.~(\ref{eq1}) is 
the Coulomb energy density from the direct part of the proton-proton, 
electron-electron, and  proton-electron interactions. Assuming that the 
electrons are uniformly distributed, this term can be written as 
\begin{eqnarray}
{\cal E}_{\rm coul}\left(\rnum_{p},\rnum_{e}\right) &=& 
\frac{1}{2} \left(\rnum_{p}({\bf r})-\rnum_{e}\right)
\, \left( V_{p}({\bf r})-V_{e}({\bf r}) \right) ,
\label{eq4}
\end{eqnarray}
with
\begin{equation}
V_{p}({\bf r}) = \int \frac{e^{2} \rnum_{p}({\bf r'}) }{|{\bf r- r'}|} d{\bf r'},
\quad
V_{e}({\bf r}) = \int \frac{e^{2} \rnum_{e}}{|{\bf r- r'}|} d{\bf r'} .
\label{eq4b}
\end{equation}
We did calculations where we allowed the profile of the electron density to 
depend locally on position. However, we found this to have a marginal influence 
on our results for compositions and energies and decided to proceed with a 
uniform distribution for the electrons.
Lastly, the term ${\cal E}_{ex}(\rnum_{p},\rnum_{e})$ in 
Eq.~(\ref{eq1}) is the exchange part of the proton-proton and electron-electron 
interactions treated in Slater approximation:
\begin{eqnarray}
{\cal E}_{ex}\left(\rnum_{p},\rnum_{e}\right) &=& 
-\frac{3}{4}\left(\frac{3}{\pi}\right)^{\!1/3} \!e^2
\left(\rnum_{p}^{4/3}({\bf r}) +\rnum_{e}^{4/3}\right) .
\label{eq41}
\end{eqnarray}

We consider the matter within a WS cell of radius $R_{c}$ and perform a fully 
variational calculation of the total energy $E(A,Z,R_{c})$ of Eq.~(\ref{eq1})
imposing charge neutrality and an average baryon density $\nb=A/V_{c}$. The 
fact that the nucleon densities are fully variational differs from some other TF 
calculations carried out with non-relativistic nuclear models  
\citep{oyamatsu93,gogelein07,onsi08,pearson12} that use parametrized trial 
neutron and proton densities in the minimization. TF calculations of the inner 
crust with fully variational densities using relativistic nuclear mean-field 
models have been reported in the literature 
\citep{cheng97,shen98a,shen98b,avancini08,grill12}.

Taking functional derivatives of Eq.~(\ref{eq1}) with respect to the particle 
densities and including the conditions of charge neutrality and constant average 
baryon density with suitable Lagrange multipliers, leads to the variational
equations
\begin{eqnarray}
\frac{\delta{\cal H}\left(\rnum_{n},\rnum_{p}\right)}{{\delta}{\rnum_{n}}}
+ m_{n}- \mu_{n} = 0,
\label{eq5}
\end{eqnarray}
\begin{eqnarray}
\frac{\delta{\cal H}\left(\rnum_{n},\rnum_{p}\right)}{{\delta}{\rnum_{p}}}
&+& V_{p}({\bf r}) - V_{e}({\bf r})
-\left(\frac{3}{\pi}\right)^{\!1/3} \!e^2 \rnum_{p}^{1/3}({\bf r}) \nonumber \\
&+& m_{p}- \mu_{p} = 0,
\label{eq6}
\end{eqnarray}
\begin{equation}
\sqrt{k_{Fe}^{2}+m_{e}^{2}}
+ V_{e}({\bf r}) - V_{p}({\bf r})
 -\left(\frac{3}{\pi}\right)^{\!1/3} \!e^2 \rnum_{e}^{1/3} - \mu_{e} = 0 ,
\label{eq7}
\end{equation}
plus the $\beta$-equilibrium condition 
\begin{equation}
\mu_{e}= \mu_{n}-\mu_{p},
\label{eq8}
\end{equation}
which is imposed along with the constraints of charge neutrality and fixed average 
baryon density in the cell.  We note that in this work the chemical potentials
$\mu_{n}$, $\mu_{p}$, and $\mu_{e}$ include the rest mass of the particle.

Equations (\ref{eq5})--(\ref{eq8}) are solved self-consistently in a WS cell of 
specified size $R_c$ following the method described in \citep{sil02} and 
the energy is calculated from Eq.~(\ref{eq1}). Next, a search of the optimal 
size of the cell for the considered average baryon density $\nb$ is performed by 
repeating the calculation for different values of $R_c$, in order to find the 
absolute minimum of the energy per baryon for that $\nb$. 
This can be a delicate numerical task because the minimum of the energy
is usually rather flat as a function of the cell radius $R_c$ (or, equivalently, 
as a function of the baryon number $A$) and the energy differences may be 
between a few keV and a few eV.

In order to obtain the EoS of the inner crust we have to compute the pressure. 
The derivation of the expression of the pressure in this region of the NS is
given in Appendix~\ref{sec:press}. As shown in the appendix (also see
\citep{pearson12}), in the inner crust the pressure assumes the form
\begin{eqnarray}
P = P_{ng} + P_{el}^{\rm free} + P_{el}^{ex},
\label{eq22}
\end{eqnarray}
where $P_{ng}$ is the pressure of the gas of dripped neutrons,
$P_{el}^{\rm free}=\rnum_{e}\sqrt{k_{Fe}^{2}+m_{e}^{2}}- {\cal 
E}_{el}(\rnum_{e})$ is the pressure of free electron gas, and 
$P_{el}^{ex}= -\frac{1}{4}\left(\frac{3}{\pi}\right)^{1/3} \!e^2 
\rnum_{e}^{4/3}$ is a corrective term from the electron Coulomb exchange.
That is, the pressure in the inner crust is exerted by the neutron
and electron gases in which the nuclear structures are embedded.
On the other hand, in Appendix~\ref{sec:press2} we show explicitly that when the
minimization of the energy per unit volume with respect to the radius of the WS
cell is attained, subjected to an average density $\nb$ and charge neutrality,
the Gibbs free energy $G= PV + E$ per particle equals the neutron chemical 
potential, i.e. $G/A = \mu_n$. This result provides an alternative way to
extract the pressure from the knowledge of the neutron chemical potential and the
energy.

The same formalism described for spherical nuclei can be applied to obtain 
solutions for spherical bubbles (hollow spheres). Assuming the appropriate 
geometry, the method can be used in a similar way for other shapes 
(nuclear pasta), as discussed in the next subsection.

\subsection{Pasta phases: cylindrical and planar geometries}

The method of solving Eqs.~(\ref{eq5})--(\ref{eq8}) can be extended to deal with
WS cells having cylindrical symmetry (rods or ``nuclear spaghetti'') or planar
symmetry (slabs or ``nuclear lasagne''). The length of the rods and the area of 
the slabs in the inner crust of a neutron star are effectively infinite. The 
calculations for these non-spherical geometries are simplified if one also 
considers the unit cells as rods and slabs of infinite extent in the direction 
perpendicular to the base area of the rods and to the width of the slabs, 
respectively. That is, the corresponding WS cells are taken as rods of finite 
radius $R_c$ and length $L\to\infty$, and as slabs of finite width $2R_c$ and 
transverse area $S\to\infty$. With this choice of geometries, the number of 
particles and energy per unit length (rods) or area (slabs) are finite. By 
taking $dV = 2{\pi}Lrdr$ for rods and $dV = Sdx$ for slabs in Eq.~(\ref{eq1}), 
the energy calculations are reduced to one-variable integrals over the finite 
size $R_c$ of the WS cell (i.e. from 0 to $R_c$ in a circle of radius $R_c$ in 
the case of rods, and from $-R_c$ to $+R_c$ along the $x$ direction for slabs). 
The calculation of the Coulomb potentials is likewise simplified. In the case 
of rods the direct Coulomb potential can be written as
\begin{equation}
V_{p}\left(r\right) =
-4{\pi}e^{2}\! \left[\ln r \!\int_{0}^{r}\!\! r'\rnum_{p}(r')dr'
+ \int_{r}^{R_c}\!\! r' \ln r' \rnum_{p}(r')dr'\right]
\label{eq11}
\end{equation}
for protons, and as
\begin{equation}
V_{e}\left(r\right)=
{\pi}e^{2}\rnum_{e}R_c^{2}\left(1-\frac{r^{2}}{R_c^{2}}-2\ln R_c \right)
\label{eq12}
\end{equation}
for the uniform electron distribution. In the case of slabs these potentials 
read
\begin{equation}
V_{p}\left(x\right)=
-4{\pi}e^{2}\! \left[x \!\int_{0}^{x}\!\! \rnum_{p}(x')dx'
+ \int_{x}^{R_c}\!\! x'\rnum_{p}(x')dx'\right],
\label{eq9}
\end{equation}
and 
\begin{equation}
V_{e}\left(x\right)=
-2{\pi}e^{2}\rnum_{e}\left(x^{2} + R_c^{2}\right).
\label{eq10}
\end{equation}

The other piece of the energy that depends on the geometry of the WS cell is
the nuclear surface contribution, which is given by Eq.~(\ref{surf}) for the 
BCPM energy density functional. In the case of spherical nuclei, performing 
the angular integration, the surface energy density of Eq.~(\ref{surf}) can be 
recast as
\begin{eqnarray}
%
{\cal E}_\textrm{surf}\left(\rnum_{q},\rnum_{q'}\right) & = &
\frac{\pi}{2}\sum_{q,q'} \alpha^{2} {V}_{q,q'}\rnum_{q}(r) \nonumber \\
& \times & \bigg[\frac{1}{r}
\int_{0}^{\infty} \! \rnum_{q'}(r') \Big(e^{{-(r - r')^{2}}/{\alpha^{2}}}
-e^{{-(r + r')^{2}}/{\alpha^{2}}}\Big) ~r'~ dr'  \nonumber \\
& -& \pi^{1/2}\alpha\rnum_{q'}(r)\bigg] .
\label{eq14}
\end{eqnarray}
For rods the nuclear surface contribution to the energy density is given by
\begin{eqnarray}
%
{\cal E}_\textrm{surf}\left(\rnum_{q},\rnum_{q'}\right) & = &
\frac{\pi}{2}\sum_{q,q'} \alpha^2 {V}_{q,q'}\rnum_q(r) \nonumber \\
& \times & \bigg[ \frac{2}{\alpha} \pi^{1/2} e^{-r^{2}/\alpha^{2}}
\!\! \int_{0}^{\infty} \! \rnum_{q'}(r')e^{-r'^{2}/\alpha^{2}}I_{0}
\Big(\frac{2rr'}{\alpha^{2}}\Big) ~r'~ dr' \nonumber \\
&-&  \pi^{1/2}\alpha \rnum_{q'}(r)\bigg] ,
\label{eq15}
\end{eqnarray}
where $I_{0}$ is the modified Bessel function, while for slabs it is given by
\begin{eqnarray}
%
{\cal E}_\textrm{surf}\left(\rnum_{q},\rnum_{q'}\right) &= &
\frac{\pi}{2}\sum_{q,q'}\alpha^{2}{V}_{q,q'}\rnum_q(x) \nonumber \\
& \times & \bigg[\int_{-\infty}^{\infty}
\rnum_{q'}(x')e^{-\left(x-x'\right)^{2}/\alpha^{2}} dx' \nonumber \\
&-& \pi^{1/2}\alpha \rnum_{q'}(x)\bigg] .
\label{eq16}
\end{eqnarray}

As happens with spherical nuclei and spherical bubbles, the equations
corresponding to cylindrical rods can be similarly applied to obtain the 
solutions for tube shapes (hollow rods).

\section{The inner crust: Results}
\label{sec:innres}

\subsection{Analysis of the self-consistent solutions}
\label{solutions}

\begin{figure}
\includegraphics[width=0.92\columnwidth,clip=true]{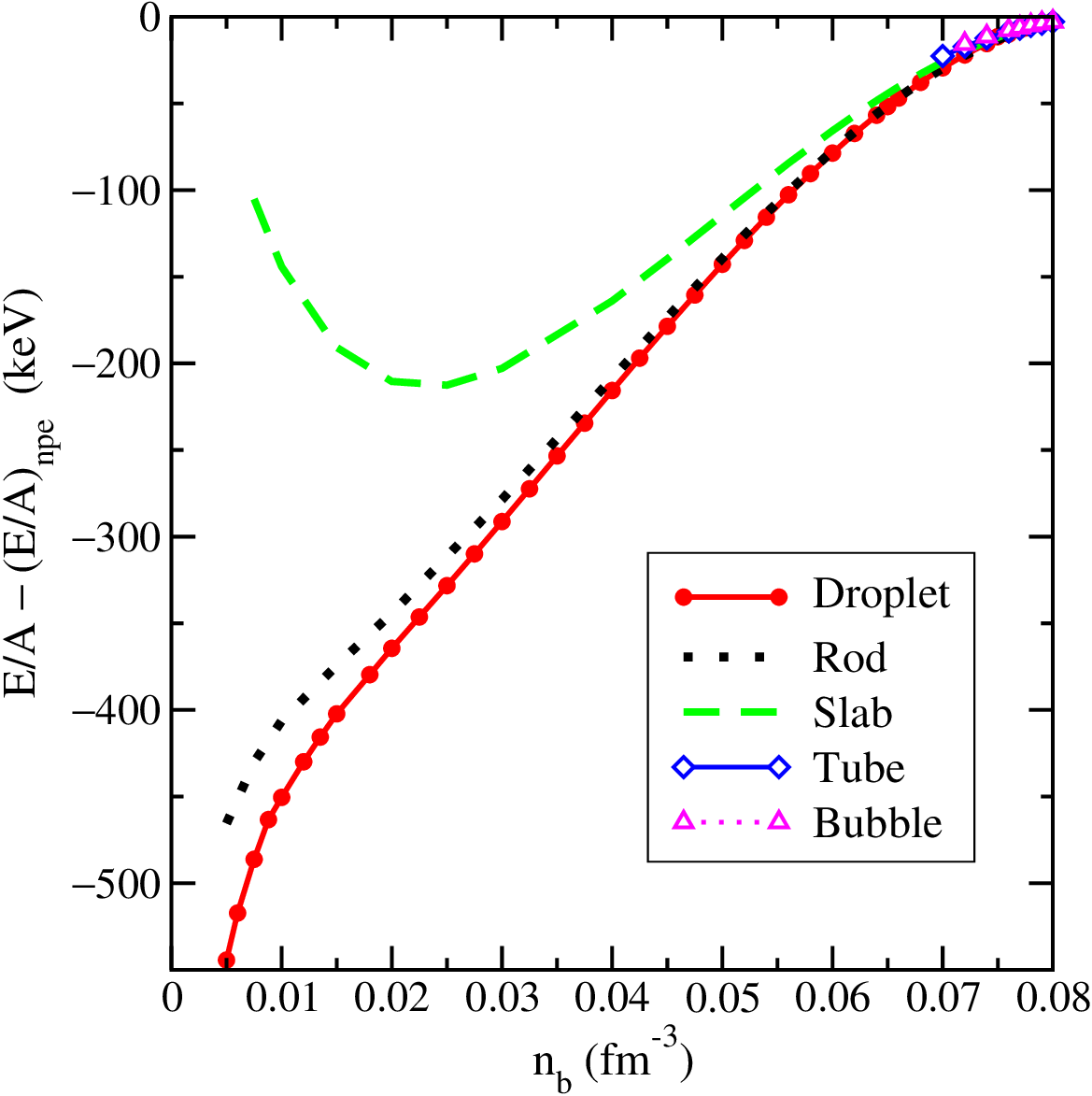}
\caption{\label{fig3}Energy per baryon of different shapes relative to uniform
$npe$ matter as a function of baryon density in the inner crust.}
\end{figure}

To compute the EoS of a neutron-star inner crust we need to determine the optimal
configuration of the WS cell, i.e. the shape and composition that yield the
minimal energy per baryon, as a function of the average baryon density $\nb$. This
is done for each $\nb$ value following the method explained in the previous
section. In Fig.~\ref{fig3} we display the results for the minimal energy per
baryon $E/A$ in the different shapes. The energy per baryon is shown relative to
the value in uniform neutron-proton-electron ($npe$) matter in order to be able to
appreciate the energy separations between shapes. Our set of considered shapes
consists of spherical droplets, cylindrical rods, slabs, cylindrical tubes, and
spherical bubbles.

As can be seen in Fig.~\ref{fig3}, it was possible to obtain solutions for rods
and slabs starting at densities as low as $\nb\sim0.005$ fm$^{-3}$. Solutions
at lower crustal densities, i.e. from the transition density between the outer 
crust and the inner crust until $\nb=0.005$ fm$^{-3}$, were obtained for spherical
droplets only. These solutions are not plotted in Fig.~\ref{fig3} for visual 
purposes of the rest of the figure. We note that at a density $\nb=0.0003$ fm$^{-3}$
immediately after the neutron drip point, i.e.  at the beginning of the inner 
crust, the difference $E/A-(E/A)_\textrm{npe}$ is of $-2.5$~MeV. For tube and 
bubble shapes, solutions could be obtained for crustal densities higher than 
about $0.07$ fm$^{-3}$, where the system is not far from uniform matter.

\begin{figure}
\includegraphics[width=0.90\columnwidth,clip=true]{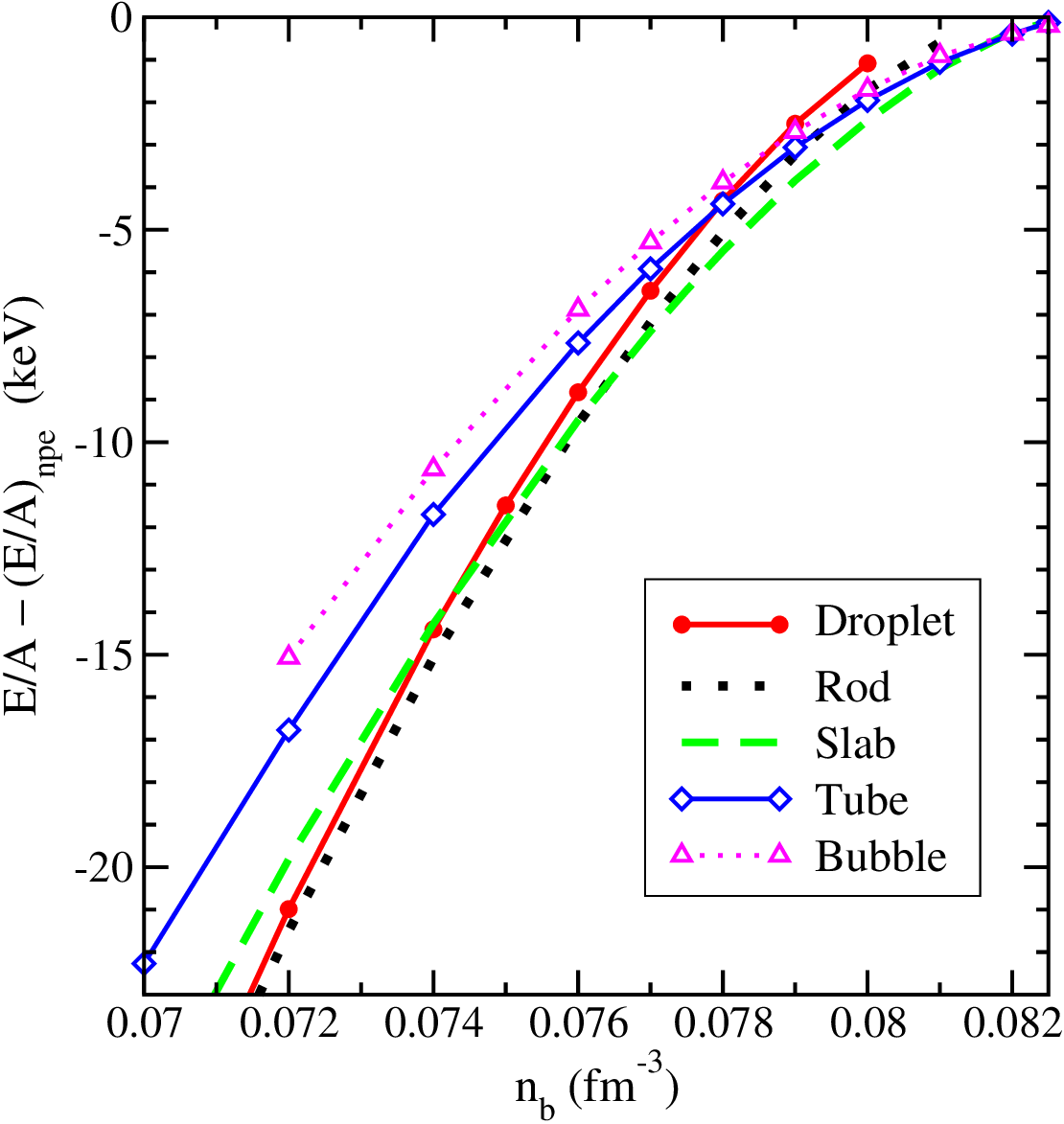}
\caption{\label{fig3b}Energy per baryon of different shapes relative to
uniform $npe$ matter as a function of baryon density in the
high-density region of the inner crust.}
\end{figure}

The spherical droplet configuration is the energetically most favourable shape all
the way up to $\nb\sim0.065$ fm$^{-3}$, see Fig.~\ref{fig3}.
When the crustal density reaches $\sim\,$0.065 fm$^{-3}$ (approximately $10^{14}$ 
g/cm$^3$), the nucleus occupies a significant fraction of the cell volume and it 
may happen that non-spherical structures have lower energy than the spherical
droplets \citep{ravenhall83,lorenz93,oyamatsu93,haenselbook,livingrev}.
We show in Fig.~\ref{fig3b} the high-density region between
$\nb=0.07$~fm$^{-3}$ and $\nb=0.0825$~fm$^{-3}$, where the TF-BCPM model
predicts the appearance of non-spherical shapes as optimal configurations. Indeed, 
the first change of nuclear shape occurs at $\nb= 0.067$~fm$^{-3}$ from droplets 
to rods. It can be seen in Fig.~\ref{fig3b} that the cylindrical shape is 
the energetically favoured configuration up to a crustal density of 
0.076~fm$^{-3}$ where a second change takes place to the planar slab shape.
As the density of the crustal matter increases further, the energy per baryon of 
tubes and bubbles becomes progressively closer 
to that of the slabs. Very close to the crust-core transition, estimated to occur 
in the TF-BCPM model at $\nb=0.0825$~fm$^{-3}$, there are successive shape changes
from slabs to tubes at $\nb=0.082$~fm$^{-3}$ and to spherical bubbles at almost
$0.0825$~fm$^{-3}$. At this latter density, the energy per baryon of the 
self-consistent TF bubble solution and that of uniform matter differ by barely 
$-200$~eV (in comparison, the value of $E/A-m_n$ is of 8.3 MeV). In summary, as 
compiled in Table~\ref{Tableshapes}, the TF-BCPM model predicts that the sequence
of shapes in the inner crust changes from spherical droplets to rods, slabs,
tubes, and spherical bubbles. This is in consonance with the ordering of pasta
phases reported in previous TF calculations
\citep{oyamatsu93,avancini08,grill12}. It can be noticed that the consideration
of pasta shapes renders the transition to the core somewhat smoother. The energy
differences between the most bound shape and the spherical solution at the same
density are, though, small and do not exceed 1--1.5 keV per nucleon.


\begin{table}[tb]
\caption{\label{Tableshapes} Baryon density of the successive changes of 
energetically favourable topology in the inner crust. Units are fm$^{-3}$.}
\small
\centering
\begin{tabular}{ccccc}
\hline\hline
{drop/rod} & rod/slab & slab/tube & tube/bubble & {bubble/uniform}\\
\hline\\
0.067 & 0.076 & 0.082 & $\lessapprox$0.0825 & $\approx$0.0825\\[1mm]
\hline
\end{tabular}
\end{table}

\begin{figure}[t]
\includegraphics[width=1.00\columnwidth,clip=true]{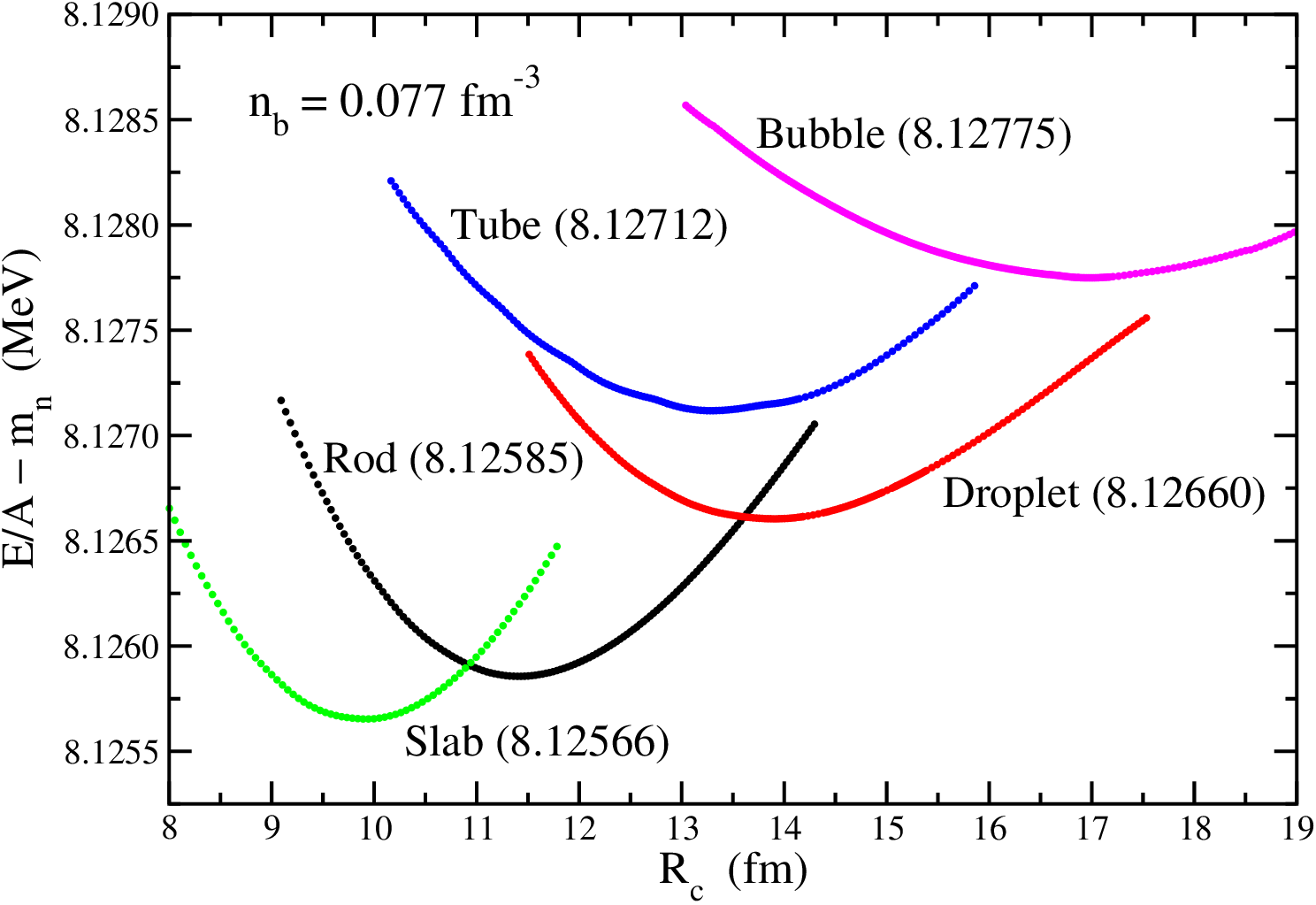}
\caption{\label{fig4} The minimum energy per baryon relative to the neutron mass
for different shapes as a function of the cell size $R_{c}$ at an average
baryon density $\nb = 0.077$ fm$^{-3}$. The value of the absolute minimum for
each shape is shown in MeV in brackets.}
\end{figure}

A typical landscape of solutions is illustrated in Fig.~\ref{fig4} where, for
a density $\nb=0.077$ fm$^{-3}$, the energy per baryon relative to the neutron
rest mass is shown as a function of the size of the WS cell for the various 
shapes. Usually, the curves for a given shape are flat around their minimum. 
For example, in the case of spherical droplets in Fig.~\ref{fig4}, a 1\% 
deviation of $R_c$ from the value of the minimum implies a shift of merely 25~eV 
in $E/A$, but it changes the baryon number by as much as 25 units.
Sufficiently close points have to be computed in order to determine precisely 
the $R_c$ value (equivalently, the $A$ value) that corresponds to the minimal 
energy per baryon. On the other hand, at high crustal densities the differences 
among the energy per baryon of the minima of the various shapes (quoted in 
brackets in Fig.~\ref{fig4}) are small; e.g. the minima of the 
slab and rod solutions in Fig.~\ref{fig4} differ by~190~eV.

\begin{figure}[b]
\includegraphics[width=1.00\columnwidth,clip=true]{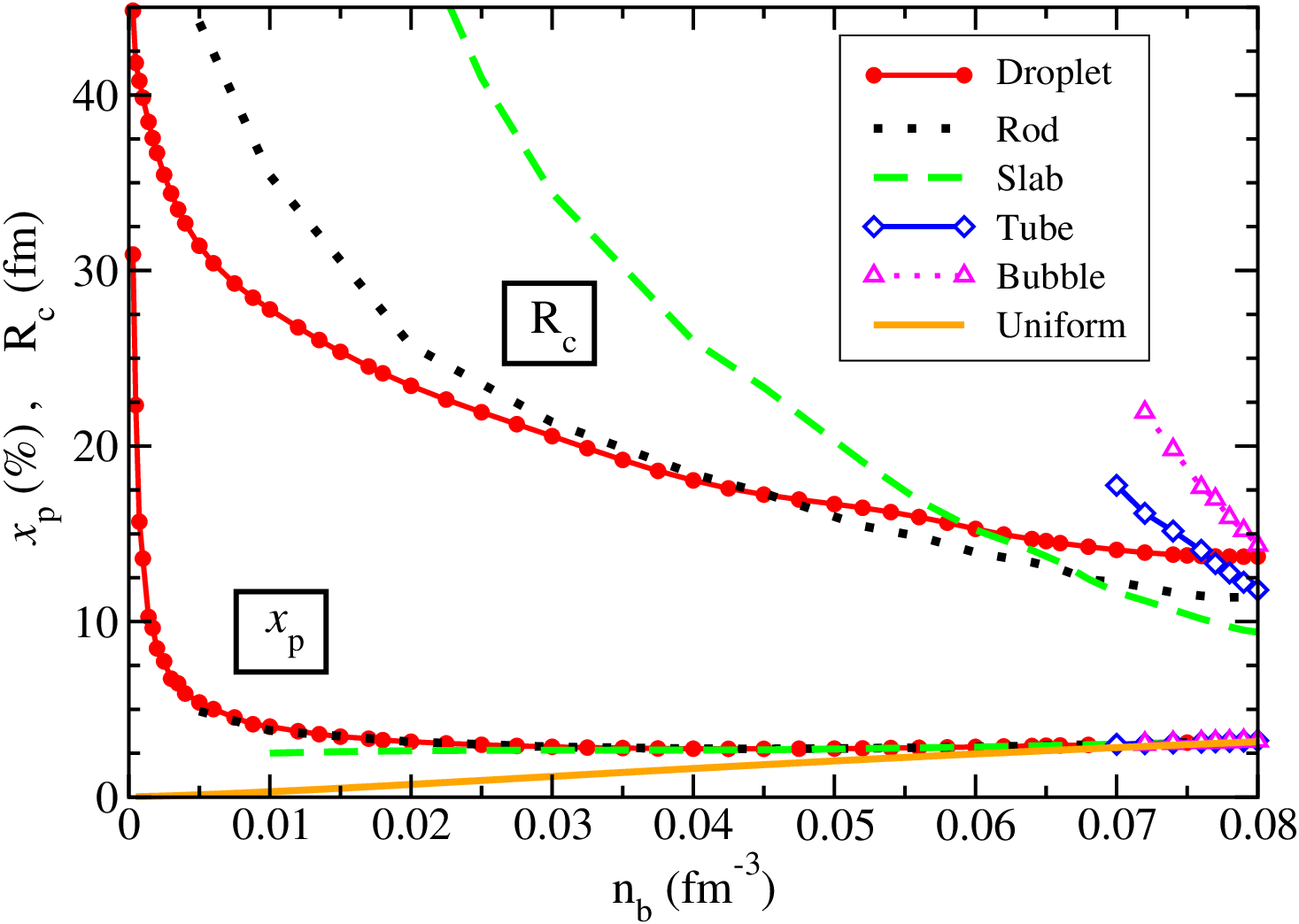}
\caption{\label{fig5}Radius $R_{c}$ of the Wigner-Seitz cell and proton
fraction $x_{p}=Z/A$ (given in percentage) in different geometries with respect 
to the baryon density.}
\end{figure}

Figure~\ref{fig5} displays the cell size $R_c$ and the proton fraction 
$x_{p}=Z/A$ of the equilibrium configurations against $\nb$ for the different 
shapes in our calculation. $R_c$ shows a nearly monotonic downward 
trend when the density increases, in agreement with other studies of NS inner 
crusts \citep{oyamatsu93,onsi08,pearson12,avancini08,grill12}. The size of the 
cell has a significant dependence on the geometry of the nuclear structures, as 
seen by comparing $R_c$ of the different shapes. In the spherical solutions, the 
cell radius decreases from almost 45~fm at $\nb=0.0003$~fm$^{-3}$ to 13.7~fm at 
$\nb=0.08$~fm$^{-3}$ near the transition to the core. As regards the proton 
fraction $x_{p}$, it takes quite similar values for the various cell geometries 
in the ranges of densities where we obtained solutions for the respective 
shapes. Beyond a density of the order of 0.02~fm$^{-3}$ the proton fraction 
shows a weak change with density, and after $\nb \sim 0.05$~fm$^{-3}$ it 
smoothly tends to the value in uniform $npe$ matter. For the spherical droplet 
solutions, we find that $x_p$ rapidly decreases from 31\% at $\nb = 
0.0003$~fm$^{-3}$ to $\sim\,$3\% at $\nb = 0.02$~fm$^{-3}$. It afterward 
remains almost constant, presenting a minimum value of 2.75\% at $\nb = 
0.045$~fm$^{-3}$ and then a certain increase up to 3.2\% at the last densities 
before the core.

\begin{figure}
\includegraphics[width=1.00\columnwidth,clip=true]{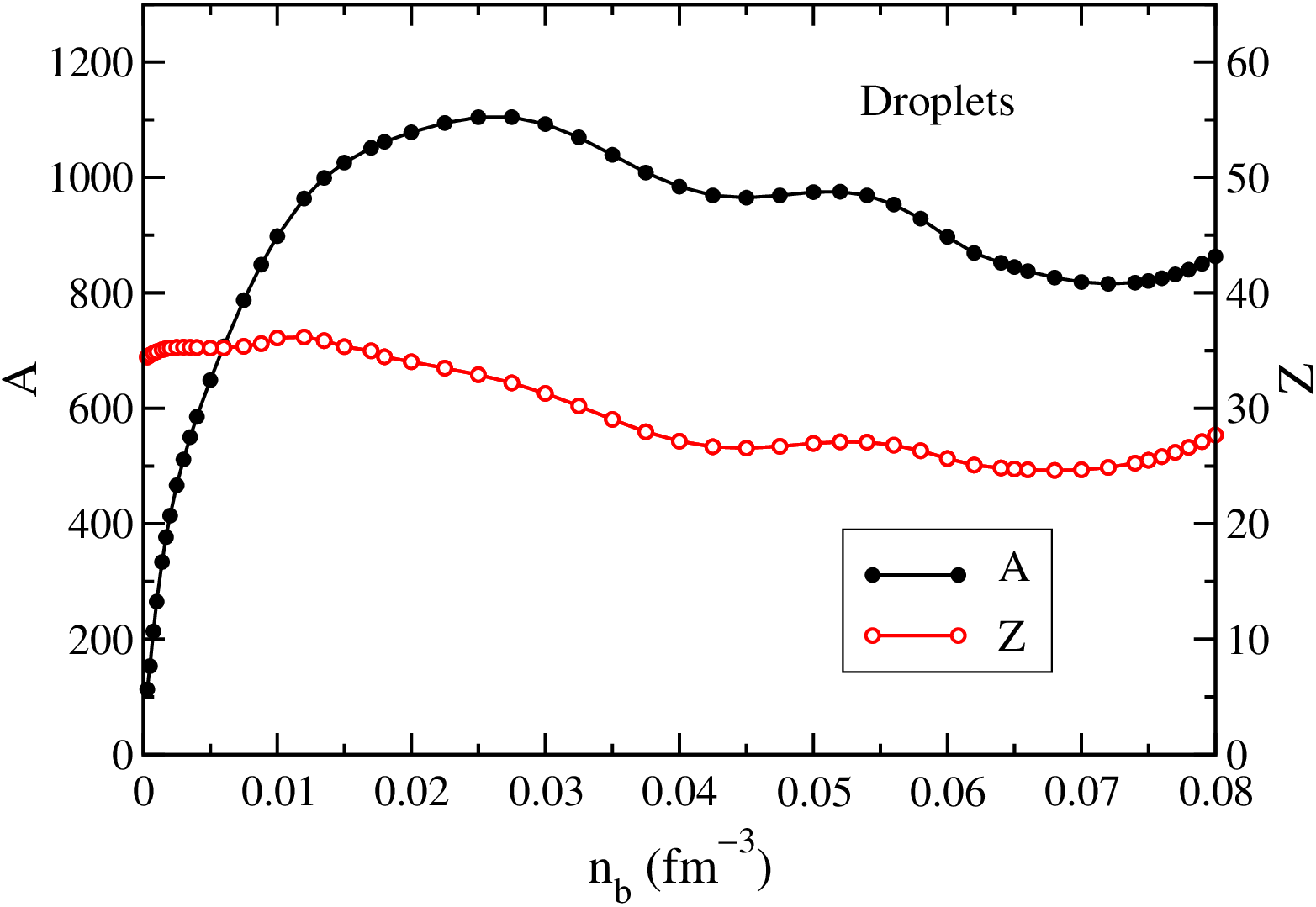}
\caption{\label{fig6}Mass number $A$ (left vertical scale) and atomic number 
$Z$ (right vertical scale) corresponding to spherical nuclei with respect to the
baryon density.}
\end{figure}

Taking into account that except for densities close to the crust-core transition 
the spherical shape is the energetically preferred configuration, and when it 
is not, the energy differences are fairly small, we restrict the discussion 
about the $A$ and $Z$ values in the WS cells to spherical nuclei.
Figure \ref{fig6} depicts the evolution with $\nb$ of $A$ and $Z$ of
the equilibrium spherical solutions.
The numerical values are collected in Table~\ref{t:compos.inner}.
When free neutrons begin to appear in the crust the number of nucleons
contained in a cell quickly increases from $A=113$ up to a maximum $A 
\simeq 1100$ at $\nb=0.025$~fm$^{-3}$. From this density onwards, $A$ shows a
slowly decreasing trend (with a relative plateau around $\nb\sim0.05$ fm$^{-3}$)
till $A \simeq 820$ at $\nb\simeq0.07$~fm$^{-3}$, and finally it presents an
increase approaching the core.
We see in Fig.~\ref{fig6} that the results for the atomic number $Z$ in the 
inner crust show an overall smooth decrease as a function of $\nb$ and a final 
slight increase before the transition to the core in agreement with the 
behaviour of~$A$. The range of $Z$ values lies between an upper value $Z \simeq 
36$ at $\nb \simeq 0.01$~fm$^{-3}$ and a lower value $Z \simeq 25$ at $\nb 
\simeq 0.07$~fm$^{-3}$.
We note that in our calculation the mass and atomic numbers vary continuously and 
are not restricted to integer values because we have used the TF approximation 
which averages shell effects.
The same fact explains that the atomic number $Z$ of the optimal configurations
does not correspond to proton magic numbers, as it happens when proton shell
corrections are included in the calculations 
\citep{negele73,baldo07,chamel11,pearson12,fantina13,potekhin13}.
However, as we have seen from the flatness of $E/A$ with respect to $A$ 
(or~$R_c$) around the minima in the discussion of Fig.~\ref{fig4}, the resulting 
EoS is robust against the details of the composition.

\begin{table}
\caption{\label{t:compos.inner} Composition of the inner crust.}
\centering
\begin{tabular}{|c c c r c c|}
\hline\hline
& $\nb$ & $Z$ & $A$ \ \ \ \ & $R_{c}$ & \\
& (fm$^{-3}$) & & & (fm) & \\
\hline
&    0.0003  &  34.934  &   112.991  & 44.8000 & \\
&    0.0005  &  34.237  &   153.293  & 41.8300 & \\
&\ \,0.00075 &  33.479  &   213.369  & 40.8000 & \\
&    0.0010  &  36.012  &   264.978  & 39.8450 & \\
&    0.0014  &  34.265  &   333.809  & 38.4675 & \\
&    0.0017  &  36.291  &   376.721  & 37.5400 & \\
&    0.0020  &  35.091  &   414.026  & 36.6975 & \\
&    0.0025  &  36.104  &   466.725  & 35.4550 & \\
&    0.0030  &  34.519  &   511.212  & 34.3925 & \\
&    0.0035  &  35.645  &   550.067  & 33.4775 & \\
&    0.0040  &  34.549  &   585.320  & 32.6900 & \\
&    0.0050  &  34.990  &   648.872  & 31.4075 & \\
&    0.0060  &  35.472  &   707.137  & 30.4150 & \\
&    0.0075  &  35.711  &   787.198  & 29.2625 & \\
&    0.0088  &  35.252  &   848.825  & 28.4500 & \\
&    0.0100  &  36.094  &   898.261  & 27.7825 & \\
&    0.0120  &  36.181  &   963.496  & 26.7625 & \\
&    0.0135  &  35.863  &   999.069  & 26.0450 & \\
&    0.0150  &  35.339  &  1025.682  & 25.3675 & \\
&    0.0170  &  34.982  &  1051.388  & 24.5325 & \\
&    0.0180  &  34.461  &  1061.641  & 24.1475 & \\
&    0.0200  &  34.036  &  1078.235  & 23.4350 & \\
&    0.0225  &  33.477  &  1094.430  & 22.6450 & \\
&    0.0250  &  32.910  &  1104.446  & 21.9300 & \\
&    0.0275  &  32.204  &  1104.566  & 21.2450 & \\
&    0.0300  &  31.290  &  1092.541  & 20.5625 & \\
&    0.0325  &  30.203  &  1069.599  & 19.8800 & \\
&    0.0350  &  29.036  &  1039.295  & 19.2100 & \\
&    0.0375  &  27.959  &  1008.341  & 18.5850 & \\
&    0.0400  &  27.152  &   984.099  & 18.0425 & \\
&    0.0425  &  26.665  &   968.891  & 17.5900 & \\
&    0.0450  &  26.549  &   965.017  & 17.2350 & \\
&    0.0475  &  26.701  &   968.928  & 16.9500 & \\
&    0.0500  &  26.955  &   974.581  & 16.6950 & \\
&    0.0520  &  27.096  &   975.352  & 16.4825 & \\
&    0.0540  &  27.065  &   968.814  & 16.2400 & \\
&    0.0560  &  26.808  &   953.172  & 15.9575 & \\
&    0.0580  &  26.322  &   928.561  & 15.6350 & \\
&    0.0600  &  25.650  &   897.063  & 15.2825 & \\
&    0.0620  &  25.080  &   869.075  & 14.9575 & \\
&    0.0640  &  24.833  &   852.005  & 14.7025 & \\
&    0.0650  &  24.750  &   844.737  & 14.5850 & \\
&    0.0660  &  24.672  &   837.603  & 14.4700 & \\
&    0.0680  &  24.613  &   826.389  & 14.2625 & \\
&    0.0700  &  24.674  &   818.891  & 14.0825 & \\
&    0.0720  &  24.875  &   815.658  & 13.9325 & \\
&    0.0740  &  25.249  &   817.728  & 13.8175 & \\
&    0.0750  &  25.502  &   820.707  & 13.7725 & \\
&    0.0760  &  25.810  &   825.326  & 13.7375 & \\
&    0.0770  &  26.190  &   832.083  & 13.7150 & \\
&    0.0780  &  26.615  &   840.127  & 13.7000 & \\
&    0.0790  &  27.111  &   850.432  & 13.6975 & \\
&    0.0800  &  27.677  &   863.085  & 13.7075 & \\
\hline
\end{tabular}
\end{table}

\begin{figure}
\includegraphics[width=1.00\columnwidth,clip=true]{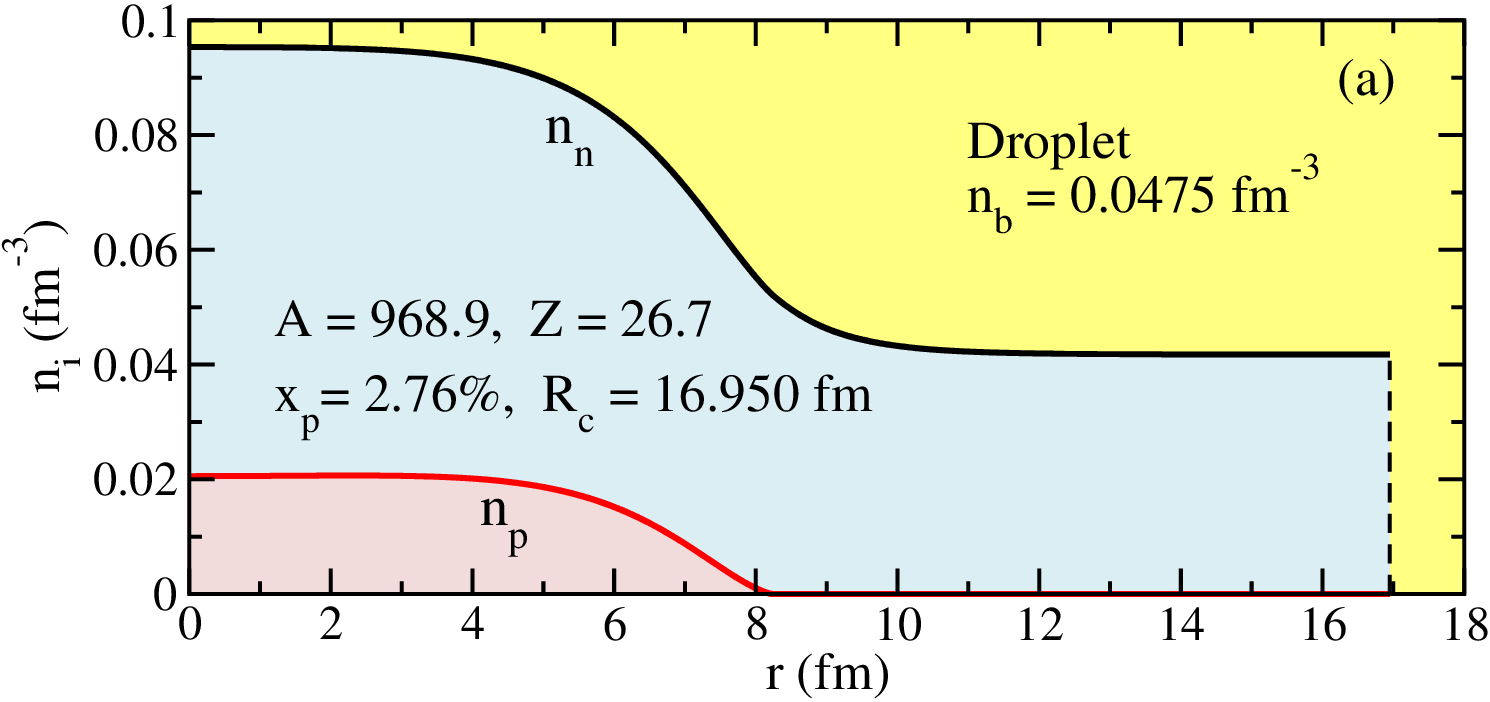}
\includegraphics[width=1.00\columnwidth,clip=true]{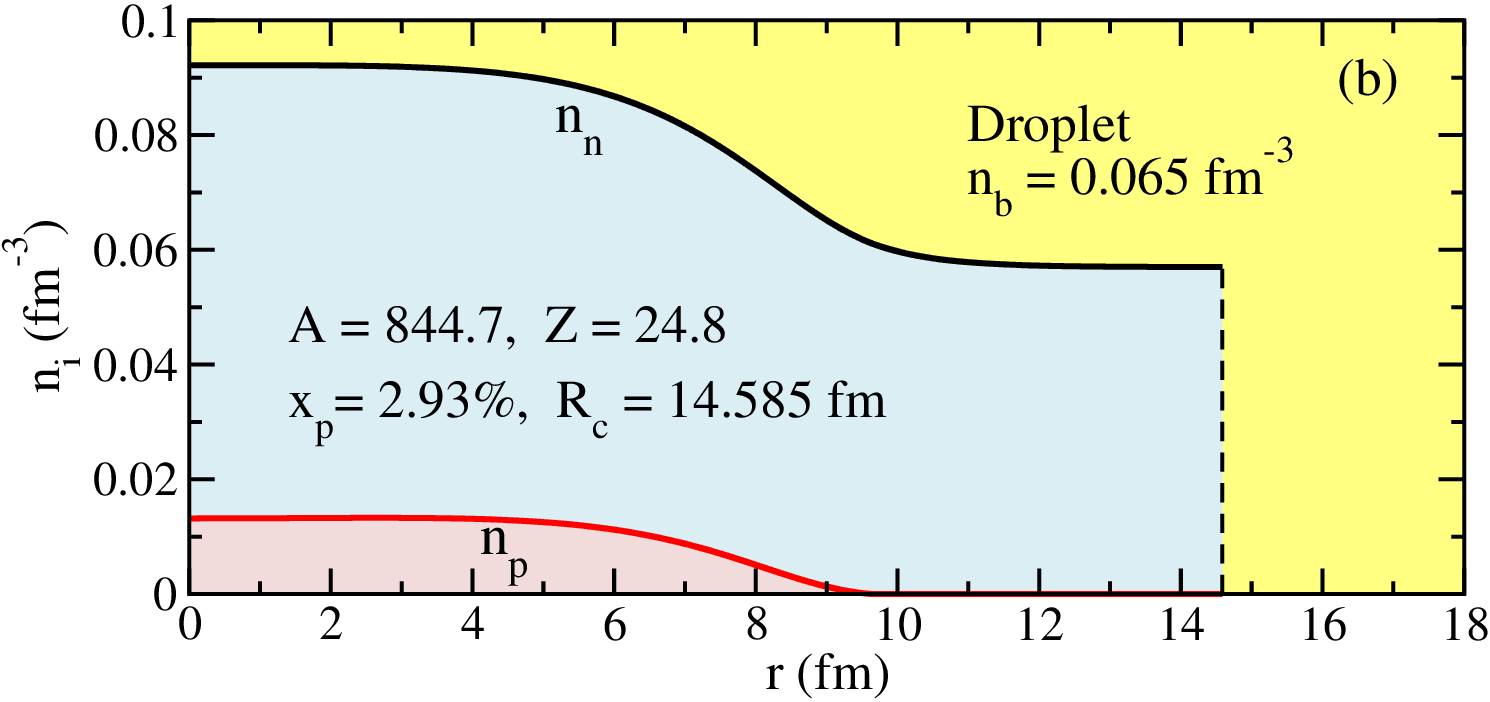}
\includegraphics[width=1.00\columnwidth,clip=true]{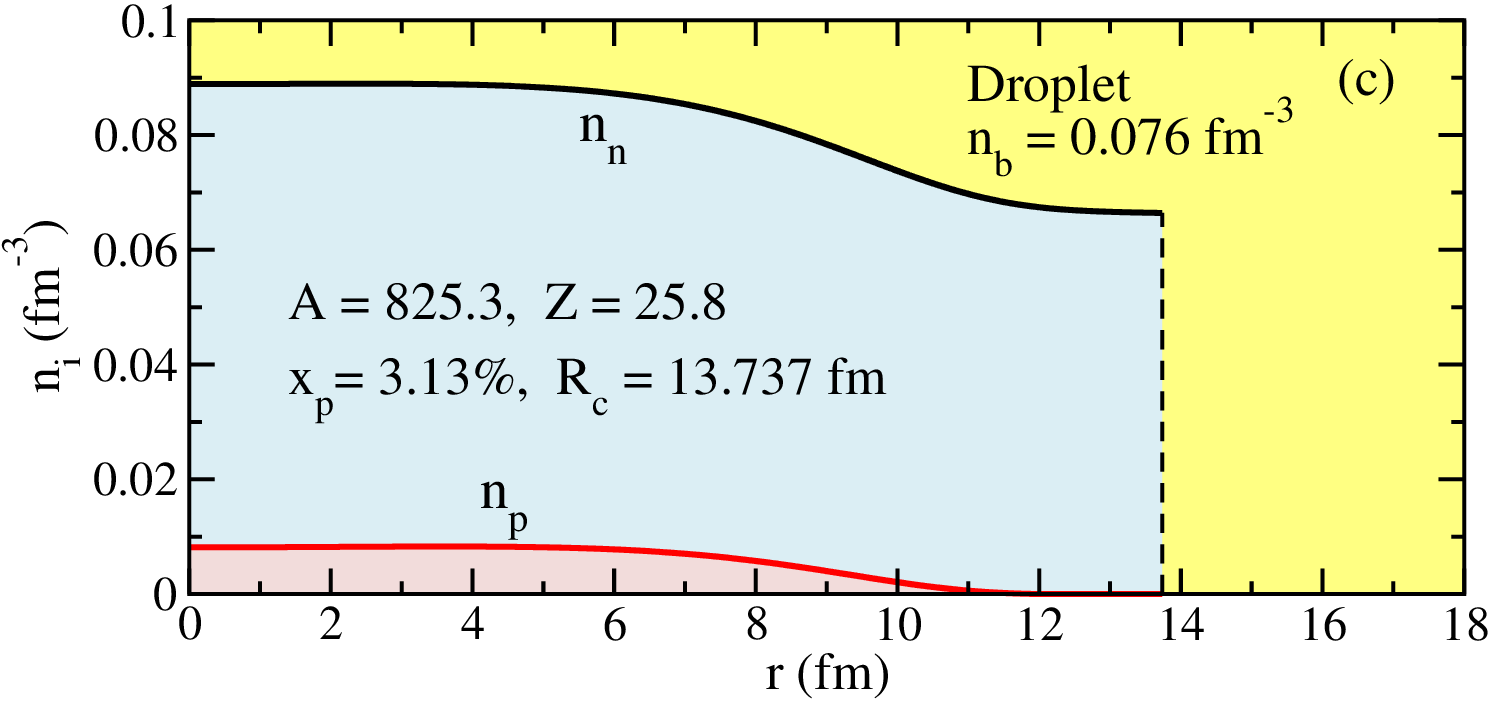}
\caption{\label{fig7} (a) Optimal density profile of neutrons $n_{n}$ and 
protons $n_{p}$ for spherical nuclear droplets at average baryon density $\nb
= 0.0475$ fm$^{-3}$. The associated baryon and proton numbers,
proton fraction $x_p=Z/A$ in percentage, and radius of the cell are shown. The 
vertical dashed line marks the location of the end of the WS cell. (b) The same 
for $\nb = 0.065$ fm$^{-3}$. (c) The same for $\nb = 0.076$ fm$^{-3}$.}
\end{figure}

\begin{figure}
\includegraphics[width=0.98\columnwidth,clip=true]{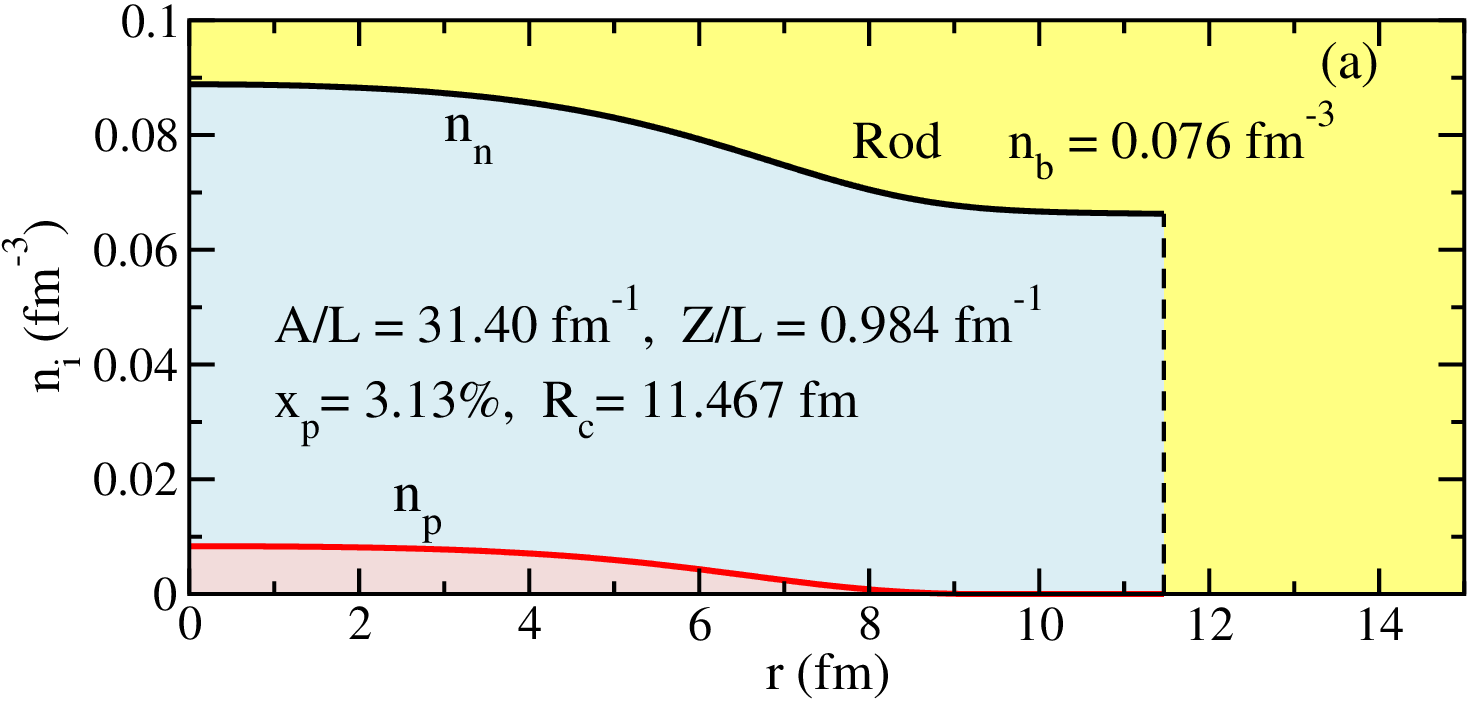}
\includegraphics[width=0.98\columnwidth,clip=true]{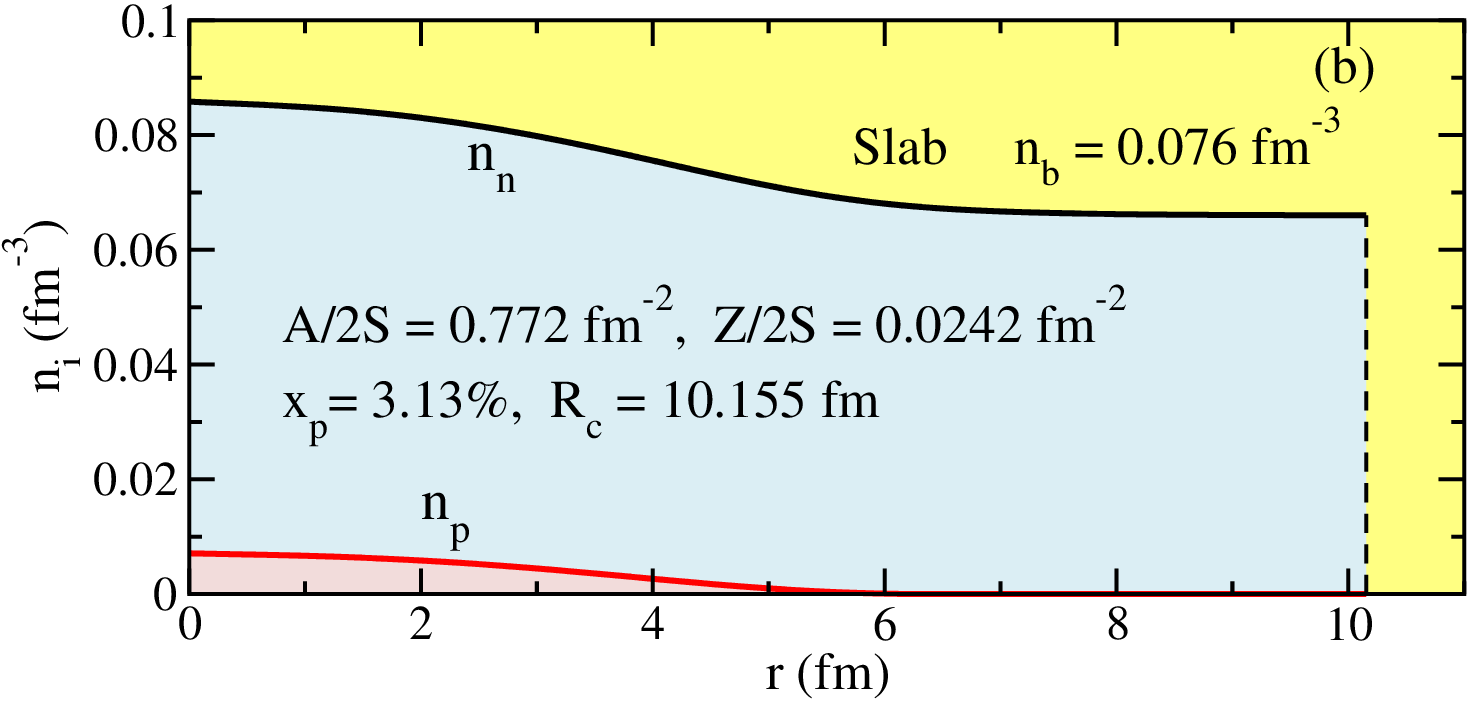}
\includegraphics[width=0.98\columnwidth,clip=true]{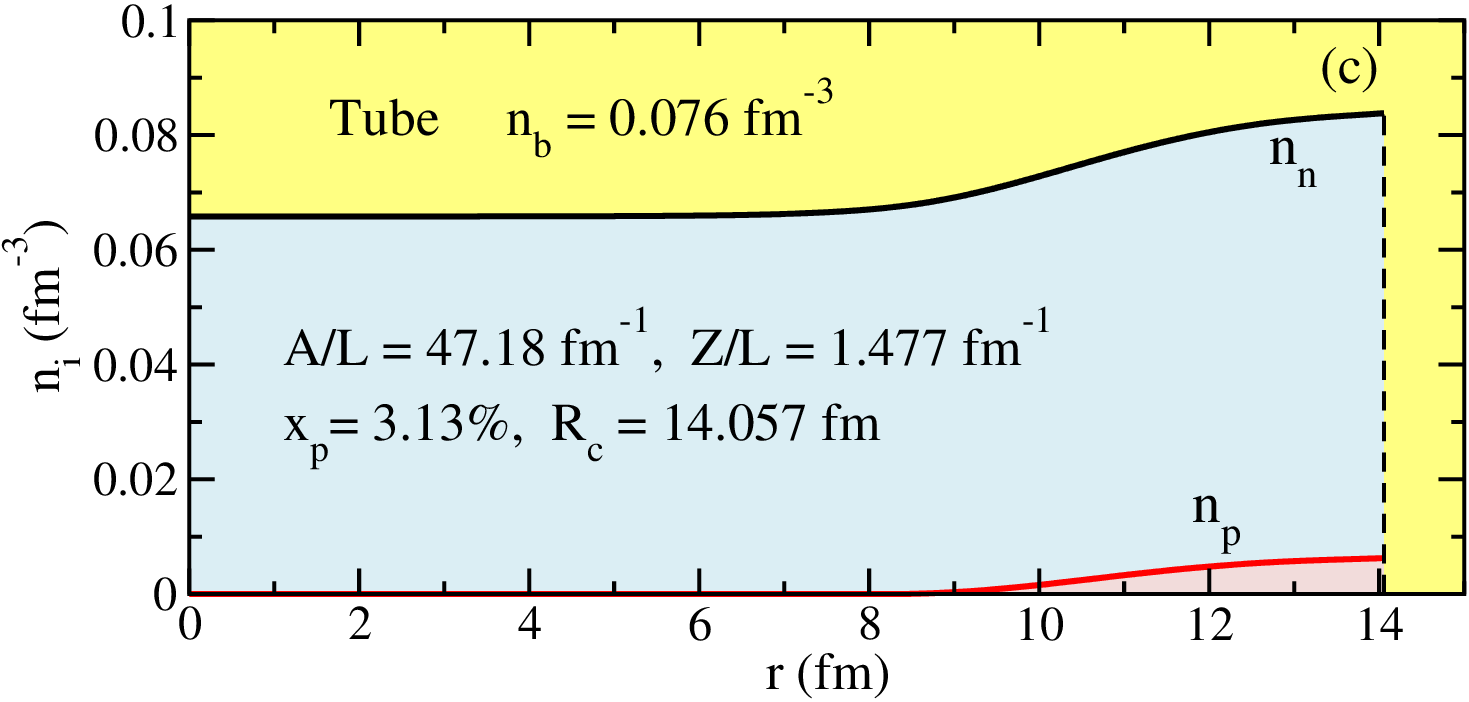}
\includegraphics[width=1.00\columnwidth,clip=true]{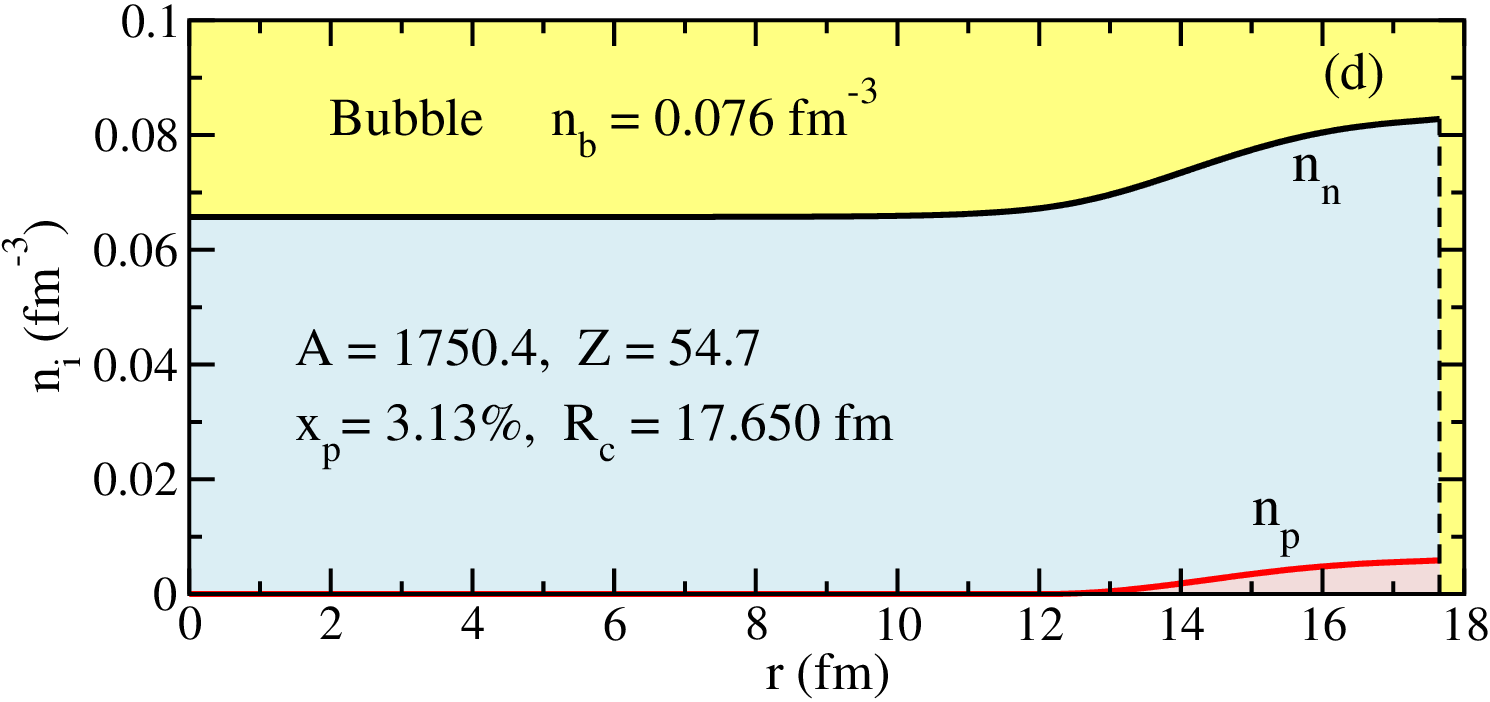}
\caption{\label{fig8} (a) Optimal density profile of neutrons $n_{n}$ and 
protons $n_{p}$ for rod shapes at average baryon density $\nb = 0.076$
fm$^{-3}$. The associated baryon and proton numbers, proton
fraction $x_p=Z/A$ in percentage, and radius of the cell are shown. The vertical 
dashed line marks the location of the end of the WS cell. (b) The same for slab 
shapes. (c) The same for tube shapes. (d) The same for bubble shapes. We note that 
the scale on the horizontal axis is the same in Figs.~\ref{fig8}c 
and~\ref{fig8}a, and in Figs.~\ref{fig8}d and~\ref{fig7}c.}
\end{figure}

Before leaving this section, in Fig.~\ref{fig7} we display the spatial dependence
of the self-consistent neutron and proton density profiles for the optimal 
solutions in spherical WS cells with average baryon densities $\nb=0.0475$
fm$^{-3}$, 0.065 fm$^{-3}$, and 0.076 fm$^{-3}$. It is observed that in denser 
matter the size of the WS cell decreases, as we discussed previously, and that the 
amount of free neutrons in the gas increases, as expected. It can be seen that the 
nuclear surface is progressively washed out with increasing average baryon density 
as the nucleon distributions become more uniform. At high $\nb$ the 
density profile inside the WS cell extends towards the edge of the cell, 
pointing out that the WS approximation may be close to its limits of validity 
\citep{negele73,chamel07,baldo07,pastore11,gogelein07,newton09}.
Although the proton number $Z$ is similar for the 
three average baryon densities of Fig.~\ref{fig7}, the local 
distribution of the protons is very different in the three cases.
In Fig.~\ref{fig7}c the proton density profile extends more than 3 fm farther from 
the origin than in Fig.~\ref{fig7}a, while the central value of the proton density 
has decreased by more than a factor 2, hinting at the fact that the neutrons have
a strong drag effect on the protons.
Figure~\ref{fig8} presents the nucleon density profiles obtained for  
cylindrical and planar geometries at the same average density $\nb=0.076$
fm$^{-3}$ as in Fig.~\ref{fig7}c. From Figs.~\ref{fig7}c (droplets), 
\ref{fig8}a (rods), and~\ref{fig8}b (slabs) we see that the size of the WS 
cells decreases with decreasing dimensionality, i.e. $R_{c,\rm droplet} >
R_{c,\rm rod} > R_{c,\rm slab}$. At high average densities near the crust-core
transition, nucleons inside the WS cell can arrange themselves in such a way that
the region of higher density is concentrated at the edge of the cell, leaving the
uniform region of lower density in the inner part of the cell. This
distribution of nucleons corresponds to the cylindrical tube and spherical bubble 
configurations. In Figs.~\ref{fig8}c and \ref{fig8}d, we plot the neutron and 
proton density profiles of the optimal solution for tubes and bubbles at 
$\nb=0.076$ fm$^{-3}$. At equal average density, the size of the cells
containing tubes and bubbles is larger than the size of the cells accommodating 
rods and droplets, respectively, as can be appreciated by comparing 
Fig.~\ref{fig8}a for rods with Fig.~\ref{fig8}c for tubes, and Fig.~\ref{fig7}c 
for droplets with Fig.~\ref{fig8}d for bubbles.
As a consequence of this fact and of the effectively larger value of the
integration factors $2\pi r$ and $4\pi r^2$ when the densities are accumulated 
near the edge of the cell, the total number of nucleons and the atomic number 
in the tube and bubble cells is about 1.5--2 times larger than in their rod and 
droplet counterparts. The proton fraction $x_p=Z/A$ is, however, practically the 
same for all geometries.

\subsection{Equation of state of the inner crust}
\label{sec:inneos}

\begin{figure}
\includegraphics[width=1.00\columnwidth,clip=true]{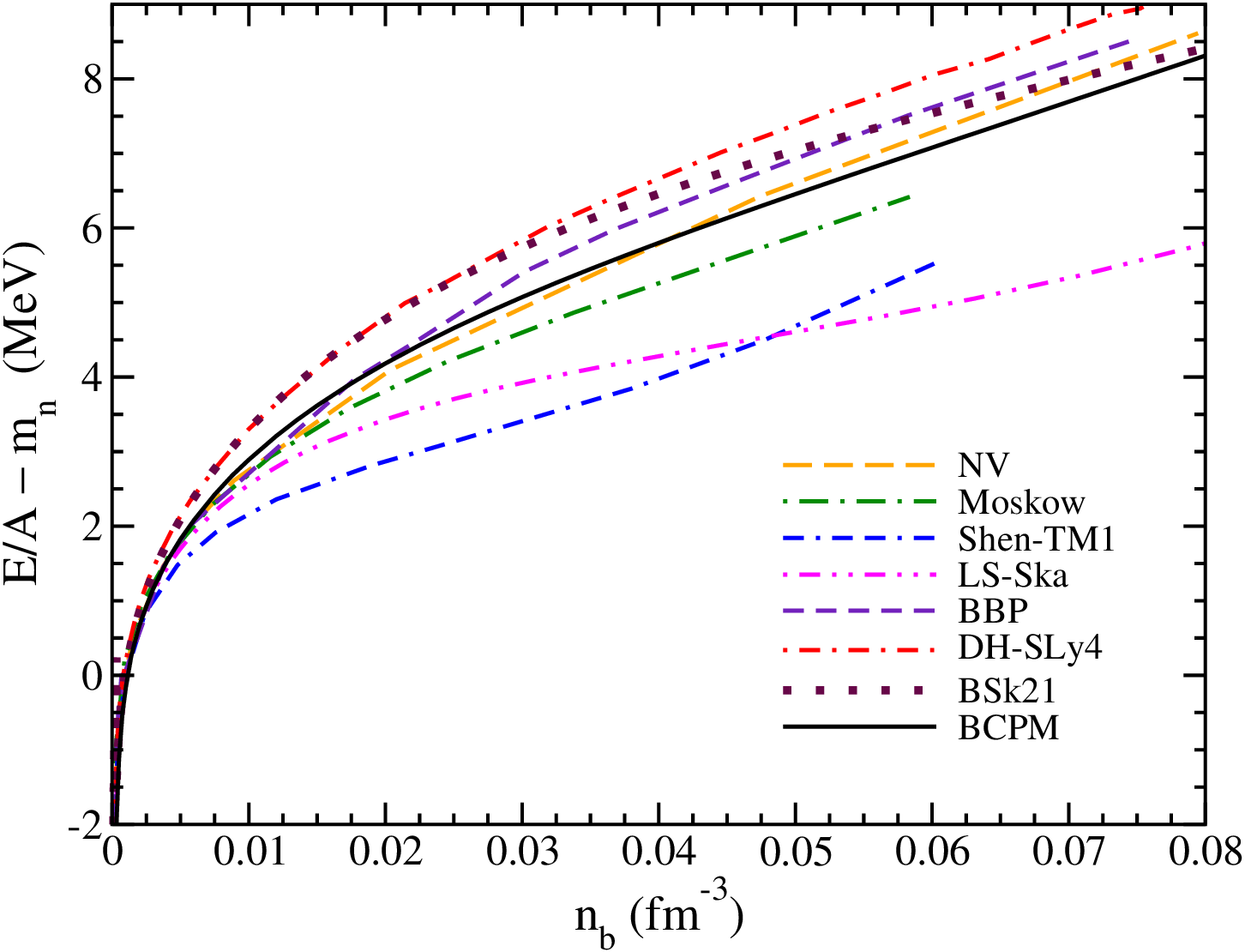}
\caption{\label{fig9} Energy per baryon relative to the neutron rest mass as a 
function of the average baryon density in the inner crust for the BCPM functional 
and other EoSs.}
\end{figure}

The energy per baryon in the inner crust predicted by the BCPM functional is 
displayed against the average baryon density in Fig.~\ref{fig9}. The result
is compared with other calculations available in the literature. This comparison
will be, at the same time, useful to discriminate popular EoSs used in
neutron-star modeling among each other.
We show in Fig.~\ref{fig9} the results of the quantal calculations of Negele and 
Vautherin \citep{negele73} (label NV) and of Baldo et al.\ \citep{baldo07} (label 
Moskow). These two EoSs of the inner crust
include shell effects and in the case of Moskow also pairing correlations. In addition to
these models, we compare with some of the EoSs that have been devised to describe
the complete neutron-star structure. The EoS by Baym, Bethe, and Pethick
\citep{bbp71,bps71} (label BBP), the EoS by Lattimer and Swesty \citep{ls91} in its
Ska version \citep{LSweb} (label LS-Ska), and the EoS by Douchin and Haensel
\citep{douchin01} (label DH-SLy4) were all obtained using the CLDM model to
describe the inner crust. The results by Shen et al.\
\citep{shen98a,shen98b,Shenweb} (label Shen-TM1) were computed in the TF approach
with trial nucleon density distributions. Finally, in the recent EoS of the
Brussels-Montreal BSk21 force \citep{pearson12,fantina13,potekhin13,goriely10} the
inner crust was calculated in the extended TF approach with trial nucleon density
profiles and with proton shell corrections incorporated by means of the Strutinsky
method.

The energy in the inner crust is largely influenced by the properties of the
neutron gas and, therefore, the EoS of neutron matter of the different
calculations plays an essential role. The NV calculation \citep{negele73} is based
on a local energy density functional that closely reproduces the
Siemens-Pandharipande EoS of neutron matter \citep{siemens71} in the
low-density regime. The Moskow calculation \citep{baldo07} employs a
semi-microscopic energy density functional obtained by combining the
phenomenological functional of Fayans et al.\ \citep{fayans00} inside the nuclear
cluster with a microscopic part calculated in the Brueckner theory with the
Argonne $v_{18}$ potential \citep{v18} to describe the neutron environment in the
low-density regime \citep{baldo04}. The BBP calculation \citep{bbp71,bps71} gives
the EoS based on the Brueckner calculations for pure neutron matter of Siemens
\citep{siemens71}. The LS-Ska \citep{ls91,LSweb} and DH-SLy4
\citep{douchin01} EoSs were constructed using the Skyrme effective nuclear forces
Ska and SLy4, respectively. The SLy4 Skyrme force \citep{chabanat98} was
parametrized, among other constraints, to be consistent with the microscopic
variational calculation of neutron matter of Wiringa et al.\ \citep{wiringa88} 
above the nuclear saturation density. The Shen-TM1 EoS
\citep{shen98a,shen98b,Shenweb} was computed using the relativistic mean field
parameter set TM1 for the nuclear interaction. The calculations of 
LS \citep{ls91,LSweb} and Shen et al.\ \citep{shen98a,shen98b,Shenweb} are the 
two EoS tables in more widespread use for astrophysical simulations. The BSk21 
EoS \citep{pearson12,fantina13,potekhin13,goriely10} is based on a Skyrme force 
with the parameters accurately fitted to the known nuclear masses and
constrained, among various physical conditions, to the neutron matter EoS derived 
within modern many-body approaches which include the contribution of 
three-body forces.
 
It can be realized in Fig.~\ref{fig9} that the energies per baryon predicted in
the inner crust by the BCPM functional and by the NV calculation \citep{negele73}
lie close over a wide range of densities, as also noticed before \citep{50years}.
The result of the BBP model \citep{bbp71,bps71} agrees similarly with BCPM and NV
at low densities, while above $\nb\sim0.02$ fm$^{-3}$ it predicts somewhat larger
energies than BCPM and NV. The DH-SLy4 calculation \citep{douchin01} consistently
predicts throughout the inner crust the largest energies of all the models
analyzed in Fig.~\ref{fig9}. The energies of the BSk21 calculation
\citep{pearson12,fantina13,potekhin13,goriely10} are very close to those of DH-SLy4
up to $\nb\sim0.03$--0.04 fm$^{-3}$. When the transition to the core is 
approached, the BSk21 energies become closer to the BCPM and NV results than to 
the DH-SLy4 result. The Moskow calculation \citep{baldo07} predicts lower 
energies than the previous models. However, the most remarkable differences are 
found with the results of the LS-Ska \citep{ls91,LSweb} and Shen-TM1 
\citep{shen98a,shen98b,Shenweb} calculations. It seems evident that the BCPM 
functional, as well as the results of the models constrained by some 
information of microscopic calculations (NV, Moskow, BBP, DH-SLy4, and BSk21), 
predicts overall substantially larger energies per baryon in the inner crust 
than the LS-Ska and Shen-TM1 models that do not contain explicit information of 
microscopic neutron matter calculations.

\begin{figure}
\includegraphics[width=1.00\columnwidth,clip=true]{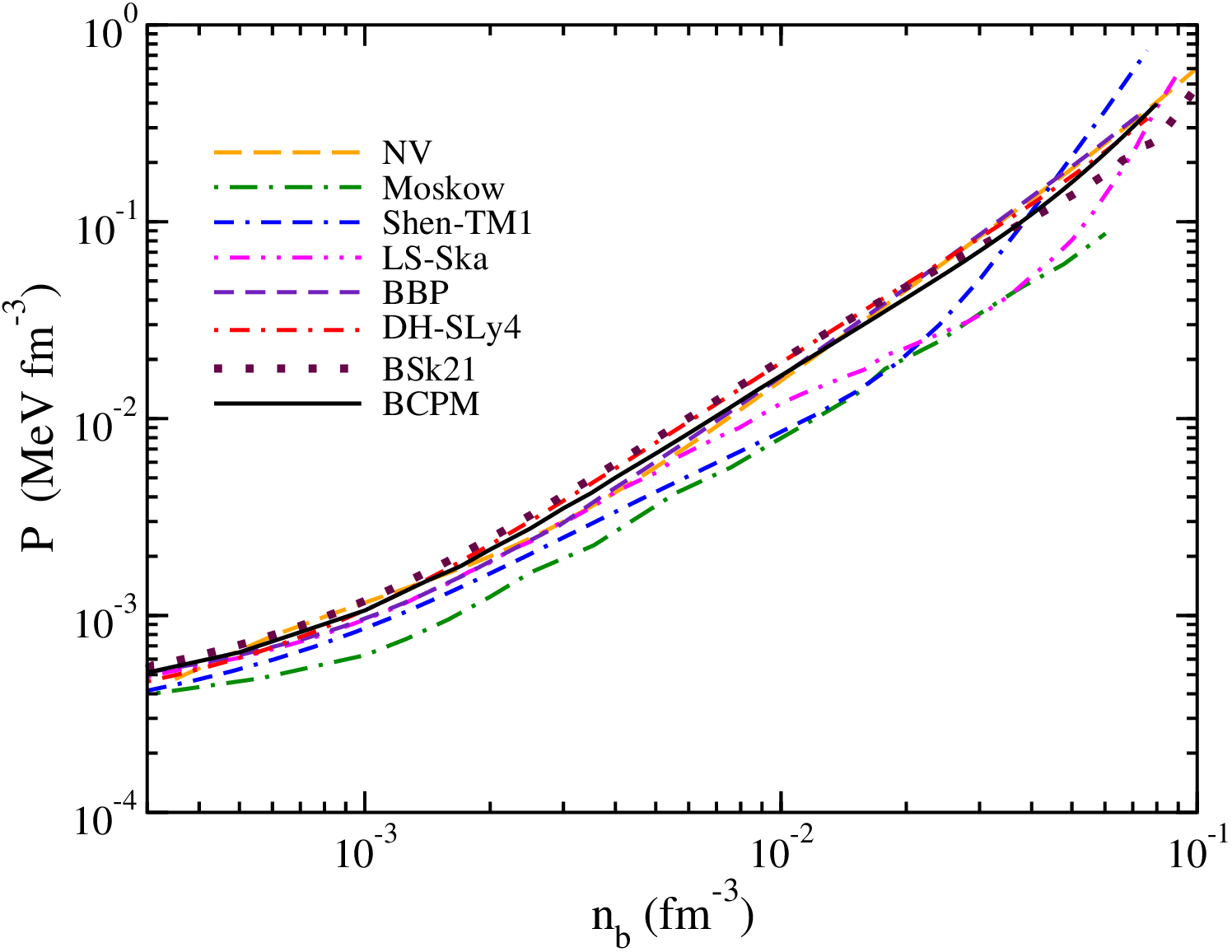}
\caption{\label{fig10} Pressure as a function of the average baryon density in the 
inner crust for the BCPM functional and other EoSs. The figure starts at 
$\nb=3\times 10^{-4}$ fm$^{-3}$ where Fig.~\ref{fig2} ended.}
\end{figure}

The pressure in the crust is an essential ingredient entering the 
Tolman-Oppenheimer-Volkoff equations \citep{shapiro} that determine the
mass-radius relation in neutron stars. The crustal pressure has also significant
implications for astrophysical phenomena such as pulsar glitches
\citep{piekarewicz2014}. As expressed in Eq.~(\ref{eq22}), the pressure in the
inner crust is provided by the free gases of the electrons and of the interacting
dripped neutrons (aside from a correction from Coulomb exchange). We note, however,
that the pressure obtained in a WS cell in the inner crust differs from the value
in homogeneous $npe$ matter in $\beta$-equilibrium at the same average density
$\nb$ owing to the lattice effects, which influence the electron and neutron
gases. The lattice effects take into account the presence of nuclear structures in
the crust and are automatically included in the self-consistent TF calculation.

\begin{table}
\caption{\label{t:eos.crust} Equation of state of the inner crust.}
\centering
\begin{tabular}{|c c c c c c|}
\hline\hline
& $\nb$ & $\rmatt$ & $P$  & $\Gamma$ & \\
& (fm$^{-3}$) & (g cm$^{-3}$) & (erg cm$^{-3}$) & & \\
\hline
&    0.0003  &  5.0138E+11  &  8.2141E+29  & 0.443 & \\
&    0.0005  &  8.3646E+11  &  1.0417E+30  & 0.560 & \\
&\ \,0.00075 &  1.2555E+12  &  1.3844E+30  & 0.747 & \\
&    0.0010  &  1.6746E+12  &  1.6984E+30  & 0.874 & \\
&    0.0014  &  2.3456E+12  &  2.3837E+30  & 1.004 & \\
&    0.0017  &  2.8488E+12  &  2.8551E+30  & 1.070 & \\
&    0.0020  &  3.3522E+12  &  3.4653E+30  & 1.121 & \\
&    0.0025  &  4.1915E+12  &  4.4319E+30  & 1.183 & \\
&    0.0030  &  5.0310E+12  &  5.6159E+30  & 1.226 & \\
&    0.0035  &  5.8706E+12  &  6.7099E+30  & 1.257 & \\
&    0.0040  &  6.7106E+12  &  8.0318E+30  & 1.280 & \\
&    0.0050  &  8.3909E+12  &  1.0646E+31  & 1.307 & \\
&    0.0060  &  1.0072E+13  &  1.3476E+31  & 1.319 & \\
&    0.0075  &  1.2594E+13  &  1.8085E+31  & 1.322 & \\
&    0.0088  &  1.4781E+13  &  2.2469E+31  & 1.318 & \\
&    0.0100  &  1.6801E+13  &  2.6490E+31  & 1.312 & \\
&    0.0120  &  2.0168E+13  &  3.3595E+31  & 1.303 & \\
&    0.0135  &  2.2694E+13  &  3.9198E+31  & 1.299 & \\
&    0.0150  &  2.5221E+13  &  4.5016E+31  & 1.297 & \\
&    0.0170  &  2.8591E+13  &  5.2957E+31  & 1.300 & \\
&    0.0180  &  3.0276E+13  &  5.7163E+31  & 1.303 & \\
&    0.0200  &  3.3648E+13  &  6.5647E+31  & 1.314 & \\
&    0.0225  &  3.7864E+13  &  7.6683E+31  & 1.334 & \\
&    0.0250  &  4.2081E+13  &  8.8299E+31  & 1.360 & \\
&    0.0275  &  4.6299E+13  &  1.0060E+32  & 1.392 & \\
&    0.0300  &  5.0519E+13  &  1.1361E+32  & 1.427 & \\
&    0.0325  &  5.4740E+13  &  1.2746E+32  & 1.466 & \\
&    0.0350  &  5.8962E+13  &  1.4221E+32  & 1.507 & \\
&    0.0375  &  6.3186E+13  &  1.5794E+32  & 1.550 & \\
&    0.0400  &  6.7411E+13  &  1.7473E+32  & 1.594 & \\
&    0.0425  &  7.1637E+13  &  1.9266E+32  & 1.638 & \\
&    0.0450  &  7.5864E+13  &  2.1181E+32  & 1.681 & \\
&    0.0475  &  8.0092E+13  &  2.3227E+32  & 1.725 & \\
&    0.0500  &  8.4322E+13  &  2.5411E+32  & 1.767 & \\
&    0.0520  &  8.7706E+13  &  2.7258E+32  & 1.800 & \\
&    0.0540  &  9.1092E+13  &  2.9200E+32  & 1.832 & \\
&    0.0560  &  9.4478E+13  &  3.1239E+32  & 1.864 & \\
&    0.0580  &  9.7865E+13  &  3.3370E+32  & 1.893 & \\
&    0.0600  &  1.0125E+14  &  3.5604E+32  & 1.922 & \\
&    0.0620  &  1.0464E+14  &  3.7946E+32  & 1.950 & \\
&    0.0640  &  1.0803E+14  &  4.0390E+32  & 1.976 & \\
&    0.0650  &  1.0973E+14  &  4.1651E+32  & 1.988 & \\
&    0.0660  &  1.1142E+14  &  4.2941E+32  & 2.000 & \\
&    0.0680  &  1.1481E+14  &  4.5601E+32  & 2.023 & \\
&    0.0700  &  1.1821E+14  &  4.8370E+32  & 2.045 & \\
&    0.0720  &  1.2160E+14  &  5.1245E+32  & 2.065 & \\
&    0.0740  &  1.2499E+14  &  5.4230E+32  & 2.083 & \\
&    0.0750  &  1.2669E+14  &  5.5763E+32  & 2.091 & \\
&    0.0760  &  1.2839E+14  &  5.7321E+32  & 2.100 & \\
&    0.0770  &  1.3008E+14  &  5.8903E+32  & 2.107 & \\
&    0.0780  &  1.3178E+14  &  6.0502E+32  & 2.114 & \\
&    0.0790  &  1.3348E+14  &  6.2077E+32  & 2.121 & \\
&    0.0800  &  1.3518E+14  &  6.3548E+32  & 2.127 & \\
\hline
\end{tabular}
\end{table}

The pressure predictions in the inner crust by the BCPM functional are shown in
Fig.~\ref{fig10} in comparison with the predictions by the same models discussed
in Fig.~\ref{fig9}. The initial baryon density in Fig.~\ref{fig10} corresponds to
the last density shown in Fig.~\ref{fig2} when we studied the EoS of the outer
crust in Sec.~\ref{sec:outer}. In the inner crust, the pressure from the BCPM
functional is comparable in general to the results of the NV, BBP,
DH-SLy4, and BSk21 calculations. Particular agreement is observed in the inner
crust regime between the BCPM and BSk21 pressures up to relatively high crustal
densities. In contrast, large differences are found when the BCPM pressure is
compared with the values from the Moskow, LS-Ska, and Shen-TM1 models. As the
crust-core transition is approached, these differences can be as large as a factor
of two, which may have an influence on the predictions of the mass-radius
relationship of neutron stars, particularly in small mass stars.
In addition to the spherical shape, we have evaluated the pressure for the 
non-spherical WS cells and the hollow shapes in the regime of high average baryon
densities using the BCPM functional. However, on the one hand, as noted before,
the pasta phases appear as the preferred configuration only in a relatively
narrow range of densities between $\nb \simeq 0.067$~fm$^{-3}$ and $\nb \simeq
0.08$~fm$^{-3}$. On the other hand, the differences between the pressure of the 
spherical shape and the pressure from the successively favoured shapes are small,
generally of the order of 1--2~keV/fm$^{3}$ or less. Therefore, as we did in
\citep{baldo13a}, we have taken as a representative result for the whole inner
crust the pressure calculated in the spherical droplet configurations. The
corresponding EoS data are collected in Table~\ref{t:eos.crust}.

\begin{figure}
\includegraphics[width=1.00\columnwidth,clip=true]{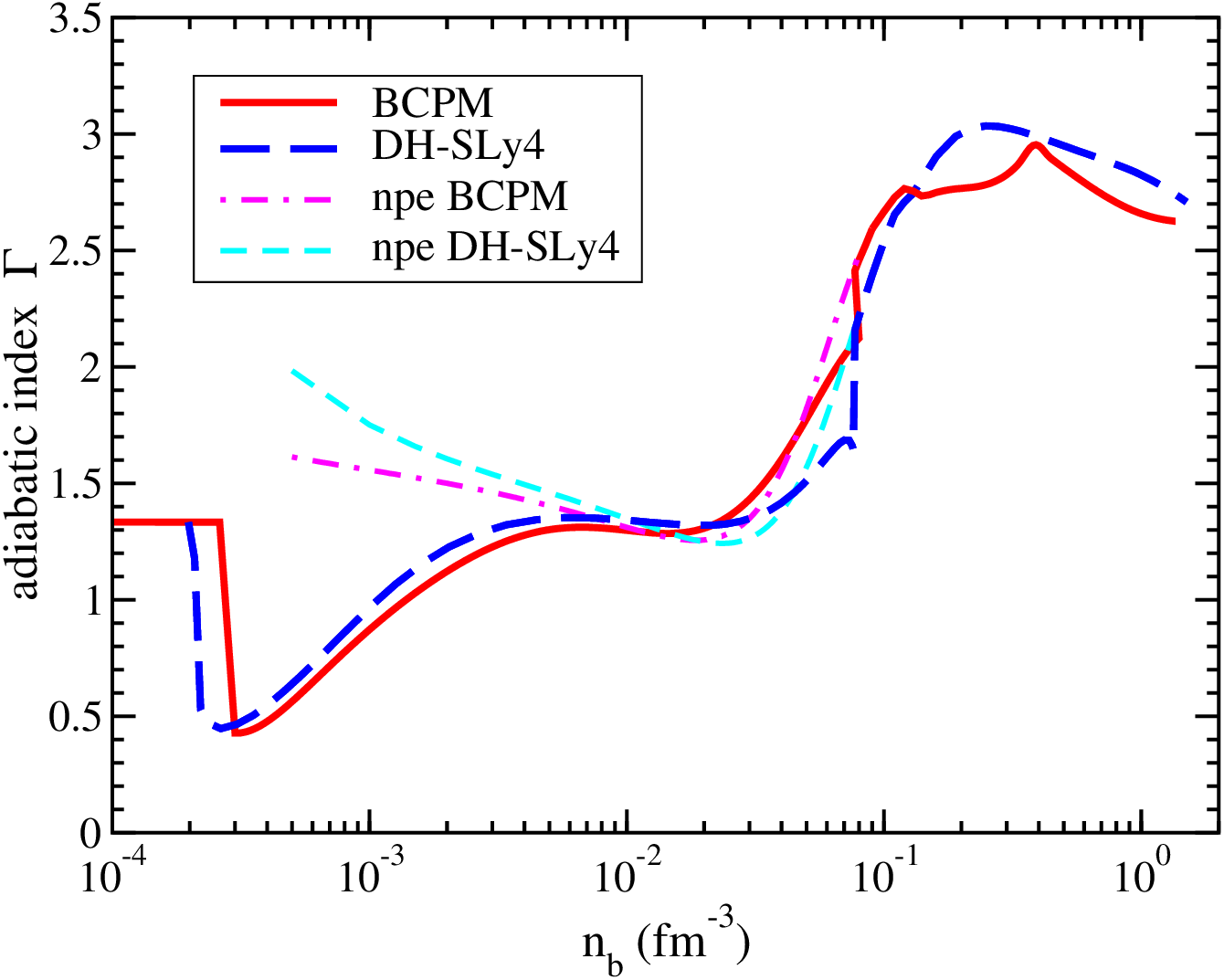}
\caption{\label{fig12} The adiabatic index from the  EoSs of BCPM and DH-SLy4 
\citep{douchin01}. The result calculated in homogeneous $npe$ matter is also
shown.}
\end{figure}

An important quantity, which actually determines the response of the crust
to the compression or decompression of matter, is the so-called adiabatic index 
\begin{equation}
\Gamma = \frac{\rmatt+P}{P} \frac{d P}{d\rmatt} = \frac{\nb}{P} \frac{d P}{d \nb},
\end{equation}
where $P$ is the pressure, $\rmatt$ the mass density of matter, and $\nb$ the
average baryon density. In Fig.~\ref{fig12} we plot the adiabatic index from the 
BCPM and DH-SLy4 \citep{douchin01} EoS in the region from the last layers of the 
outer crust until a density well within the NS core (discussed in the next 
section).
In the bottom layers of the outer crust the pressure is governed almost entirely 
by the ultra-relativistic electron gas, so that the value of $\Gamma$ is quite
close to $4/3$. At the neutron drip point, $\Gamma$ sharply decreases by
more than a factor of two due to the dripped neutrons that strongly soften the 
EoS. The just dripped neutrons contribute to the average baryon
density but exert very little pressure. In the inner crust region, when the 
density increases the adiabatic index grows because the pressure of the 
neutron gas also increases. In our EoS the adiabatic index of the inner crust 
changes from $\Gamma \simeq 0.45$ after the neutron drip point up to $\Gamma 
\simeq 2.1$  near the crust-core transition ($\nb \simeq 0.08$~fm$^{-3}$).

In the same Fig.~\ref{fig12} we report $\Gamma$ computed in a single phase of 
homogeneous $npe$ matter in $\beta$-equilibrium. It is observed that $\Gamma$ in 
the inner crust almost coincides with the result in a single phase for 
densities between $\nb \sim 0.01$~fm$^{-3}$ and 0.05~fm$^{-3}$.
Above 0.05~fm$^{-3}$, the adiabatic index of homogeneous $npe$ matter grows 
faster than in the inner crust, because the latter is softened by the presence 
of nuclear structures and the coexistence between the two phases. The 
predictions of the DH-SLy4 EoS \citep{douchin01} for $\Gamma$ in the inner crust 
show a similar qualitative behaviour but differ quantitatively from the BCPM EoS. 
For example, at the bottom of the inner crust the adiabatic index in 
the DH-SLy4 EoS is $\Gamma \sim 1.6$, while it takes a value of 2.1 in BCPM. This 
difference seems to indicate that the interacting part of the neutron gas at 
densities near the crust-core transition is weaker in SLy4 than in the BCPM 
functional (also see in this respect Fig.~1 of \citep{baldo04}). 
When the transition to the core is reached, $\Gamma$ increases in a discontinuous 
way from 2.1 to 2.5 in the BCPM EoS, due to the change from two phases to 
a single phase. With higher density in the core, the adiabatic index stiffens 
from the increase of the repulsive contributions in the nucleon-nucleon 
interaction. It is interesting to note that $\Gamma$ exhibits a small sharp 
drop at the muon onset, which in BCPM is located at a density $\nb \simeq 
0.13$~fm$^{-3}$. It arises from the appearance of muons that replace some 
high-energy electrons and effectively reduce the pressure at this density. At 
higher densities $\Gamma$ remains roughly constant, which is partly due to the 
increasing proton fraction in the $npe\mu$ matter of the core.

\begin{figure}
\includegraphics[width=1.00\columnwidth,clip=true]{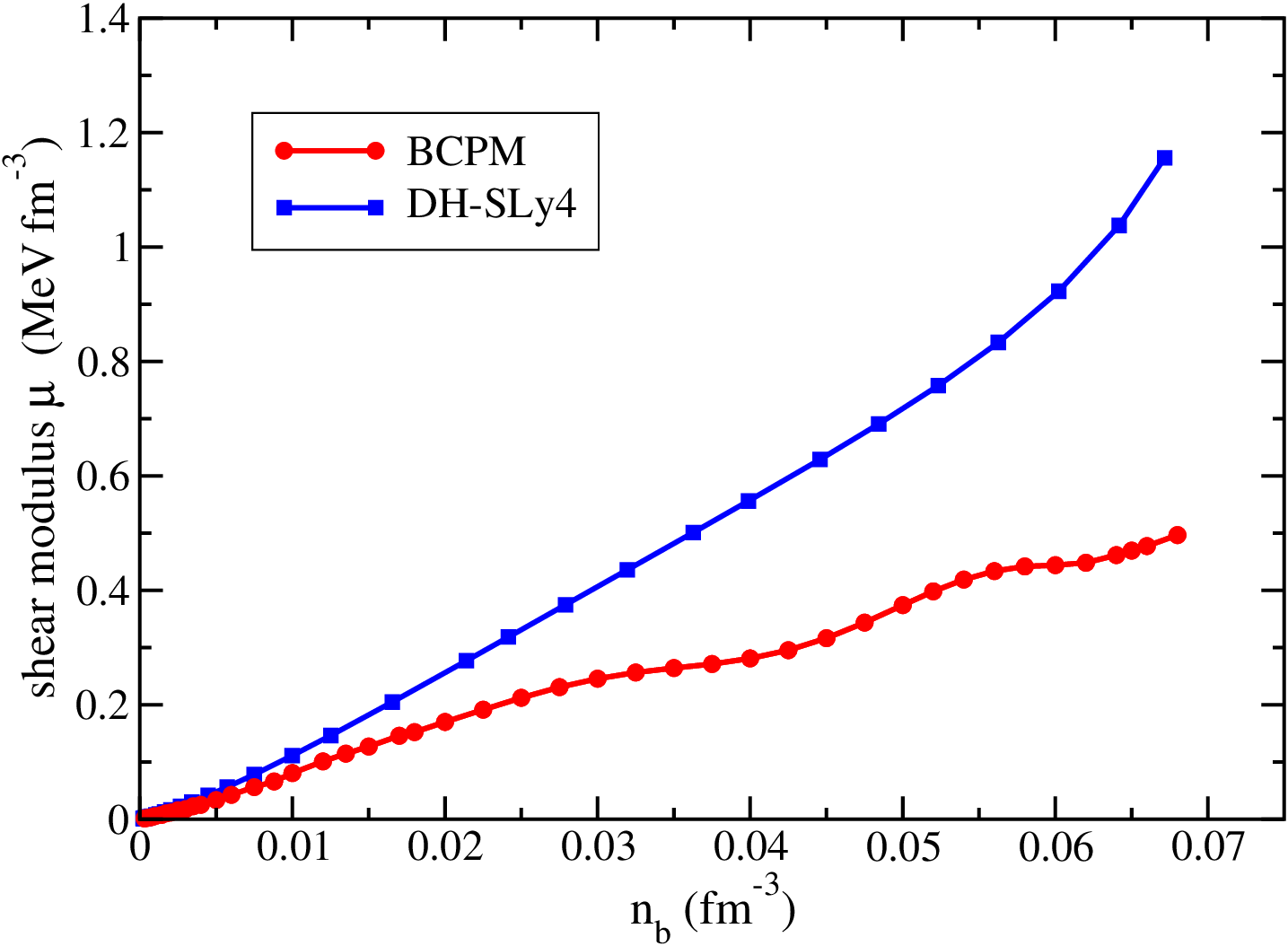}
\caption{\label{fig14} The shear modulus from the EoS of BCPM and
DH-SLy4~\citep{douchin01}.}
\end{figure}

Quasi-periodic oscillations in giant flares emitted by highly-magnetized 
neutron stars are signatures of the fundamental seismic shear mode, which is 
specially sensitive to the nuclear physics of neutron-star crusts 
\citep{steiner09,sotani12}. An important quantity for describing shear modes is 
the effective shear modulus $\mu$. It can be estimated from the known formula 
for a bcc Coulomb crystal at zero temperature 
\citep{steiner09,sotani12,strohmayer91}:
\begin{equation}
\mu = 0.1194 \,\nb \frac{(Ze)^2}{R_c} ,
\label{shear}
\end{equation}
where $Z$ and $R_c$ are the proton number and the radius of a spherical WS cell 
having average baryon density $\nb$.
In Fig.~\ref{fig14} we display the effective shear modulus from our 
calculation with the BCPM functional along with the result from DH-SLy4 
\citep{douchin01}. Because the elasticity for pasta phases, with the 
exception of spherical bubbles, is expected to be much lower than for spherical 
nuclei~\citep{sotani12}, we restrict the plot in Fig.~\ref{fig14} to spherical 
configurations, i.e. up to an average density $\nb=0.067$~fm$^{-3}$ where pasta 
phases start to be the most favourable configuration in the BCPM calculation. 
The effective shear modulus (\ref{shear}) depends on the composition of the 
crust through the proton number $Z$, which has a rather smooth variation along 
the inner crust with BCPM (see \ Fig.~\ref{fig6}), and on the size of the WS 
cells that decreases from the neutron drip till the bottom layers of the 
inner crust (see \ Fig.~\ref{fig5}). The difference between the predictions of 
BCPM and SLy4 for $\mu$ in Fig.~\ref{fig14} increases with density. At 
the higher densities of Fig.~\ref{fig14} the DH-SLy4 result is about twice the
BCPM result, pointing out the different composition and sizes of the WS cells in 
the DH-SLy4 \citep{douchin01} and BCPM models. Lower values of $\mu$ for the
BCPM case point in the direction of lower frequencies of the fundamental shear
mode, but a complete analysis is beyond the scope of the present paper.

\section{The liquid core}
\label{sec:core}

The EoS for the liquid core is derived in the framework of the 
Brueckner-Bethe-Goldstone theory~\citep{book} as described in 
Sec.~\ref{sec:micro}. The  Argonne $v_{18}$ potential \citep{v18} is used as the 
NN interaction and three-body forces based on the so-called Urbana model are 
included in the calculation to reproduce the nuclear matter saturation point
\citep{schia85,bbb,polls, Taranto:2013gya}.

In order to study the structure of the NS core, we have to calculate the 
composition and the EoS of cold, neutrino-free, catalyzed matter. As we
discussed in the Introduction, we consider a NS with a core of nucleonic matter
without hyperons or other exotic particles. We require that it contains charge
neutral matter consisting of neutrons, protons, and leptons ($e^-$, $\mu^-$) in
beta equilibrium, and compute the EoS for charge neutral and beta-stable matter
in the following standard way \citep{bbb,shapiro}. The Brueckner calculation
yields the energy density of lepton/baryon matter as a function of the different
partial densities,
\begin{eqnarray}
 \rmcore(\rnum_n,\rnum_p,\rnum_e,\rnum_\mu) &=& 
 (\rnum_n m_n +\rnum_p m_p) 
 + (\rnum_n+\rnum_p) {E\over A}(\rnum_n,\rnum_p)
\nonumber\\ 
 &+ &\, \rmcore(n_\mu) + \rmcore(n_e) \:,
\label{e:epsnn}
\end{eqnarray}
where we have used ultrarelativistic and relativistic approximations
for the energy densities of electrons and muons \citep{shapiro}, respectively.
In practice, it is sufficient to compute only the binding energy of
symmetric nuclear matter and pure neutron matter,
since within the BHF approach it has been verified \citep{hypns1,hypns2,bom1,bom2,bom3} 
that a parabolic approximation for the binding
energy of nuclear matter with arbitrary proton fraction 
$x_p=\rnum_p/\rcore$, $\rcore=\rnum_n+\rnum_p$,
is well fulfilled,
\begin{equation}
 {E\over A}(\rcore,x_p) \approx 
 {E\over A}(\rcore,x_p=0.5) + (1-2x_p)^2 E_{\rm sym}(\rcore) \:,
\label{e:parab}
\end{equation}
where the symmetry energy $E_{\rm sym}$ can be expressed in
terms of the difference of the energy per particle between pure neutron 
($x_p=0$) and symmetric ($x_p=0.5$) matter:
\begin{equation}
  E_{\rm sym}(\rcore) = 
  - {1\over 4} {\partial(E/A) \over \partial x_p}(\rcore,0)
  \approx {E\over A}(\rcore,0) - {E\over A}(\rcore,0.5) \:.
\label{e:sym}
\end{equation}

\begin{table}[t]
\caption{\label{t:compos.core} Populations of the liquid core.}
\centering
\begin{tabular}{|c c c c |}
\hline\hline
$\rcore$ &  $x_p$ & $x_e$ & $x_\mu$ \\
$(\rm {fm^{-3}})$ & (\%) & (\%)& (\%) \\
\hline
\ \,0.0825 &  2.950 & 2.950 & 0.000   \\
0.085  &  3.023 & 3.023 & 0.000       \\   
0.090  &  3.165 &  3.165 &  0.000     \\
0.100   & 3.429 &  3.429 &   0.000    \\
0.110  &  3.667 &  3.667 &   0.000    \\
0.120  &  3.882 &  3.882 &  0.000     \\
0.130  &  4.082 &  4.075 &  0.007     \\
0.160   &  4.864 &  4.480  & 0.384    \\
0.190  &  5.598 &  4.771 &  0.827     \\
0.220    &  6.274  & 5.022 & 1.253    \\
0.250   & 6.935  & 5.273  & 1.662     \\
0.280   & 7.615 &  5.544  & 2.071     \\
0.310   &  8.336 &  5.847 & 2.489     \\
0.340   & 9.105 & 6.183  & 2.922     \\
0.370   & 9.918   & 6.548 & 3.371    \\
0.400  &  10.762&  6.933  & 3.829     \\
0.430   &  11.622 & 7.330  & 4.292    \\
0.460   & 12.361  & 7.665  & 4.697    \\
0.490  & 13.108 &  8.005 &  5.102     \\
0.520  & 13.859 &  8.352 &  5.507     \\
0.550  &  14.614 &  8.702 &  5.912    \\
0.580  &  15.371 &  9.055 &  6.315    \\
0.610  &  16.126 &  9.409 & 6.717    \\
0.640  & 16.879 & 9.764 &  7.115     \\
0.670  &  17.628 &  10.117 &  7.510   \\
0.700  & 18.371 & 10.469 &   7.902    \\
0.750  & 19.591 &  11.048 &  8.542    \\
0.800  & 20.784  &  11.616 &   9.167  \\
0.850  &  21.944 &  12.170 & 9.775   \\
0.900  & 23.069 &   12.707 & 10.363   \\
0.950  &  24.156 &   13.226 & 10.930  \\
1.000  & 25.201 &   13.726 &   11.475 \\
1.100  &  27.167 &   14.667 &  12.501 \\
1.200  &  28.966 &   15.527 &  13.439 \\
1.300  & 30.601 &   16.309 &   14.292 \\
\hline
\end{tabular}
\end{table}

\begin{table}[t]
\caption{\label{t:eos.core} Equation of state of the liquid core.}
\centering
\begin{tabular}{|c c c c |}
\hline\hline
$\rcore$ & $\rmcore$ & $P$ & $\Gamma$ \\
$\rm (fm^{-3})$ & $\rm (g~cm^{-3})$ & $\rm (erg~cm^{-3})$  & \\
\hline
%
\ \,0.0825&  1.394E+14 & 6.916E+32 & 2.494 \\
0.085 &  1.437E+14 & 7.454E+32 & 2.526       \\   
0.090 &  1.522E+14 & 8.626E+32 & 2.593      \\
0.100  &  1.692E+14 & 1.138E+33 & 2.666      \\
0.110 &   1.863E+14  & 1.472E+33 &  2.728    \\
0.120 &  2.034E+14  & 1.869E+33  & 2.765     \\
0.130 &   2.206E+14 &  2.333E+33 &  2.755    \\
0.160  & 2.723E+14 &  4.120E+33  & 2.749     \\
0.190 &   3.244E+14 &  6.618E+33 &  2.764    \\
0.220   & 3.770E+14 & 9.930E+33  & 2.771     \\
0.250  & 4.303E+14  & 1.416E+34  & 2.783     \\
0.280  & 4.842E+14 &  1.943E+34  & 2.808     \\
0.310  &  5.385E+14 & 2.592E+34  & 2.856     \\
0.340  & 5.938E+14  & 3.425E+34  & 2.909     \\
0.370  & 6.501E+14  & 4.407E+34  & 2.946     \\
0.400 &  7.073E+14  & 5.546E+34  & 2.951     \\
0.430  & 7.655E+14  & 6.851E+34  & 2.926    \\
0.460  & 8.247E+14  & 8.323E+34  & 2.890     \\
0.490 & 8.852E+14 & 9.986E+34  & 2.861       \\
0.520 &  9.467E+14 & 1.183E+35 &  2.840      \\
0.550 &   1.009E+15 & 1.386E+35 &  2.821     \\
0.580 &  1.073E+15  & 1.609E+35 & 2.803     \\
0.610 &  1.139E+15 & 1.853E+35 &  2.786      \\
0.640 & 1.206E+15 &  2.118E+35 & 2.771       \\
0.670 & 1.274E+15 &  2.404E+35 &  2.756      \\
0.700 &  1.344E+15 &  2.712E+35 & 2.743      \\
0.750 &  1.464E+15 &  3.275E+35 & 2.723      \\
0.800 & 1.588E+15  & 3.902E+35 &  2.706      \\
0.850 &  1.717E+15 &  4.595E+35 &  2.691     \\
0.900 & 1.850E+15 &  5.357E+35 &  2.678     \\
0.950 & 1.988E+15  & 6.190E+35  & 2.667      \\
1.000 &  2.132E+15  & 7.095E+35 &  2.658     \\
1.100 &  2.435E+15  & 9.135E+35 &  2.643     \\
1.200 &  2.760E+15 & 1.149E+36  & 2.634     \\
1.300 & 3.108E+15  & 1.418E+36  & 2.628      \\
\hline
\end{tabular}
\end{table}

Knowing the energy density Eq.~(\ref{e:epsnn}), 
the various chemical potentials (of the species $i=n,p,e,\mu$)
can be computed straightforwardly,
\begin{equation}
 \mu_i = {\partial \rmcore \over \partial \rnum_i} \:,
\end{equation}
and the equations for beta-equilibrium,
\begin{equation}
\mu_i = b_i \mu_n - q_i \mu_e \:,
\end{equation}
($b_i$ and $q_i$ denoting baryon number and charge of species $i$)
and charge neutrality,
\begin{equation} 
 \sum_i \rnum_i q_i = 0 \:,
\end{equation}
allow one to determine the equilibrium composition $\rnum_{i}$
at given baryon density $\rcore$ and finally the EoS,
\begin{equation}
 P(\rcore) = \rcore^2 {d\over d\rcore} 
 {\rmcore(\rnum_i(\rcore))\over \rcore}
 = \rcore {d\rmcore \over d\rcore} - \rmcore 
 = \rcore \mu_n - \rmcore \:.
\end{equation}

\begin{figure}[t]
\centering
\includegraphics[width=8.5cm,angle=0,clip]{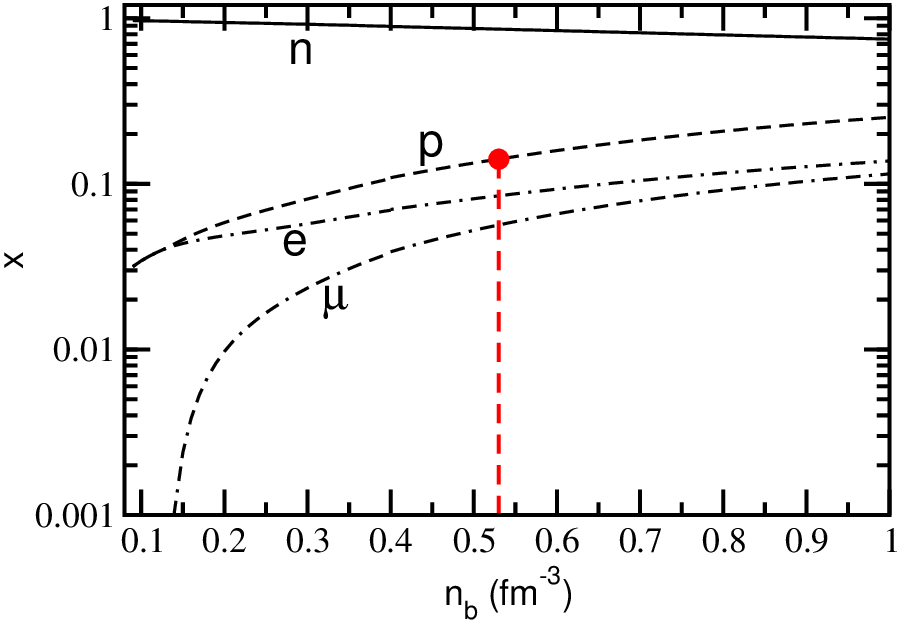}
\caption{\label{f:pop_core} The populations are displayed vs.. the nucleon density 
for the BCPM EoS discussed in the text. The full red dot indicates the value of the 
nucleon density at which direct Urca processes set in.}
\end{figure}

\begin{figure}[t]
\centering
\includegraphics[width=8.cm,angle=0,clip]{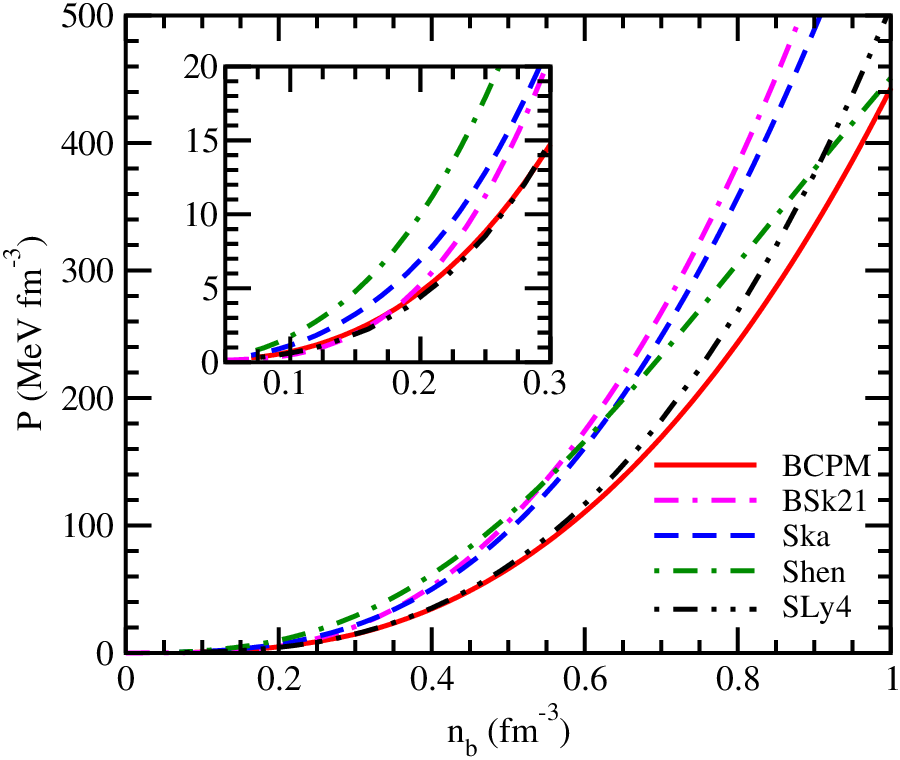}
\caption{\label{f:eos_core} The pressure is displayed vs.. the nucleon density 
for the several
EoSs discussed in the text, i.e. the BCPM (solid, red), the BSk21 (dot-dashed-dashed, magenta), 
the Lattimer-Swesty (Ska, dashed, blue), the Shen (dot-dashed, green), 
and the Douchin-Haensel (SLy4, dot-dot-dashed, black). A detail of the region
between $\rcore=0.05$~fm$^{-3}$ and 0.30~fm$^{-3}$ is shown in the inset. The
incompressibility coefficients at nuclear saturation density for these models
are $K=214$ MeV (BCPM), 230 MeV (SLy4), 246 MeV (BSk21), 263 MeV (Ska), and 281
MeV (Shen).}
\end{figure}

In Table \ref{t:compos.core} the populations are reported for a fixed nucleon density,
and are plotted in Fig.~\ref{f:pop_core}. The full red dot indicates the value of the 
nucleon density at which direct Urca processes set in. We remind that Urca processes
play an important role in the neutron star cooling \citep{1991Urca,2001Yak}. We notice that the BCPM EoS predicts
a density onset value close to 0.53 $\rm fm^{-3}$, and therefore
with our EoS medium mass NS can cool very quickly. 
In Table \ref{t:eos.core} we report the corresponding EoS, which is represented in
Fig.~\ref{f:eos_core} by a red solid curve.
We notice a remarkable similarity with the EoS derived by 
\citep{douchin01} (black curve), based on the effective nuclear interaction SLy4 
of Skyrme type. On the other hand, a strong difference with the 
Lattimer-Swesty EoS (dashed blue), the Shen EoS (dot-dashed, green), and the 
BSk21 EoS (dot-dashed-dashed, magenta) is observed at high densities. 
We recall that the pressures from the Lattimer-Swesty EoS and the Shen EoS had
already been found to differ significantly from BCPM and SLy4 for the matter at
subsaturation density in the inner crust (see \ discussion of Fig.~\ref{fig10}).
However, the BCPM and SLy4 pressures in the inner crust showed a concordance
with BSk21 that remains within the core region up to about 0.2 fm$^{-3}$ (see
inset of Fig.~\ref{f:eos_core}) but is not maintained in the extrapolation to
higher densities, where the BSk21 and Lattimer-Swesty models predict the
stiffest EoSs of Fig.~\ref{f:eos_core}.

\begin{figure}[t]
\centering
\includegraphics[width=9.cm,angle=0,clip]{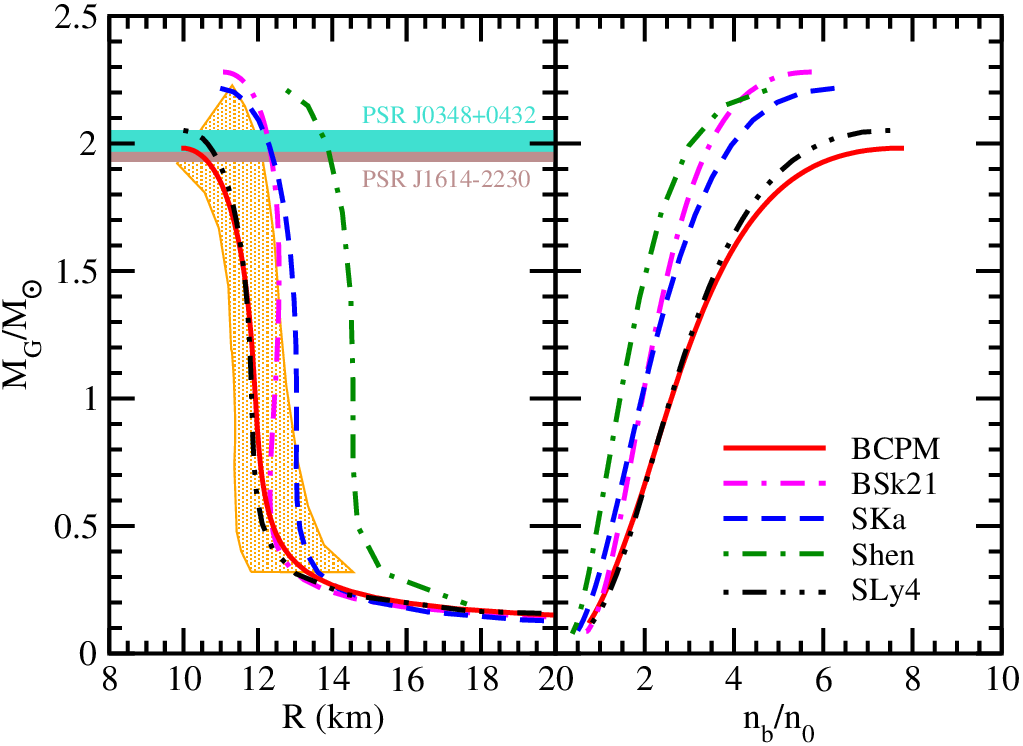}
\caption{\label{f:MR} The gravitational mass, in units of the solar mass $\rm 
M_\odot = 2 \times 10^{33} g$ is displayed 
vs.. the radius (left panel) and the central density (right panel)  in units of 
the saturation density 
$\rm \rnum_0=0.16$~fm$^{-3}$. See text for details. } 
\end{figure}

Once the EoS of the nuclear matter is known, one can solve the 
Tolman-Oppenheimer-Volkoff \citep{shapiro} equations for spherically symmetric NS:
\begin{eqnarray}
{d P\over d r}&=&- G\, {\rmcore m \over r^2} \left(1 + {P \over \rmcore} \right) \left(1 + {4\pi P r^3\over m } \right) \left(1 - {2 G m \over r }\right)^{-1} \nonumber \\
{d m\over d r}&=&4\pi r^2 \rmcore \,,
\label{eq:OV} 
\end{eqnarray}
\noindent where $G$ is the gravitational constant, $P$ the pressure, $\rmcore$ the energy density,  
$m$ the mass enclosed within a radius $r$, and $r$ the (relativistic) radius coordinate. To close the 
equations we need the relation between pressure and energy density, $P \,
=\, P(\rmcore)$, i.e. the EoS. The integration of these equations produces the mass and radius of the star 
for given central density. It turns out that the mass of the NS has a
maximum value as a function of radius (or central density), above which the star
is unstable against collapse to a black hole. The value of the maximum mass
depends on the nuclear EoS, so that the observation of a mass higher than the
maximum mass allowed by a given EoS simply rules out that EoS.
We display in Fig.\ref{f:MR} the relation between mass and radius (left
panel) and central density (right panel).  The observed trend is
consistent with the equations of state displayed in Fig.\ref{f:eos_core}. 
As expected, when the stiffness increases the NS central density decreases for a given mass. 
The considered EoSs are compatible with the largest mass observed up to now,
 i.e. $M_G=2.01 \pm 0.04 \,M_{\odot}$ in PSR J0348+0432 \citep{Antoniadis:2013en}, 
and displayed in Fig.\ref{f:MR}, along with the previously observed mass of PSR J1614-2230
\citep{2010Demo} having $M_G=1.97 \pm 0.04 \,M_{\odot}$. 
We also notice that the maximum 
mass calculated with the BCPM and the SLy4  EoSs is characterized by a radius 
of about 10~km, which is somewhat smaller than the radius calculated with the
other considered EoSs. Recent analyses of observations on quiescent low-mass 
X-ray binaries (QLMXB) \citep{Guillot14} and X-ray bursters \citep{Guver13} seem to point 
in this direction, though more studies could be needed \citep{Lattimer14}.
For a NS of 1.5 solar masses, the BCPM EoS predicts a radius of 11.63 km (see 
Table~\ref{Table10}), in line with the recent analysis shown in 
\citep{ozel2015}; see also \citep{Pieka}.
For completeness, we also display in the orange hatched band the probability distribution for
$\rm M_G$  and R deduced from five PRE (Photospheric Radius Expansion) burst sources and five
QLMXB sources, after a Bayesian analysis \citep{Stei2014}.
We see that, except the Shen EoS, all EoSs are compatible with the observational data.
High-precision determinations of NS radii that may be achieved in the future 
by planned observatories such as the Neutron star Interior Composition ExploreR (NICER)
\citep{2014Arzou}, should prove a powerful complement to maximum masses for 
resolving the equation of state of the dense matter of compact stars.

\begin{table}[tb]
\caption{\label{Table10} Properties of the maximum mass configuration for 
a given EoS. The value of the radius R is given, as well as the radius for a 
star of mass equal to 1.5$\rm M_\odot$.}
\centering
\begin{tabular}{cccc}
\hline\hline
EoS & $M_{max}/M_{\odot}$ & R (km) & $\rm R_{1.5}$ (km) \\
\hline
BCPM &1.982 & 9.95 & 11.63 \\
SLy4 & 2.05 & 10. & 11.62 \\
Ska & 2.21 & 10.98 & 12.9 \\
Shen & 2.2 & 12.77 & 14.35 \\
BSk21 & 2.28 & 11.03 & 12.57 \\
\hline
\end{tabular}
\end{table}

\begin{figure}[t]
\centering
\includegraphics[width=9.cm,angle=0,clip]{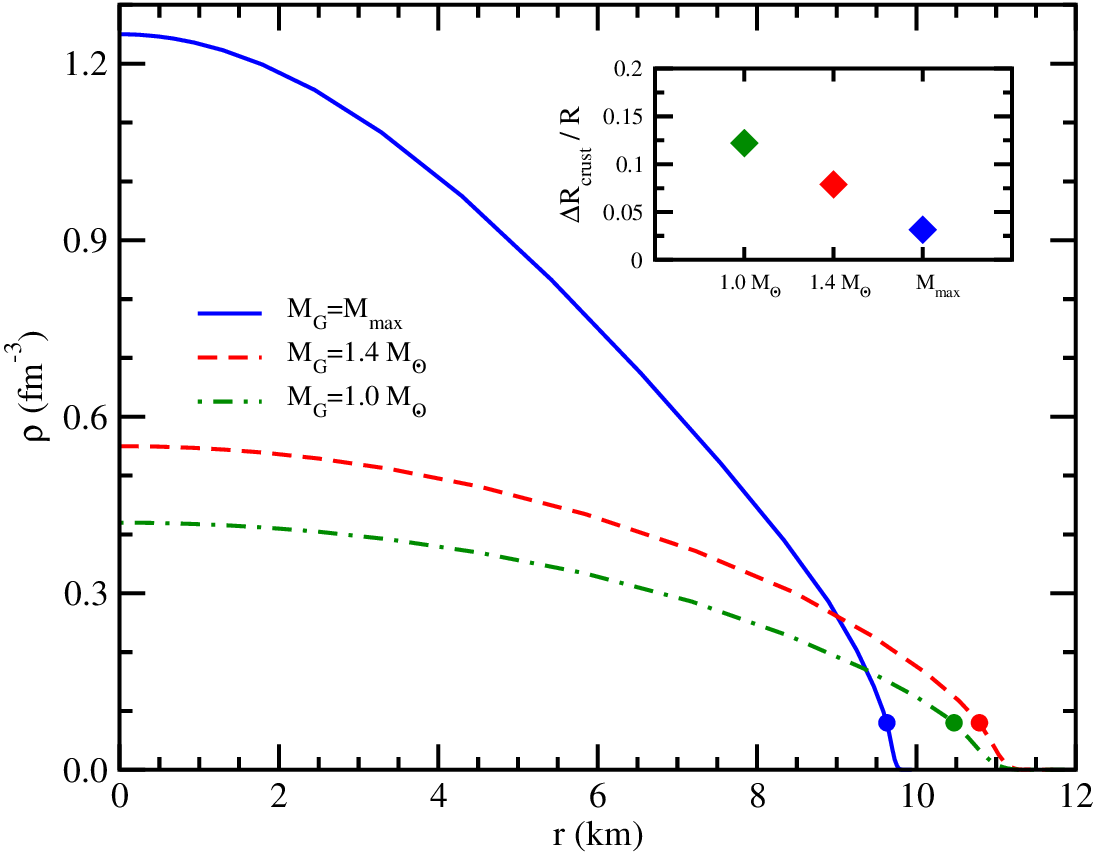}
\caption{\label{f:thick} The density profiles are shown for fixed mass configurations, i.e. 
$\rm M_G=M_{max}$ (solid blue curve), $\rm M_G=1.4~ M_{\odot}$ (dashed red curve), 
and $\rm M_G=1.0~M_{\odot}$ 
(dot-dashed green curve). Full dots indicate the onset of the crust. In the inset the crust thickness
is displayed for each fixed gravitational mass.  See text for details. } 
\end{figure}

In Fig.~\ref{f:thick}  we report the density profiles for three values of the mass calculated
with the EOS of the present work. The transition density between the inner crust and the core 
is indicated by a dot along the curves. Accordingly, in the inset we report the ratio of the crust 
thickness and the total radius of the star. These informations
can be relevant for phenomena occurring in the star, like glitches and deep crustal heating.

\section{Summary and outlook} 
\label{sec:sum}

We have derived a unified equation of state for neutron stars with a microscopic model
which is able to describe on the same physical framework, both  the core and the crust
regions. We describe the neutron star structure based on modern Brueckner-Hartree-Fock 
calculations performed in symmetric and neutron matter. These microscopic calculations
are also the basis of the Barcelona-Catania-Paris-Madrid energy density functional,
devised to reproduce accurately the nuclear binding energies throughout the nuclear 
mass table. This functional is used to describe the finite nuclei present in the crust of neutron
stars. To our knowledge, this is the first time that a whole equation of state directly 
connected to microscopic results has been reported in the literature.   

The equation of state in the outer crust is obtained using the 
Baym-Pethick-Sutherland model, which requires the knowledge of atomic masses near 
the neutron drip line. In our calculation we use the experimental masses, when 
they are available, together with the values provided by a deformed Hartree-Fock-Bogoliubov
calculation performed with the Barcelona-Catania-Paris-Madrid energy density functional
to estimate the unknown masses. 
We find that for average densities above 
$\rmatt \simeq 5\times 10^{10}$ gcm$^{-3}$, where the experimental masses are unknown, 
the composition of the outer crust is similar to the one predicted by the Finite Range 
Droplet Model of M\"oller and Nix.

We describe the structure of the inner crust in the Wigner-Seitz approximation
using the selfconsistent Thomas-Fermi method together with the Barcelona-Catania-Paris-Madrid energy 
density functional. Electrons are considered as a relativistic Fermi gas with a constant 
density, which fill up the whole Wigner-Seitz cell. To obtain the optimal configuration
in a cell for a given average density and size, the energy per baryon is minimized by 
solving self-consistently the coupled Euler-Lagrange equations for the neutron, proton 
and electron densities. To obtain the most stable configuration for a given average 
density, an additional minimization respect to the size of the Wigner-Seitz cell is 
required. 
 
Because the Thomas-Fermi model does not include shell corrections,  
the mass and atomic numbers corresponding to the configuration of minimal energy 
vary smoothly as a function of the average density. For spherical shapes, the mass number 
along the whole inner crust lie in the range between $A$=100 and $A$=800 with a maximum 
around $A$=1100 for an average density $\nb \simeq$ 0.025 fm$^{-3}$. The atomic number 
shows a roughly decreasing tendency from $Z \simeq$30 at the neutron drip up to 
$Z \simeq$25 at the crust core transition.

Using the same Thomas-Fermi model, we have also investigated the possible existence of
pasta phases. To this end we have computed the minimal energy 
per baryon in Wigner-Seitz cells with planar and cylindrical geometries 
for average densities above $\nb \simeq$ 0.05 fm$^{-3}$ and for tube and bubble
configurations from  $\nb$=0.07 fm$^{-3}$ and $\nb$=0.072 fm$^{-3}$, respectively,
until the crust-core transition density reached in our model at $\nb$=0.0825 fm$^{-3}$.
Our model predicts that up to average densities of $\nb =$ 0.067 fm$^{-3}$, spherical nuclei
are the minimal energy configurations. With growing average densities, our model predicts
the successive appearance of rods, slabs, tubes and spherical bubbles as the most stable shape.
   
To describe the core of neutron stars within our model we consider uniform matter containing 
neutrons, protons, electrons and eventually muons in $\beta$-equilibrium, which 
determines the asymmetry of the homogeneous system. To derive the Equation of State
in the core, we use directly the microscopic Brueckner-Hartree-Fock results
in symmetric and neutron matter, which allow us to easily obtain the pressure as 
a function of the density in this regime. The $npe\mu$ model is expected to be valid at least up to 
densities of about three times the saturation density, above which exotic matter could appear.
However, we restricted ourselves to a 
NS nucleonic core and extrapolated the $npe\mu$ matter to higher densities as 
was done in earlier literature \citep{wiringa88,douchin01}.

We have compared the predictions of our Equation of State, derived on a microscopic basis
from the outer crust to the core, with the results obtained using some well known Equations
of State available in the literature. Our Equation of State clearly differs from the results
provided by the Lattimer and Swesty and Shen models, obtained in a more phenomenological way
and widely used in astrophysical calculations.
Our calculation agrees reasonably well in the crust with the predictions of the
Equation of State of Douchin and Haensel and with the results of the BSk21 model of the
Brussels-Montreal group, both based on Skyrme forces but including some microscopic information,
and also shows a remarkable similarity with the former in the core but differs 
more of the latter in this region of the neutron stars.

The mass and radius of non-rotating neutron stars are obtained by solving the 
Tolman-Oppenheimer-Volkov equations. 
Our model predicts a maximal mass of about two solar masses, compatible with the largest
mass measured up to now, and a radius of about 10 km. The radius obtained with our model 
is within the range of values estimated from observations of quiescent low-mass X-ray binaries
and from type I X-ray bursts. The mass-radius relationship computed with our model is
comparable to the results obtained using the Equation of State of Douchin and Haensel above 
the standard mass of neutron stars (1.4 solar masses) and differs from the predictions 
of the BSk21, Lattimer and Swesty and Shen models in this domain of neutron star masses.
The maximal masses of the neutron stars are determined by the stiffness of the Equation of State
at high densities. As can be observed in Figures 14 and 15, where the Equation of State,
the maximal mass and the central density obtained using the different models considered in this work
are displayed, when the stiffness increases the maximum value of the mass increases and, for a fixed mass, 
the central density decreases.

However, there are some theoretical caveats to be considered.
It can be expected that quark matter appears in the centre of massive NS.
To describe these ``hybrid" NS one needs to know the quark matter EOS.
Many models for the deconfined quark matter produce
a too soft EOS to support a NS of mass compatible with observations
\citep{bag1,bag2,NJL,CDM,bag3,FCM,DS}.
The quark-quark interaction in the deconfined phase must be more
repulsive in order to stiffen the EOS, and indeed,
with a suitable quark-quark interaction,
mixed quark-nucleon matter can have an EOS compatible with two solar masses
or more \citep{Alford2013}.
An additional problem arises if strange matter is introduced in the NS matter.
It turns out that BHF calculations using realistic hyperon-nucleon interactions
known in the literature produce a too soft NS matter EOS and the maximum mass
is reduced to values well below the observational limit \citep{H1,H2}.
Although relativistic mean field models can be adjusted to
accomodate NS masses larger than two solar masses \citep{Oertel}, 
and additional terms in the hyperon-nucleon interaction can
provide a possible solution \citep{Rijken}, this ''hyperon puzzle"
has to be considered still open \citep{Haensel}. 
In any case the nuclear EOS with the inclusion of quarks and/or hyperons
at high density must reach a stiffness at least similar to the EOS with 
only nucleons if the two solar masses problem has to be overcome.
         
The full BCPM EoS from the outer crust to the core in a tabulated form as a function of the
baryon density as well as some other useful information is given in the text
as well as supplementary material.
There are different improvements that would be welcome but are left 
for a 
future work. In particular,
we shall take into account shell effects and pairing correlations for protons
in order to obtain a more accurate
information about the compositions of the different WS cells along the inner crust. This can be done 
in a perturbative way on top of the TF results by using different 
techniques such as 
the Strutinsky integral method developed by Pearson and coworkers. 
On the other hand, the self-consistent TF-BCPM
model is well suited for performing calculations at finite
temperature, which can be of interest to investigate the melting point
of the crystal structure. Another natural extension of this work 
using the TF formalism at finite temperature                                                                             
can be to obtain the EoS in the conditions of supernova matter.

\begin{acknowledgements}
We are indebted to Prof. L. M. Robledo for providing the deformed HFB results 
needed in the outer crust calculations.
M. C., B. K. S., and X. V. have been partially supported by Grants No.
FIS2011-24154 and No. FIS2014-54672-P from the Spanish MINECO and FEDER, Grant 
No. 2014SGR-401 from Generalitat de Catalunya, the Spanish Consolider-Ingenio 2010
Programme CPAN CSD2007-00042, and the project MDM-2014-0369 of ICCUB (Unidad de 
Excelencia Mar\'{\i}a de Maeztu) from MINECO.  B. K. S. greatly acknowledges the 
financial support from Grant No. CPAN10-PD13 from CPAN (Spain).  M.B. and G.F.B. warmly
thank Dr. P. Shternin for pointing us out some inaccuracies in the manuscript.
The authors also wish to acknowledge the ``NewCompStar'' COST Action MP1304.
\end{acknowledgements}

\appendix

\section{Expression of the pressure in the inner crust}
\label{sec:press}

In this appendix we derive the expression of the pressure in the model of the
inner crust described in Sec.~\ref{sec:inncrust}. To this end we follow closely
Appendix B of \citep{pearson12}. For the sake of simplicity we work in
spherical symmetry, although the calculation can be extended to the cylindrical
and planar geometries and to tube and bubble configurations as well. The
thermodynamical definition of the pressure allows one to write
\begin{eqnarray}
P = -\left(\frac{\partial E}{\partial V}\right)_{T,A,Z}= 
- \frac{1}{4\pi R_{c}^{2}}
\left(\frac{\partial{E}}{\partial{R_{c}}}\right)_{A,Z} ,
\label{eq21}
\end{eqnarray}
where the volume $V$ is identified with the volume $V_c$ of the WS cell by
treating the inner crust as a perfect crystal. With use of Leibniz's rule for
differentiation of a definite integral, the derivative of the energy
[see ~Eq.~(\ref{eq1})] with respect to the radius $R_c$ of the WS cell is
\begin{eqnarray}
\frac{\partial{E}}{\partial{R_{c}}} &=&
4{\pi}R_{c}^{2} \Big[ {\cal H}\left(\rnum_{n},\rnum_{p}\right)
+ m_{n}\rnum_{n} + m_{p}\rnum_{p} + {\cal E}_{el}\left(\rnum_{e}\right)
\nonumber \\[2mm]
&& \mbox{}
+ {\cal E}_{\rm coul}\left(\rnum_{p},\rnum_{e}\right)
+ {\cal E}_{ex} \left(\rnum_{p},\rnum_{e}\right) \Big]_{r=R_{c}}
\nonumber \\[2mm]
&& \mbox{}
+ 4{\pi}\int_{0}^{R_{c}}\left[\mu_{n}\frac{\partial{\rnum_{n}}}{\partial{R_{c}}}
+ \mu_{p}\frac{\partial{\rnum_{p}}}{\partial{R_{c}}}
+ \mu_{e}\frac{\partial{\rnum_{e}}}{\partial{R_{c}}}\right]r^{2}dr.
\qquad
\label{eq17}
\end{eqnarray}
For the calculation of the pressure this derivative is to be taken at constant
mass and atomic numbers. If the neutron, proton, and electron numbers in the WS 
cell are fixed, we have
\begin{eqnarray}
0 &=& 
\frac{{\partial}}{\partial{R_{c}}}\int_{0}^{R_{c}}4{\pi}r^{2}\rnum_{i}(r) dr 
\nonumber \\
&=& 4{\pi}R_{c}^{2}\, \rnum_{i}(R_{c})+ 4{\pi} \!\int_{0}^{R_{c}}\!
\frac{{\partial\rnum_{i}}}{\partial{R_{c}}} r^{2} dr
\label{eq18}
\end{eqnarray}
for $i = n,p,e$, which in view of Eq.~(\ref{eq17}) allows writing
\begin{eqnarray}
\left(\frac{\partial{E}}{\partial{R_{c}}}\right)_{A,Z} &=&
4{\pi}R_{c}^{2} \Big[ {\cal H}\left(\rnum_{n},\rnum_{p}\right)
+ m_{n}\rnum_{n} + m_{p}\rnum_{p} 
+ {\cal E}_{el}\left(\rnum_{e}\right)
\nonumber \\[2mm]
&& \mbox{}
+ {\cal E}_{\rm coul}\left(\rnum_{p},\rnum_{e}\right)
+ {\cal E}_{ex} \left(\rnum_{p},\rnum_{e}\right)
\nonumber \\[2mm]
&& \mbox{}
-\mu_{n}\rnum_{n}-\mu_{p}\rnum_{p}-\mu_{e}\rnum_{e} \Big]_{r=R_c}.
\label{eq19}
\end{eqnarray}

In the inner crust of a neutron star it is assumed that there are no protons in
the gas of dripped neutrons and, consequently,  $\rnum_{p}\left(R_{c}\right) = 
0$. Therefore, at the edge of the cell we have 
${\cal H}\left(\rnum_{n},\rnum_{p}\right)$ =
${\cal H}\left(\rnum_{n}(R_c),0\right)$
and ${\cal E}_{\rm coul}\left(R_{c}\right)= -\frac{1}{2} \rnum_{e}
\, \left(V_{p}\left(R_{c}\right)-V_{e}\left(R_{c}\right)\right)$.
Charge neutrality implies ${\cal E}_{\rm coul}\left(R_{c}\right)= 0$,
because $V_{p}\left(R_{c}\right) = V_{e}\left(R_{c}\right)$.
Taking into account these constraints, it is easy to show that the nuclear 
contribution to the pressure is provided by only the interacting neutron gas at 
the edge of the WS cell. This neutron gas pressure is given by
$P_{ng}= \mu_n \rnum_n(R_c) - \left[{\cal H}\left(\rnum_{n}(R_c),0\right) + m_n 
\rnum_n(R_c)\right]$.
The variational equation (\ref{eq7}) taken at $r=R_{c}$ implies
$\mu_{e}=\sqrt{k_{Fe}^{2}+m_{e}^{2}}
-\left(\frac{3}{\pi}\right)^{1/3} \!e^2 \rnum_{e}^{1/3}$, and therefore the 
total pressure from the electrons is
$P_{el}= \rnum_{e}\sqrt{k_{Fe}^{2}+m_{e}^{2}}- {\cal E}_{el}(\rnum_{e})
- \frac{1}{4}\left(\frac{3}{\pi}\right)^{1/3} \!e^2 \rnum_{e}^{4/3}$.
We see that the free electron pressure
$P_{el}^{\rm free}=\rnum_{e}\sqrt{k_{Fe}^{2}+m_{e}^{2}}- {\cal 
E}_{el}(\rnum_{e})$
is modified by the electronic Coulomb exchange by an amount
$P_{el}^{ex}= -\frac{1}{4}\left(\frac{3}{\pi}\right)^{1/3} \!e^2 
\rnum_{e}^{4/3}$ (that is, $P_{el}^{ex}= {\cal E}_{el}^{ex}/3$, where 
${\cal E}_{el}^{ex}$ is the contribution to the energy density due to 
electronic Coulomb exchange).

Putting together the previous results, Eq.~(\ref{eq19}) can be written as
\begin{eqnarray}
\left(\frac{\partial{E}}{\partial{R_{c}}}\right)_{A,Z} = 
-4{\pi}R_{c}^{2}\left[P_{ng} + P_{el}^{\rm free} + P_{el}^{ex}\right] .
\label{eq20}
\end{eqnarray}
Thus, comparing with Eq.~(\ref{eq21}), we see that the pressure of the system
takes the form $P = P_{ng} + P_{el}^{\rm free} + P_{el}^{ex}$, as stated in
Sec.~\ref{sec:inncrust}.

\section{The minimization procedure in the inner crust revisited}
\label{sec:press2}

As alluded to in Sec.~\ref{sec:inncrust}, in this appendix we show explicitly
that when the minimization of the energy per unit volume with respect to the 
radius $R_c$ of the WS cell is attained, the Gibbs free energy per particle 
$G/A$ equals the neutron chemical potential $\mu_{n}$. For simplicity we 
restrict the derivation to the case of the spherical shape.

The optimal size $R_{c}$ of the WS cell that minimizes the energy per unit volume 
under the constraints of a given average baryon density $\nb$ and charge
neutrality is the solution of the equation 
\begin{eqnarray}
\frac{d}{d R_c}\bigg[\frac{E}{V}\bigg] =
\frac{1}{V}\bigg[-\frac{dV}{dR_c} \frac{E}{V} +
\frac{\partial E}{\partial R_c}\bigg]=0,
\label{eq23}
\end{eqnarray}
which results in the condition
\begin{eqnarray}
4 \pi R_c^2\, \frac{E}{V} = \frac{\partial{E}}{\partial{R_{c}}} .
\label{eq23b}
\end{eqnarray}
The total number of neutrons, protons, and electrons in the WS cell is given by
\begin{eqnarray}
&&\frac{4 \pi}{3}R_c^3 \nb (1 - x_{p}) = 4 \pi \int^{R_c}_0 \rnum_n(r) r^2 dr, \nonumber \\
&&\frac{4 \pi}{3}R_c^3 \nb x_{p} = 4 \pi \int^{R_c}_0 \rnum_p(r) r^2 dr, \nonumber \\
&&\frac{4 \pi}{3}R_c^3 \nb x_{p} = \frac{4 \pi}{3} R_c^3 \rnum_e,
\label{eq24}
\end{eqnarray}
where $x_p$ is the proton fraction. Next we take the derivative of (\ref{eq24})
with respect to $R_c$. We note that in the present calculation, where we are
looking for the minimum of $E/V$ vs.. $R_c$, the number of neutrons and
protons in the WS cell is not to be assumed fixed, and therefore
Eq.~(\ref{eq18}) does not apply. From the derivative of (\ref{eq24}) with
respect to $R_c$, and recalling Leibniz's rule for differentiation of integrals,
one has
\begin{eqnarray}
&&4 \pi R_c^2 \big[ \nb (1 - x_{p}) - \rnum_n(R_c) \big]
= 4 \pi \int^{R_c}_0 \frac{\partial{\rnum_n(r)}}{{\partial R_c}} r^2 dr, \nonumber \\
&&4 \pi R_c^2 \big[ \nb x_{p} - \rnum_p(R_c) \big]
= 4 \pi \int^{R_c}_0 \frac{\partial{\rnum_p(r)}}{{\partial R_c}} r^2 dr, \nonumber \\
&&4 \pi R_c^2 \big[ \nb x_{p} - \rnum_e \big]
= 4 \pi \int^{R_c}_0 \frac{\partial{\rnum_e}}{{\partial R_c}} r^2 dr.
\label{eq25}
\end{eqnarray}
The expression for ${\partial E}/{\partial R_c}$ has been given in
Eq.~(\ref{eq17}). Using Eq.~(\ref{eq25}) in  Eq.~(\ref{eq17}) and taking into
account that $\rnum_p(R_c)=0$ because there are no drip protons in the gas, one
obtains
\begin{eqnarray}
\frac{\partial{E}}{\partial{R_{c}}} &=&
4\pi R_{c}^{2}\Big[{\cal H}\left(\rnum_{n},0\right) + m_{n}\rnum_{n}
+ {\cal E}_{el}\left(\rnum_{e}\right)
-\frac{3}{4}\left(\frac{3}{\pi}\right)^{1/3} \! \rnum_{e}^{4/3}
\nonumber \\[2mm]
&& \mbox{}
- \mu_n\rnum_{n} + \mu_n\nb 
- \mu_{e}\rnum_{e} - \nb x_{p} (\mu_n - \mu_p -\mu_e) \Big]_{r=R_c}.
\nonumber \\[1mm]
\label{eq26}
\end{eqnarray}
Recalling that the pressure of the system is $P= P_{ng} + P_{el}^{\rm free} +
P_{el}^{ex}$, the result (\ref{eq26}) can be recast as
\begin{equation}
\frac{\partial{E}}{\partial{R_{c}}} =
4{\pi}R_{c}^{2} 
\left[- P + \mu_n \nb - \nb x_{p} (\mu_n - \mu_p - \mu_e) \right].
\label{eq27}
\end{equation}
The minimization condition (\ref{eq23b}) implies
\begin{equation}
4 \pi R_c^2 \frac{E}{V} = 4{\pi}R_{c}^{2}\left[- P + \mu_n \nb -
\nb x_{p} (\mu_n - \mu_p - \mu_e) \right],
\label{eq28}
\end{equation}
and consequently,
\begin{equation}
P + \frac{E}{V} = \mu_n \nb - \nb x_{p} (\mu_n - \mu_p - \mu_e).
\label{eq29}
\end{equation}

When the system is in $\beta$-equilibrium, the chemical potentials fulfill $\mu_n 
- \mu_p - \mu_e=0$, and thus Eq.~(\ref{eq29}) gives
\begin{equation}
PV + E = \mu_n A \;\Rightarrow\;   G = \mu_n A,
\label{eq30}
\end{equation}
which confirms that at the minimum of the energy per unit volume with respect
to the size of the cell, the Gibbs free energy equals the neutron chemical 
potential times the total number of baryons inside the WS cell.

\bibliographystyle{aa} 
\bibliography{crust} 

\end{document}